\documentclass[12pt,cite]{report}
\usepackage{amsmath,amsfonts}
\usepackage{amssymb}
\usepackage{epsfig}
\usepackage{psfig}
\newsavebox{\fmbox}                       

\newcommand\be{ \begin{equation}}
\newcommand\ee{ \end{equation} }
\newcommand\bea{ \begin{eqnarray} }
\newcommand\eea{\end{eqnarray} }
\newcommand{\state}[1]{|#1\rangle}
\def\PRL#1{{ Phys.\ Rev.\ Lett.} {\bf #1}}
\def\PRD#1{{ Phys.\ Rev.\ D} {\bf #1}}
\def\NPB#1{{ Nucl.\ Phys.\ B} {\bf #1}}
\def\PLB#1{{Phys.\ Lett.\ B} {\bf #1}}

{\newcommand{\lsim}{\mbox{\raisebox{-.6ex}{~$\stackrel{<}{\sim}$~}}}
{\newcommand{\gsim}{\mbox{\raisebox{-.6ex}{~$\stackrel{>}{\sim}$~}}}  
\def\sss{\scriptscriptstyle}

\def\GD{\Gamma_{\sss D}}
\def\GS{\Gamma_{\sss S}}
\def\TBL{T_{B-L}}
\def\non{N_1}
\def\Ls{{\sss L}}
\def\sL{\mathcal{L}}
\def\ssM{\scriptstyle M}
\def\ssC{\scriptstyle C}
\def\sC{\mathcal{C}}
\def\dsC{\mathcal{C}^\dagger}
\def\ssS{\scriptstyle S}

\def\sP{\mathcal{ P}}
\def\ev{\,{\rm eV}}
\def\gev{\,{\rm GeV}}

\begin{document}
\pagenumbering{none}
\def\basellinestrech{1.5cm}
\begin{titlepage}
\begin{center}
\vspace{5cm}
{\LARGE \bf{ Bounds on Neutrino Masses from Baryogenesis}}
\vspace{0.8cm}\\
{\LARGE \bf{ in }} \vspace{0.8cm}\\
{\LARGE \bf{ Thermal and Non-thermal Scenarios}} \\
\vspace{1.9cm}
{\em Submitted in partial fulfillment of the requirements\\
\vspace{0.1cm}
for the degree of\\
\vspace{0.1cm}
Ph. D. (Physics)}\\ 
\vspace{1.9cm}
by\\
\vspace{0.1cm}
{\large{\bf Narendra Sahu}} \\
\vspace{0.1cm}
{\it 00412907}\\
\vspace{2cm}
Under the guidance of\\
\vspace{0.2cm}
{\bf Prof.~Urjit.~A.~Yajnik}\\
\vspace{0.6cm}
\centerline{\psfig{file=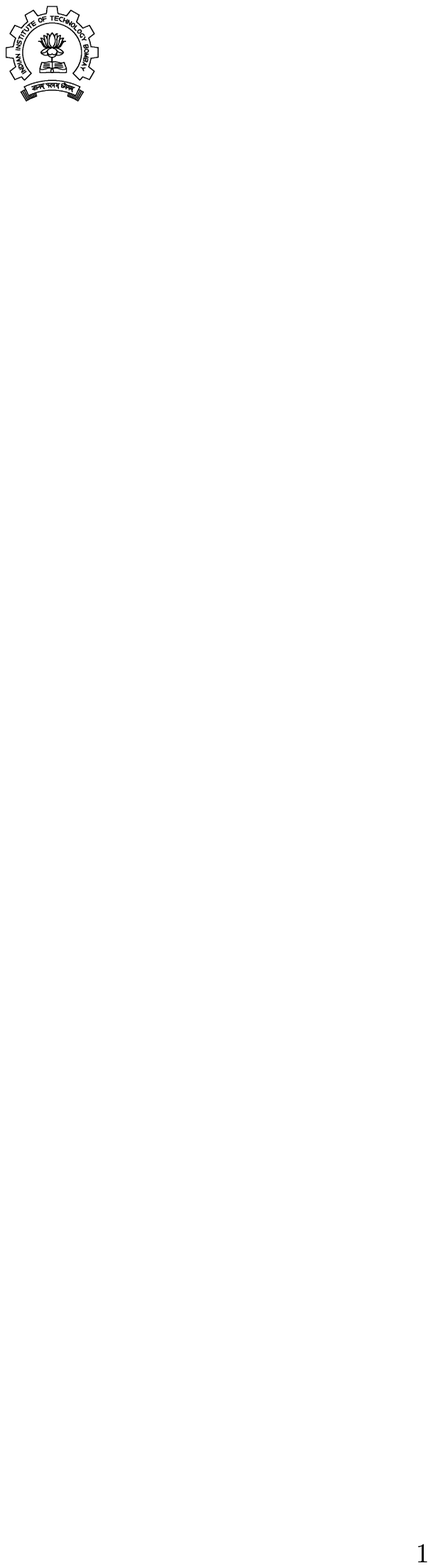,width=0.16\textwidth}} 
\vspace{0.6cm}
{\em Department of Physics}\\
{\em Indian Institute of Technology, Bombay, Mumbai, 400076, India.}
\end{center}
\end{titlepage}
\newpage
\pagenumbering{roman}
\newpage
\begin{center}
{\large {\bf ACKNOWLEDGMENTS}}
\end{center}
\vskip 2\baselineskip
 
\begin{normalsize}
This piece of work is a result of constant guidance and 
support of my supervisor Prof. Urjit A Yajnik. I have 
no words to thank him. Without his help it would not be 
possible to bring up to the present form of the manuscript. 

I am very much indebted to Prof. S Uma Sankar and Prof. P 
Ramadevi for their encouragement and support. Without their 
help it would not be possible to reach at the present status 
of the thesis. 

I am also thankful to Prof. Pijushpani Bhattacharjee 
for his collaboration and allowing me to work at Indian 
Institute of Astro Physics (IIAP). At this juncture, I 
would like to extend my thanks to staffs of IIAP for 
their co-operation and providing me all the facilities to 
work there.

I would like to give my sincere thanks to Prof. R.N. Mohapatra, 
Prof. M.K. Parida, Prof. Utpal Sarkar and Dr. M. Plumacher 
for their helps and suggestions.  

It is my great pleasure to thank Prof. Shiva Prasad, the 
present Head of the Department (HOD) of Physics and 
Prof. S.S. Major, the former HOD of Physics, who have provided 
me a studious atmospehere to work here. 

My sincere thanks to Prof. P.P. Singh, Prof. D.S. Mishra, 
Prof. A. Shukla, Prof. B.P. Singh who have taught me various 
subjects during my course work. I would also like to thank 
Prof. S.H. Patil and Prof. S.N. Bhatia with whom I enjoy 
discussing general physics. 

I would like to thank Ameeya for his helps in learning the 
computational packages, like ``LINUX" , ``LATEX" and ``XMGR". 
Without his helps it would be too difficult for me to learn 
these packages. At this juncture I would like to extend 
my thanks to Ranjit who has helped me a lot in learning 
``FORTRAN". My special thanks to Rabi, Bipin Singh and 
Mohan with whom I enjoy discussing physics of neutrinos. 

I would like to thank my friends Vinod, Joseph, Niharika, Ajay, 
Biswajit and Ramesh who have comeforwarded to share their 
thoughts and providing me the moral supports. At this juncture 
I would like to extend my thanks to my lab-mates Pravina, 
Anjishnu, Ashutosh and Poonam for their helps in many aspects.  

The help and excellent logistic support from the office staff 
of the Department of Physics, especially Mrs. B. Jose and 
Mr. Dilip Kalambate is gratefully acknowledged.

Lastly, but not least, I would like to thank my family members
for their patience and providing me constant support to
work at IIT Bombay.

\vskip 3.5\baselineskip
\begin{flushright} Narendra Sahu \end{flushright}
\end{normalsize}
\newpage
\vspace{5cm}
\begin{center}{\huge {\bf Approval Certificate}}\end{center}
\vspace{1.0cm}
\begin{large}
\hspace{5mm}This is to certify that the thesis entitled 
``{\bf Bounds on Neutrino Masses from Baryogenesis in Thermal 
and Non-thermal scenarios''} by {\bf Narendra Sahu, 00412907} 
is approved for the the degree of Doctor of Philosophy.
\end{large}
\vspace{1.5cm}
\begin{large}
\begin{flushright}
\begin{minipage}{0.3\textwidth}
{\bf Examiners}
\vskip 1.5\baselineskip
\hrule{~}
\vskip 1.5\baselineskip
\hrule{~}
\vskip 1.5\baselineskip  
{\bf Supervisor}
\vskip 1.7\baselineskip
\hrule{~}
\vskip 1.3\baselineskip  
{\bf Chairman}
\vskip 1.5\baselineskip
\hrule{~}
\end{minipage}
\end{flushright}
\vskip 1.5\baselineskip
\begin{flushleft}
\begin{minipage}{0.3\textwidth}

Date: \dotfill
\vskip\baselineskip   
Place: \dotfill

\end{minipage}
\end{flushleft}
\end{large}
\newpage
\begin{center}
{\large {\bf ABSTRACT}}
\end{center}
\vskip 2\baselineskip
                                                                                
\begin{normalsize}

Present low energy neutrino oscillation data are elegantly 
explained by neutrino oscillation hypothesis with very small 
masses ($\leq 1 eV$) of the light neutrinos. These masses can 
be either Dirac or Majorana. Small Majorana masses of the light 
neutrinos, however, can be generated through the seesaw 
mechanism without any fine tuning. This can be achieved 
by introducing right handed 
neutrinos into the electroweak model which are invariant under 
all gauge transformations. The Majorana masses of these right 
handed neutrinos are free parameters of the model and are 
expected to be either at $TeV$ scale or at a higher scale.
This indicates the existence of new physics beyond 
Standard Model ($SM$) at some predictable high energy scale. 

Beyond $SM$ baryogenesis via leptogenesis is an attractive 
scenario that links the physics of right handed neutrino sector 
with the low energy neutrino data. Majorana mass of the neutrino 
violates lepton ($L$) number and thus provides a natural path to 
generate $L$-asymmetry. The leptogenesis occurs via the 
out of equilibrium decay of heavy right handed Majorana 
neutrinos to $SM$ leptons and Higgses. Assuming a normal 
mass hierarchy in the right handed heavy Majorana neutrino 
sector the final $L$-asymmetry is given by the $CP$-violating 
decays of lightest right handed Majorana neutrino, $N_1$. 
A partial $L$-asymmetry is then transformed to the baryon ($B$) 
asymmetry via the non perturbative sphaleron processes. 
 
We divide the thesis into two parts. In part-I, we study 
baryogenesis via leptogenesis in a {\it thermal} scenario, while 
part-II of the thesis is devoted to study the same in a 
{\it non-thermal} scenario. In both scenarios we discuss bounds 
on the mass scale of right handed heavy Majorana neutrinos from 
the leptogenesis constraint. Moreover, we divide the 
phenomenological models into two categories, {\it type-I} and 
{\it type-II}, depending on the seesaw mechanism used to generate 
the light Majorana neutrino masses.

Part-I of the thesis begins with a brief introduction to type-I 
seesaw models. In this model, the Majorana mass matrix of the 
light neutrinos is given by $m_\nu=m_D M_R^{-1}m_D^T$, where $m_D$ 
is the Dirac mass matrix of the neutrinos. On the other hand, in 
the type-II seesaw 
models the Majorana mass matrix of the light neutrinos is given by 
$m_\nu=M_L-m_DM_R^{-1}m_D^T$, where the additional mass $M_L$, in 
contrast to type-I case, is provided by the vacuum expectation value 
of the triplet $\Delta_L$. The two terms, $M_L$ and 
$m_DM_R^{-1}m_D^T$, 
contributing the neutrino mass matrix $m_\nu$ are called type-II and 
type-I respectively. Irrespective of the magnitudes of type-II 
and Type-I terms, it is shown that in a hierarchical mass basis 
of right handed Majorana neutrinos the leptogenesis constraint 
gives rise a lower bound on the mass scale of $N_1$ to be $M_1\geq 
O(10^8)\gev$, assuming that the $CP$-violating decays of $N_1$ 
produces the observed $B$-asymmetry via the leptogenesis route. 
Numerically we check the compatibility of this bound with the 
low energy neutrino oscillation data. 

As a specific example of type-II seesaw models, we consider 
Left-Right symmetric model with spontaneous $CP$-violation. The 
Lagrangian of this model is $CP$-invariant and the Yukawa 
couplings are real. Due to spontaneous breaking of the gauge 
symmetry, some of the neutral Higgses acquire complex vacuum 
expectation values, which lead to $CP$-violation.
In this model, we identify the neutrino Dirac mass matrix with 
that of charged lepton mass matrix. We assume a 
hierarchical spectrum of the right handed neutrino masses and 
derive a bound on this hierarchy by assuming that the decays 
of $N_1$ produces the observed $B$-asymmetry via the leptogenesis 
route. It is shown that the mass hierarchy 
we obtain is compatible with the current neutrino oscillation 
data.

The bound on the mass scale of $N_1$, in production of $L$-asymmetry 
through it's $CP$-violating decays to $SM$ Higgs and lepton, is 
$\geq 10^8 GeV$ which is far above the current accelerator 
energy range and beyond the reach of the next generation 
accelerators. However, these scenarios are well motivated by 
the current status of low energy neutrino oscillation data. An 
elegant alternative is to consider mechanisms which work at 
$TeV$ scale and be consistent with the low energy neutrino data. 
We assume that the required asymmetry of the present Universe 
is produced during the $B-L$ gauge symmetry breaking phase 
transition. Below $T_{\rm B-L}$, the scale of $B-L$ symmetry 
breaking phase transition, the preservation of $L$-asymmetry 
constrains the mass scale of $N_1$, to be $O(10) TeV$ if the Dirac 
masses of the light neutrinos are of $10^2$ order smaller than 
the charged lepton masses. By solving the required Boltzmann 
equations we check the compatibility of $10 TeV$ scale right 
handed neutrino with the low energy neutrino oscillation data. 
We discuss a scenario for the production of large $L$-asymmetry 
during the $B-L$ gauge symmetry breaking phase transition.

In part-II of the thesis, we discuss soliton-fermion systems 
in gauge theories. Solitons emerge as the time independent 
solutions of non linear wave equations in classical gauge 
theories. However, their interactions with fermions lead to 
a curious phenomenon of fractional fermion number. We have 
considered the possibility of fermion fractionization in 
various toy models and its implication for stabilizing otherwise 
metastable solitons. 

A typical solitonic solution in 3+1 dimensional gauge theory 
is cosmic string. It is a 1+1 dimensional extended object. 
During the early Universe phase transitions such objects 
are formed as topological defects. These objects are highly 
{\it non-thermal} and carry a fraction of energy of the 
Universe in their core called false vacuum. The decay of 
these objects produces quanta of massive particles, which 
may survive for long times and hence can provide a link between 
the early Universe and recent cosmology. In particular, we 
study baryogenesis via the route of leptogenesis. 

We study the contribution to the baryon asymmetry of the 
Universe ($BAU$) due to decay of heavy right handed Majorana 
neutrinos released from closed loops of $B-L$ cosmic strings 
in the light of current ideas on light neutrino masses and 
mixings implied by atmospheric and solar neutrino measurements. 
We have estimated the contribution to $BAU$ from cosmic string 
loops which disappear through the process of (a) slow 
shrinkage due to energy loss through gravitational radiation --- 
which we call slow death (SD), and (b) repeated self-intersections 
--- which we call quick death (QD). We find that for reasonable 
values of the relevant parameters, the SD process dominates over 
the QD process as far as their contribution to BAU is concerned. 
It is shown that for the $B-L$ symmetry breaking scale 
$\eta_{\rm B-L}\gsim 1.7\times 10^{11}\gev$ the SD process of 
cosmic string loops contribute significantly to the present $BAU$. 


\end{normalsize}
\vspace{1cm}
\tableofcontents
\listoffigures
\addcontentsline{toc}{chapter}{List of Figures}
\listoftables
\addcontentsline{toc}{chapter}{List of Tables}
\newpage
\pagenumbering{arabic}
\chapter{Introduction}           
One of the challenging problems in theoretical physics is 
{\it baryogenesis}. Since we live in a baryonic Universe it 
is worth investigating. Moreover, baryogenesis plays an 
important role in the interface of particle physics and 
cosmology and thus provides a scope to link them. Though 
there are a few evidences regarding the presence of antibaryons, 
they are tenuous. In particular, the presence of 
antiproton in the cosmic ray shower is one in O($10^{4}$). 

The present Universe is electrically neutral. This is 
an indication of the $U(1)_{em}$ symmetry of the 
present Universe. Therefore, without loss of generality we 
assume it holds since the Big-Bang. If baryon ($B$) number 
and lepton ($L$) number were absolutely 
conserved by all possible interactions occurring in the 
early Universe, then the total $B$ and $L$ numbers of the 
present Universe must simply reflect their apparently 
arbitrarily imposed initial values. A plausible guess 
would be that the initial $B$ and $L$ numbers were exactly 
zero. That means each particle has its own antiparticle carrying 
an equal and opposite quantum number thus maintaining a charge 
neutrality since the Big-Bang. Then the questions arise 
``where are the antibaryons (antimatter) ?'' and  `` why the 
present Universe is baryonic (matter) ?''. 

Assuming a highly symmetric state in the early Universe, a 
matter-antimatter asymmetry can be dynamically generated 
in an expanding Universe if the particle interactions and 
the cosmological evolution satisfy Sakharov criteria~\cite
{sakharov.67}, i.e. 
\begin{itemize}
\item{Baryon number violation.} 
\item{$C$ (charge conjugation) and $CP$ (charge 
conjugation plus parity)-violation.} 
\item{Out of thermal equilibrium.}
\end{itemize} 

Based on these criteria, several mechanisms have been 
put forwarded since late seventies. One of the early proposals 
is Grand Unified Theory ($GUT$)-baryogenesis~\cite{yoshimura,
weinBgen}. Since $B-L$ is a gauge symmetry of most of the 
known $GUT$ models, any asymmetry produced at the $GUT$ scale 
will be erased. Thus the $GUT$ solution of baryogenesis is 
unlikely to be true. During the early Universe phase transitions 
the last opportunity to produce the baryon asymmetry is the 
electroweak ($EW$) phase transition. The strategy is to assume 
a first order phase transition to ensure an epoch of 
non-equilibrium evolution~\cite{cknrev,trodrev,yajbgen}, 
during which the $B$, $C$ and $CP$ violating effects must take 
place, satisfying the Sakharov criteria~\cite{sakharov.67}. 
However, the thermodynamics of $EW$ phase transition indicates 
that a first order phase transition is unlikely~\cite{jansen.96} 
thus making baryogenesis unfeasible within the Standard Model 
($SM$). 

In the thermal era of the early Universe a plausible 
explanation of the observed $B$-asymmetry of the Universe 
($BAU$) is that it arose from a $L$-asymmetry
~\cite{fukugita.86, luty.92, mohapatra.92, plumacher.96}.
The conversion of the $L$-asymmetry to the $B$-asymmetry
then occurs via the high temperature behavior of the $B+L$
anomaly of the $SM$~\cite{krs.86,arn_mac.88,aaps.91}. This 
is an appealing route for several reasons. First, the extremely 
small neutrino masses, suggested by the solar~\cite{solar_data} 
and atmospheric ~\cite{atmos_data} neutrino anomalies and the 
KamLAND experiment~\cite{kamland_data}, point to the possibility of 
Majorana masses for the neutrinos generated by the seesaw mechanism
~\cite{gel-ram-sla} that involves the right handed heavy Majorana 
neutrinos. This suggests the existence of new physics at a 
predictable high energy scale. Since the Majorana mass terms 
violate lepton number they can generate $L$-asymmetry in a natural 
way. Second, most particle physics models incorporating the above 
possibility demand new Yukawa couplings and also possibly scalar 
self-couplings; these are the kind of couplings which, unlike 
gauge couplings, can naturally accommodate adequate $CP$ violation, 
one of the necessary ingredients~\cite{sakharov.67} for generating 
the $BAU$.
 
Most proposals along these lines rely on out-of-equilibrium
decay of the heavy Majorana neutrinos to generate the
$L$-asymmetry. In the simplest scenario a right handed
neutrino per generation is added to the $SM$~\cite{fukugita.86,
luty.92}. They are coupled to left handed neutrinos via Dirac 
charged lepton mass matrix~\cite{gel-ram-sla}.
Since the right handed neutrino is a singlet under $SM$ gauge
group a Majorana mass term ($M_R$) can be added to the Lagrangian.
Diagonalization of neutrino mass matrix leads to two
Majorana neutrino states per generation: a light neutrino
state (mass $\sim m_D^2/M_R$) which is almost left handed
and a heavy neutrino state (mass $\sim M_R$) which is almost
right handed. This is called type-I seesaw mechanism~\cite{
gel-ram-sla} in which there are no Majorana mass terms for the 
left handed fields. The wide class of models in which the light 
neutrinos derive their masses via type-I seesaw mechanism are 
called {\it type-I seesaw models}. In such models the right 
handed neutrinos do not possess any gauge interaction. An 
appealing way to solve this problem is to extend the $SM$ by 
the inclusion of an extra $U(1)_{\rm B-L}$ gauge symmetry~
\cite{plumacher.96}. At a high scale, the singlet Higgs field 
$\chi$ acquires a vacuum expectation value ($VEV$) and thus 
breaking the $B-L$ gauge symmetry. The $VEV$ of $\chi$ provides 
heavy Majorana masses to the singlet right handed neutrinos 
through the Yukawa couplings. However, the $B-L$ gauge symmetry 
in such models is quite ad hoc.

An alternative is to consider models in which the $B-L$ gauge 
symmetry emerges naturally. One of the possibilities is the 
Left-Right symmetric model $SU(2)_L\times SU(2)_R\times 
U(1)_{\rm B-L}$. It can be embedded in the higher gauge 
groups of most of the known Grand Unified Theories (GUTs)~\cite{
slansky_rep}. In such models Majorana masses, $M_L$, for 
left handed neutrinos occur in general, through the 
$VEV$ of the triplet $\Delta_L$~\cite{magg-wet.80,wett.81,
moh-senj.81,laz-shf-wet.81,moha-susy-book.92}. The 
diagonalization of neutrino mass matrix in such models gives 
also a light and a heavy neutrino state per generation. The 
heavy neutrino state has mass $\sim M_R$ but the light neutrino 
mass is $\sim(M_L-m_D^2/M_R)$. The two contributions to the 
light neutrino mass matrix, $m_D^2/M_R$ and $M_L$ are called 
type-I and type-II terms respectively. Such models in which 
both type-I and type-II terms contributing the light neutrino 
mass matrix are called {\it type-II seesaw models}. 
Because of $B-L$ is a gauge charge of such models, no primordial 
$B-L$ can exist. Further, the rapid violation of the $B+L$ 
conservation by the anomaly due to the high temperature sphaleron 
fields erases any $B+L$ generated earlier. Thus the $L$-asymmetry
must have been produced entirely during or after the $B-L$ gauge 
symmetry breaking phase transition.

The goal of the present neutrino oscillation experiments
is to determine the nine parameters in the leptonic mixing 
matrix assuming that the neutrino masses are Majorana in nature. 
The set of parameters include three light neutrino masses, three 
mixing angles and three phases which include one Dirac and two 
Majorana. At present the neutrino oscillation experiments able 
to measure the two mass square differences, the solar and the 
atmospheric, and three mixing angles with varying degrees of 
precision, while there is no information about the phases. The 
Majorana phases can be investigated in neutrinoless double beta decay
experiments, while the Dirac phase can be investigated in the long
base line neutrino oscillation experiments. Moreover, it is
difficult to constrain the parameters of the right handed neutrinos
from the low energy neutrino data. However, several attempts~\cite{
smirnov_jhep} have been made by inverting the neutrino
mass matrix in type-I seesaw models.

Baryogenesis via leptogenesis provides an attractive scenario 
to link the physics of right handed neutrino sector with the low 
energy neutrino data~\cite{lep_phase}. We assume that the 
mass basis of right handed Majorana neutrinos, 
$N_i$, $i=1,2,3,$ is diagonal. In this basis, we further assume that 
the mass spectrum of right handed Majorana neutrinos is in normal 
hierarchy, $M_1<M_2<M_3$. In this scenario, while the heavier 
neutrinos $N_2$ and $N_3$ decay, the lightest right handed Majorana 
neutrino $N_1$ is still in thermal equilibrium. Thus any 
$L$-asymmetry produced by the decay of $N_2$ and $N_3$ is erased 
by the lepton number violating interactions mediated by $N_1$. 
Hence the final $L$-asymmetry is given by the $CP$-violating decays 
of $N_1$. The required $L$-asymmetry constrains the mass scale 
of $N_1$, in the type-I seesaw models, to be $M_{1}\geq 10^{9} GeV$
~\cite{davidson&ibarra.02,buch-bari-plum.02,hamaguchi-etal.02,
hamby.03}. On the other hand, in type-II seesaw models, 
where the Majorana mass term for left handed fields also contribute 
to the neutrino mass matrix, this bound can be reduced by an order 
of magnitude~\cite{antusch.04,sahu&uma_prd.04}. This bound is 
well compatible with the low energy neutrino oscillation data.

As a specific example of type-II seesaw models, we study 
Left-Right symmetric model. In this model, we~\cite{sahu&uma.05} 
consider a special case in which $CP$-asymmetry arises through 
the spontaneous symmetry breaking~\cite{scpv}. 
The Lagrangian of the model is $CP$-invariant which demands that 
all the Yukawa couplings should be real. In this
scenario the vacuum expectation values ($VEV$s) of the neutral
Higgses are complex which lead to $CP$-violation. In the 
Left-Right symmetric model, there are four complex neutral scalars
which acquire $VEV$s. However, the unbroken global $U(1)$ symmetries
associated with $SU(2)_L$ and $SU(2)_R$ gauge groups allow
two of the phases to be set to zero. Using the remnant $U(1)$ 
symmetry after the breaking of $SU(2)_R$, one phase choice is 
made to make the $VEV$ of $\Delta_R$, and hence the mass matrix 
of right handed neutrinos, real. The phase associated with the 
other $U(1)$ symmetry can be chosen to achieve two different 
types of simplification of neutrino mass matrix. In the
{\it type-II choice}, the $m_\nu^I$ is made real leaving
the $CP$-violating phase purely with $m_\nu^{II}$.
In this phase convention, we derive a lower bound on the mass
scale of $N_1$ from the leptogenesis constraint by assuming a
normal mass hierarchy in the right handed neutrino sector. 
It is shown that the mass scale of $N_1$ satisfy the
constraint $M_1\geq 10^8 GeV$~\cite{sahu&uma.05}, which is in
good agreement with the lower bound on $M_1$ in generic type-II
seesaw models~\cite{antusch.04,sahu&uma_prd.04}. In the 
{\it type-I phase choice}, only the type-I term contains 
$CP$-violating phase leaving type-II term real. This allows us 
to derive an upper bound on the heavy neutrino mass hierarchy 
from the leptogenesis constraint. In order to achieve the
observed baryon asymmetry of the present Universe, it is
found that the mass hierarchy of right handed neutrinos
satisfy the constraint $M_2/M_1\leq 17$ and $M_3/M_1\leq 289$
simultaneously~\cite{sahu&uma.05}. Numerically we verified that
these bounds are compatible with the low energy neutrino
oscillation data for all values of $M_1\geq 10^8 GeV$ as implied
by the lower bound on $M_1$ in type-II phase convention.

The bound on $M_1$, in production of $L$-asymmetry
through the $CP$-violating decays of thermally generated $N_1$,
is $\geq 10^8 GeV$ which is far above the current accelerator
energy range and beyond the reach of the next generation
accelerators. However, these scenarios are well motivated by
the current status of low energy neutrino oscillation data.
An alternative is to consider mechanisms which  work at the 
$TeV$ scale and may rely on the new particle content implied in
supersymmetric extensions of the $SM$~\cite{susy_tev_group}. 
The Minimal 
Supersymmetric $SM$ ($MSSM$) holds only  marginal possibilities 
for baryogenesis.
The Next to Minimal or $NMSSM$ possesses robust mechanism for
baryogenesis~\cite{huber&schimdt.01} however the model has
unresolved issues vis a vis the $\mu$ problem due to domain
walls~\cite{abel_sar_white.95}. However its restricted version,
the $nMSSM$ is reported~\cite{pan_tam_PLB1.99,pan_tam_PLB2.99,
pan&pil_PRD.01,ded_hug_mor_tam_PRD.01,men_mor_wag_prd.04}
to tackle all of the concerned issues.

It is worth investigating other possibilities, whether or not
supersymmetry is essential to the mechanism. We therefore 
studied~\cite{sahu&yajnik_prd.04} the consequence of heavy 
Majorana neutrinos given the current knowledge of light neutrinos. 
The starting point is the observation~\cite{har&tur.90,fglp.91} 
that the heavy neutrinos participate in the erasure of 
any pre-existing asymmetry through scattering as well as decay 
and inverse decay processes. Estimates using general behavior of 
the thermal rates lead to a conclusion that there is an upper 
bound on the temperature $T_{\rm B-L}$ at which $B-L$ asymmetry 
could have been created. This bound is
$T_{\rm B-L}\lsim 10^{13}GeV \times(1 eV/m_\nu)^2$,
where $m_\nu$ is the typical light neutrino mass. We extend this
analysis by numerical solution of the Boltzmann equations and
obtain regions of viability in the parameter space spanned by
$\tilde{m}_1$-$M_1$, where $\tilde{m}_1$ is called {\it effective 
neutrino} mass parameter. We find that our results are in 
consonance with~\cite{fglp.91}
where it was argued that scattering processes provide a weaker
constraint than the decay processes. If the scatterings
become the main source of erasure of the primordial asymmetry
then the constraint can be interpreted to imply  $\TBL<M_1$.
Further, this temperature can be as low as $TeV$ range with
$\tilde{m}_1$ within the range expected from neutrino
observations. This is compatible with  seesaw mechanism if
the ``pivot" mass  scale is two order smaller than that 
of the charged leptons. We conjecture that the hypothesis of 
$TeV$ scale right handed neutrino can be verified in near future 
and hence an indirect evidence of generating the baryon asymmetry 
at $TeV$ scale~\cite{sahu&yajnik_prd.04}. 

Part-II of this manuscript is devoted to study the formation and 
evolution of topological defects~\cite{vilen&shell}, in 
particular cosmic strings, in the early Universe. The 
corresponding consequences have been studied in details 
along the direction of baryogenesis via the route of 
leptogenesis. 

Topological defects arise as the solitonic solutions in gauge 
theories. `Solitons' or `solitary waves' are the time independent 
solutions of non-linear wave equations in classical field theories. 
The prime among them is $\lambda \phi^4$ theory. In 1+1-dimensions 
the solitonic solutions in $\lambda \phi^4$ theory are called 
`kinks'. Note that, kink is purely a solitary wave but not a 
soliton. However, in field theory the distinction between them 
is completely blurred. Each solitonic solution is designated by 
a number called the `topological charge' or `winding number'. 
The topological charge is nothing but the boundary conditions 
imposed on the field which is conventionally different from the 
Noether charge comes from the continuous global symmetry associated 
with the theory. 

In Quantum Field Theory ($QFT$), solutions of Dirac equation in the 
presence of solitonic objects lead to a curious phenomenon of 
`fractional fermion number'~\cite{jackiw&rebbi.76}. This is 
because of the existence of degenerate zero energy modes of 
fermions while quantized in the background of a solitonic 
vacuum. In contrast to it, in the 
translational invariant vacuum there is no zero energy 
solutions of Dirac equation and therefore fermions are quantized 
by integral unit. The fractionally charged solitonic states are 
therefore superselected from the normal vacuum and are not 
allowed to decay in isolation. Note that the fermion number we are 
talking about is the eigenvalue of the number operator in $QFT$. 

An inevitable feature of the early Universe phase transitions 
is the formation of topological defects~\cite{vilen&shell}. 
In particular we shall deal with cosmic strings. The breakdown 
of any $U(1)$ gauge symmetry to $\mathbb{Z}$ ensures the 
formation cosmic strings since $\pi_1\left(U(1)/\mathbb{Z}\right)
=\pi_0(\mathbb{Z})=\mathbb{Z}$. These defects are extended 
objects and are not distributed thermally. Therefore, 
the decay of these objects can be a {\it non-thermal} source of 
massive particles. Moreover, the cosmic strings formed at a 
phase transition can also influence the nature 
of a subsequent phase transition that may have important 
implications for the generation of $BAU$~\cite{sbdanduay1, 
brandenetal}. 

An important feature of these cosmic strings is that during their
formation they trap zero modes of fermions~\cite{jackiw.81}
which are well predicted~\cite{weinberg.81}. These fermionic 
zero modes induce fractional fermion number ($|n|/2$), where 
$n$ is the winding number of a string. If $n$ is odd then the 
induced fermion number on the string is half-integral. Therefore, 
it is superselected~\cite{sahu&yajnik_plb.04} from the translation 
invariant vacuum where the eigenvalues of the number operator 
possesses integral fermion number. So a string of half integer 
fermion number can not decay in isolation. However, this 
conclusion may not be true for closed loops which are chopped 
off from the infinite straight strings. This remains an 
important open question for cosmology. 

There exist both analytical as well as numerical studies of
the evolution of cosmic string networks in the early Universe. 
These suggest that the string network quickly enters a 
scaling regime in which the energy density of the strings scales 
as a fixed fraction of the energy density of radiation in the 
radiation dominated epoch or the energy density of matter in the 
matter dominated epoch. In both cases the energy density scales 
as the inverse square of cosmic time $t$. In this regime one of the 
fundamental physical process that maintains the strings network 
to be in that configuration is the formation of sub-horizon size
{\it closed loops} which are pinched off from the network whenever 
a string segment curves over into a loop, intersecting itself. 

In many scenarios~\cite{pijush.82,brandenberger&co.91,
riotto&lewis.94,jeannerot.96, pijush.98} it has been studied 
that the decaying, collapsing, or repeatedly self-intersecting 
closed loops of such cosmic strings provide a non-thermal source 
of massive particles that ``constitute'' the string. The decay of these 
massive particles give rise to the observed $B$-asymmetry or at 
least can give a significant contribution to it. 
Assuming that the final demise of each string loop produces 
$O(1)$ right handed neutrino, $N_1$, the observed $B$-asymmetry
~\cite{bha_sahu_yaj_prd.04} requires the constraint on its mass 
$M_1\leq 2.4 \times10^{12}
\left(\eta_{\rm B-L}/10^{13}\gev\right)^{1/2}\gev$, $\eta_{\rm B-L}$ 
being the scale of $U(1)_{\rm B-L}$ gauge symmetry breaking phase 
transition. Here we have assumed that at a temperature above the 
mass scale of $N_1$ there is no lepton asymmetry. A net 
asymmetry has been produced just below the mass scale of 
$N_1$ by it's $CP$-violating decay to $SM$ Higgs and lepton. 
In order to take into account the wash out effects we solve the 
required Boltzmann equations~\cite{luty.92,plumacher.96,
buch-bari-plum.02,strumia.04} by including the right handed 
neutrinos of cosmic string origin as well of thermal origin. 
It is shown that the delayed decay of cosmic string loops over 
produce the baryon asymmetry of the Universe for 
$\eta_{\rm B-L}>O(10^{11})\gev$~\cite{sahu_bha_yaj_prep}. This, 
On the other hand, gives an upper bound on the CP-violating 
phase to produce the required asymmetry of the present Universe.

The rest of the manuscript is organized as follows. In 
chapter 2, we briefly discuss the thermal baryogenesis via the 
route of leptogenesis in type-I seesaw models. In chapter 3, 
thermal baryogenesis via the route of leptogenesis in type-II 
seesaw models is discussed in detail. In both cases we find 
that the scale of operation of leptogenesis should 
be very high, of the order $O(>10^8)\gev$. In chapter 4, we 
propose the possibility that $TeV$ scale masses for the 
right handed heavy neutrinos are consistent with the seesaw 
and may yet suffice to explain the baryon asymmetry. A 
relevant model is also summarized. The second part of the 
thesis is devoted to study the formation and evolution 
of topological defects in the early Universe. In chapter 5, we 
briefly introduce the soliton-fermion systems in $QFT$. 
We than discuss the consequences of quantization of fermions 
in the background of solitons. The same hypothesis is 
extended to the case of cosmic strings in chapter 6. In 
chapter 7, evolution of $B-L$ cosmic strings 
is discussed in greater detail. In a scaling regime the decay 
of the massive particles emitted from the cosmic string loops 
produce a baryon asymmetry via the leptogenesis route. The 
consequences for the energy scale of leptogenesis is discussed. 
The thesis ends by a summary and conclusions in chapter 8. Through 
out this manuscript we use natural units and set $c=\hbar=k_B=1$. 
\newpage
~~
\vskip 7\baselineskip
\begin{LARGE}
\pagestyle{empty}
\begin{center}
{\bf Part-I}
\vskip1cm
{\bf Baryogenesis via Leptognesis\\ 
\vskip2mm
{\bf in}\\ 
\vskip2mm
{\bf Thermal Scenario}}
\end{center}
\end{LARGE}
\addcontentsline{toc}{chapter}{Part I Baryogenesis via Leptogenesis 
in Thermal Scenario}

\chapter{Thermal leptogenesis in type-I seesaw models and 
bounds on neutrino masses}           
\section{Heavy Majorana neutrinos: The physics beyond standard model}
Within the $SM$ the left handed neutrinos are massless. 
However, the present evidence from the neutrino oscillation 
experiments~\cite{solar_data,atmos_data,kamland_data}
suggests that the left handed neutrinos possess small masses 
($\leq 1eV$). These masses can be either Dirac or Majorana. 
Small Majorana masses of the light neutrinos, however, can be 
generated via seesaw mechanism~\cite{gel-ram-sla} that involves 
the right handed heavy Majorana neutrinos. This indicates the 
existence of new physics beyond $SM$ at a predictable high 
energy scale.

In the simplest scenario a massive right handed neutrino 
$N_R$ of mass $M_R$ per generation is added to the $SM$. They 
are coupled to the left handed neutrinos via Dirac mass 
matrix $m_D$. Since the right handed neutrino is a singlet 
under $SM$ gauge group, $N_R-N_R$ coupling can be added to 
the $SM$ Lagrangian. This gives a Majorana mass to the right 
handed neutrino. On the other hand, the $\nu_L-\nu_L$ coupling 
is not allowed by the $SM$ Lagrangian as it violates the 
lepton number by two units. Therefore, the upper left $3\times 3$ 
block of the $6\times 6$ neutrino mass matrix is zero; see for 
example section 2.2. The diagonalization of the neutrino mass matrix 
thus leads to two Majorana neutrino states per generation: a light 
neutrino state of mass $\sim m_{D}^{2}/M_{R}$ which is almost left 
handed and a heavy neutrino state of mass $\sim M_{R}$ which is 
almost right handed. This is called type-I seesaw mechanism. The 
class of models in which the Majorana masses of the light 
neutrinos are obtained via this mechanism are called {\it 
type-I seesaw models}. 

\section{Type-I seesaw mechanism and neutrino masses}
To generate the light neutrino masses via type-I seesaw 
mechanism we add a massive right handed neutrino per generation 
to the $SM$ Lagrangian. For three generation of neutrinos, the 
terms in the Lagrangian for the massive right handed Majorana 
neutrinos are taken to be 
\begin{equation}
\mathcal{L}_{N_R}=i\bar{N}_{Ri}\gamma^{\mu}\partial_{\mu}N_{Ri}+
\left[ \bar{\nu}_{Li} m_{Dij} N_{Rj}+\frac{1}{2} 
\bar{N^c}_{Ri} M_{Rij} N_{Rj} +H.C.\right],
\end{equation}   
where $i,j=1,2,3$ are family indices. Without loss of generality
we can choose a basis in which the right handed Majorana neutrinos
are diagonal. In this basis the neutrino mass matrix in $6\times 
6$ block is given as   
\begin{equation}\begin{pmatrix}
0 & m_D\\
m_D^T & M_R\end{pmatrix}.
\label{mass-matrix}
\end{equation} 
Diagonalizing the neutrino mass matrix (\ref{mass-matrix}) 
into $3\times 3$ blocks we get the mass matrix for the light 
neutrinos to be 
\begin{eqnarray}
m_{\nu} &=& -m_D\frac{1}{M_R} m_D^T\nonumber\\
&=& -v^2 h\frac{1}{M_R} h^T,
\label{see-saw-1}
\end{eqnarray}
where $v$ is the vacuum expectation value of $SM$ Higgs and $h$ 
is the relevant Yukawa coupling matrix. From equation 
(\ref{see-saw-1}), it is clear that larger the mass scale of the 
right handed neutrinos the smaller is the mass scale of the left 
handed neutrinos. Diagonalization of the light neutrino mass matrix 
$m_{\nu}$, through the lepton flavor mixing matrix 
$U_{PMNS}$~\cite{mns-matrix}, gives us three light Majorana 
neutrino states. The eigenvalues are   
\begin{equation}
U_L^{\dagger} m_{\nu} U_L^*=-dia (m_1, m_2, m_3)\equiv -D_m,
\label{diag-1}
\end{equation}
where $m_1$, $m_2$, $m_3$ are chosen to be real. Combining 
equations (\ref{see-saw-1}) and (\ref{diag-1}) we get the 
diagonal mass matrix 
\begin{equation}
D_m=v^2 U_L^{\dagger}h\frac{1}{M_R}h^T U_L^*.
\label{Dm-matrix}
\end{equation}

The smallness of the light neutrino masses imply that the mass 
scales of right handed neutrinos, $N_1, N_2$ and $N_3$ exist at 
a high scale and beyond the energy range of current accelerators. 
However, their decay to $SM$ particles may have significant 
consequences. In particular, we consider baryogenesis via the 
route of leptogenesis through the out-of-equilibrium decay of the 
right handed heavy Majorana neutrinos to $SM$ particles. 

\section{Thermal leptogenesis in type-I seesaw models}
In thermal scenario we assume that the out-of-equilibrium 
decay of the heavy right handed Majorana neutrinos to $SM$ 
leptons and Higgses produce a net lepton asymmetry
~\cite{fukugita.86}. We further assume a normal mass hierarchy, 
$M_1 \ll M_2 < M_3$ among the right handed Majorana neutrinos 
$N_1, N_2$ and $N_3$. In this scenario, while the heavier 
neutrinos $N_2$ and $N_3$ decay, the lightest right handed 
neutrino $N_1$ is still in thermal equilibrium. Any $L$-asymmetry 
thus produced by the decay of $N_2$ and $N_3$ is erased by 
the lepton number violating interactions mediated by $N_1$. 
Therefore, it is reasonable to assume that the final $L$-asymmetry 
is produced by the out-of-equilibrium decays of $N_1$ only. A 
partial $L$-asymmetry is then transformed to $B$-asymmetry via the 
high temperature sphalerons which are in equilibrium at a 
temperature $10^{12}GeV$ to $10^{2}GeV$. Below the electroweak 
phase transition ($T_{ew}\sim 100 GeV$), the sphaleron transitions 
fall quickly out of thermal equilibrium. Therefore, the 
B-asymmetry produced until at the scale of $100 GeV$ is the final 
$B$-asymmetry of the universe that is observed today. 

\subsection{Upper bound on $CP$-asymmetry} 
Beyond $SM$ all the processes mediated by right handed 
neutrino, $N_R$ naturally violate lepton number. 
Out of which the dominant channel is the decays of $N_R$ 
to $SM$ lepton ($\ell$) and Higgs ($\phi$) through the 
Yukawa coupling. The required Lagrangian for the coupling is 
given by
\be
\mathcal{L}_{Y}=h_{ij}\bar{\ell}_i\phi N_{Rj}+h.c.\,,
\label{yukawa_term}
\ee
where $h_{ij}$ is the Yukawa coupling matrix, and $i,j=1,2,3$
for three flavors.
 
We shall work in a basis in which the Majorana mass matrix 
of right handed neutrinos is diagonal. In this basis the 
right handed Majorana neutrino is given by 
$N_{j}=N_{Rj}\pm N^{c}_{Rj}$, which satisfies $N_{j}^{c}=
\pm N_{j}$. The type-I seesaw mechanism then gives the 
corresponding light neutrino mass eigenstates $\nu_1$, $\nu_2$, 
$\nu_3$ with masses $m_1$, $m_2$, $m_3$, respectively; these are 
mixtures of flavor eigenstates $\nu_e$, $\nu_\mu$, $\nu_\tau$. 

The decays of the heavy right handed Majorana neutrino can 
create a non-zero $L$-asymmetry only if their decay 
violates $CP$. The $CP$-asymmetry parameter in the decay 
of $N_j$ is defined as
\begin{equation}
\epsilon_{j}=\frac{\Gamma(N_j\rightarrow \ell\phi)-
\Gamma(N_j\rightarrow \ell^c\phi^c)}{\Gamma(N_{j}
\rightarrow \ell\phi)+\Gamma(N_{j}\rightarrow \ell^{c}
\phi^{c})}\,.
\label{epsilon_def-1}
\end{equation} 
In the hierarchical scenario it is reasonable to assume that 
the decays of $N_1$ is the leading process that produces the 
final $L$-asymmetry. In this scenario, equation 
(\ref{epsilon_def-1}) gives rise to the $CP$-asymmetry, obtained 
by superposing the tree diagram with the one loop radiative 
correction~\cite{fukugita.86} and self energy correction diagrams~\cite{
selfenergy_group}, to be  
\begin{figure}[h]
\begin{center}
\epsfig{file=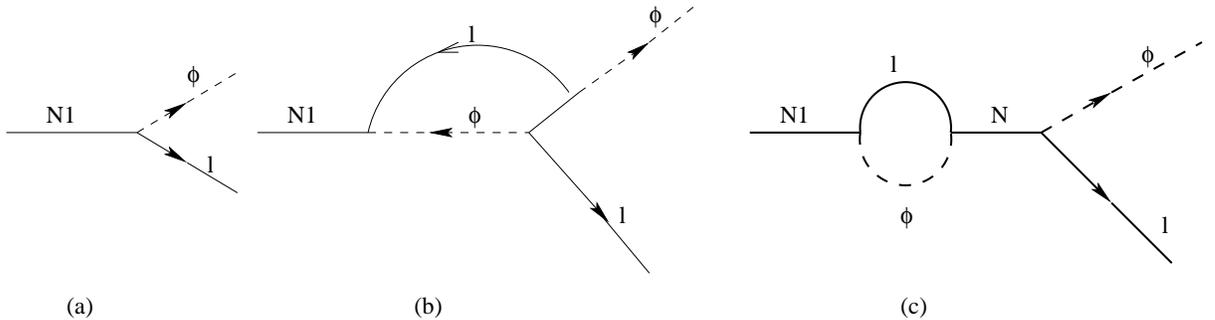, width=1.0\textwidth}
\caption{(a)tree level (b) one loop correction (c) self-energy 
correction diagrams for the decay process of $N_1$.}
\end{center}
\end{figure}                                                          
\begin{equation}
\epsilon_{1}=\frac{1}{8\pi\left[h^{\dagger}h\right)_{11}}
\sum_{j}Im\left[(hh^{\dagger})_{1j}^{2}\right]
g\left(\frac{M_{j}^{2}}{M_{1}^{2}}\right),
\label{cpasym1-1}
\end{equation}
where $g(x)=-3/2\sqrt{x}$ in the limit $x=(M_j^2/M_1^2)\gg 1$ 
(which indicates a hierarchy among right handed neutrinos). In this 
approximation equation (\ref{cpasym1-1}) translates to
\begin{eqnarray}
\epsilon_{1} &\simeq& -\frac{3M_{1}}{16\pi[h^{\dagger}h]_{11}}
\sum_{j}Im\left[(h^{\dagger}h)^2_{1j}\frac{1}{M_{j}}
\right]\nonumber\\
 &\simeq& -\frac{3M_{1}}{16\pi[h^{\dagger}h]_{11}}
\sum_{j=2,3}Im\left[(h^{\dagger}h)_{1j}\left(\frac{1}{M_R}\right)
_{jj}(h^Th^*)^T_{1j}\right].
\label{cpasym2-1}
\end{eqnarray} 
Using (\ref{see-saw-1}) and (\ref{Dm-matrix}) in equation 
(\ref{cpasym2-1}) we get 
\begin{eqnarray}
\epsilon_1 &\simeq & -\frac{3M_{1}}{16\pi v^2 [h^{\dagger}h]_{11}}
Im (h^\dagger U_L D_m U_L^T h^*)_{11}\nonumber\\
&=& -\frac{3M_{1}}{16\pi v^2 [h^{\dagger}h]_{11}} 
Im\sum_{i} \left[ (h^\dagger U_L)_{1i} (D_m)_{ii} 
(U_L^T h^*)_{i1}\right]\nonumber\\
&=& \frac{3M_{1}}{16\pi v^2 [h^{\dagger}h]_{11}}\left[
m_1 Im \left[(U_L^{\dagger}h)_{11}\right]^2+\sum_{i\neq 1}m_i 
Im \left[ (U_L^{\dagger} h)_{i1}\right]^2\right].
\label{cpasym3-1}
\end{eqnarray}    
We now define \cite{davidson&ibarra.02,buch-bari-plum.02}
\begin{equation}
R=v D_{m}^{-1/2} U_L^{\dagger}h D_M^{-1/2},
\label{orthogonal-matrix}
\end{equation}
such that $RR^{T}=1$. This implies that $Im(RR^{T})_{11}=0$.
Then it is straightforward to show that
\begin{equation}
\frac{1}{m_{1}}Im (U_L^\dagger h)_{11}^{2}=-Im \sum_{i\neq 1}
\frac{1}{m_{i}}Im (U_L^\dagger h)_{i1}^{2}.
\label{real-im-split-1}
\end{equation}
Using (\ref{real-im-split-1}) in equation (\ref{cpasym3-1}) we get 
\begin{equation}
\epsilon_1 = \frac{3M_{1}}{16\pi v^2 [\tilde{h}
^{\dagger}\tilde{h}]_{11}} \sum_{i\neq 1}\frac{(m_i^2-m_1^2)}{m_i}
Im (\tilde{h}_{i1})^2,
\label{cpasym3-2}
\end{equation}
where $\tilde{h}=U_L^{\dagger}h$. 

The recent atmospheric neutrino data~\cite{atmos_data} indicates 
$\nu_{\mu}\leftrightarrow \nu_{\tau}$ oscillation with nearly 
maximal mixing($\theta_{atm}\simeq 45^{\circ}$) and a 
mass-squared-difference 
\be
\Delta m^2_{atm}\equiv |m_3^2-m_2^2| \approx 2.6\times 10^{-3}eV^2.
\ee 
We assume a normal hierarchy among the light neutrino mass 
eigen states. In this scenario the atmospheric neutrino mass 
is 
\be
\sqrt{\Delta m^2_{atm}}= m_3 \approx 0.05 eV.
\label{atmo-mass}
\ee
Using (\ref{atmo-mass}) in equation (\ref{cpasym3-2}) we get 
the upper bound on the $CP$-asymmetry parameter to be 
\begin{equation}
|\epsilon_1|\leq \frac{3M_{1}}{16\pi v^2}\sqrt{\Delta m_{atm}^2}.
\label{cp-u-bound}
\end{equation}

\subsection{Analytical estimation of $L$-asymmetry and lower bound 
on the mass of lightest right handed neutrino}
In this section, we analytically estimated the $L$-asymmetry 
from the out-of-equilibrium decays of $N_1$. We assume that 
above the mass scale of $N_1$, all the lepton violating processes 
mediated by $N_1$ are fast enough to bring it in thermal 
equilibrium. Hence there is no net $L$-asymmetry. Below the mass 
scale of $N_1$ all these processes fall out of equilibrium. Any 
$L$-asymmetry thus produced by the decays of $N_1$ is not erased. 

In a comoving volume, the net $L$-asymmetry is defined by 
\begin{equation}
Y_{L}=\frac{n_{N_R}}{s}\epsilon_1 \delta,
\label{l-asymmetry}
\end{equation}
where $s=(2\pi^2/45)g_* T^3$ is the entropy density at any 
epoch of temperature $T$, $\epsilon_1$ is the $CP$-violating 
parameter and $\delta$ is the wash out factor due to the 
$L$-violating processes mediated by $N_1$ at a temperature 
$T\leq M_1$. A part of the produced $L$-asymmetry is then 
transferred to $B$-asymmetry via the 
sphaleron transitions and is given by~\cite{har&tur.90} 
\begin{equation}
Y_B=\frac{p}{p-1}Y_L,~~ 
\mathrm{with}~~ p=\frac{8N_F+4N_{\phi}}{22N_F+13N_{\phi}},
\label{b-asymmetry}
\end{equation}
where $N_F$ is the number of generation for right handed neutrinos 
and $N_{\phi}$ is number of doublet Higgs. For three generation of 
right handed neutrinos, $p=28/79$. Using (\ref{l-asymmetry}) and 
(\ref{cp-u-bound}) in equation (\ref{b-asymmetry}) we get a 
bound on the net $B$-asymmetry to be
\be
Y_B \leq \frac{p}{p-1}\frac{n_{N_R}}{s}\frac{3M_{1}}
{16\pi v^2}\sqrt{\Delta m_{atm}^{2}} \delta. 
\label{b-asy-ineq}
\ee

Recent observations from Wilkinson Microwave Anisotropy 
Probe ($WMAP$) show that the matter-antimatter asymmetry in 
the present Universe measured in terms of $\frac{n_{B}}
{n_{\gamma}}$ is~\cite{spergel.03}
\be
\left(\frac{n_{B}}{n_{\gamma}}\right)_{0}\equiv
\left(6.1^{+0.3}_{-0.2}\right)\times 10^{-10},
\label{b-asy-wmap}
\ee
where the subscript $0$ presents the matter-antimatter 
asymmetry that is observed today. 

To compare with the observed value, we now recast the bound 
on $B$-asymmetry (\ref{b-asy-ineq}) in terms of $(n_B/n_\gamma)$. 
This is given to be   
\be
(n_B/n_\gamma) = 7Y_B \leq  7\frac{p}{p-1}\frac{n_{N_R}}{s}
\frac{3M_{1}}{16\pi v^2}\sqrt{\Delta m_{atm}^{2}} \delta.
\label{b-asy-ineq-1}
\ee
Comparing (\ref{b-asy-wmap}) and (\ref{b-asy-ineq-1}) 
we get the constraint on the mass scale of $N_1$ to be 
\begin{equation}
M_1\geq 0.84\times 10^9 GeV \left(\frac{\eta_B}{6.4\times 10^{-10}}
\right) \left(\frac{10^{-3}}{\frac{n_{N_R}}{s}
\delta}\right)\left(\frac{v}{174GeV}\right)^2\left(\frac{0.05eV}
{\sqrt{\Delta m_{atm}^2}}\right),
\label{lower-bound-M1}
\end{equation} 
where we have used $v=174 GeV$, the scale of electroweak phase 
transition. Other physical quantities, the atmospheric neutrino 
mass and the observed baryon asymmetry, are normalized with respect 
to their observed values. 

\subsection{Numerical estimation of $L$-asymmetry}
The analytical estimation in section 2.3.2 shows that in a 
thermal scenario to create a successful $L$-asymmetry 
from the decay of $N_1$ it is required that 
$M_1\geq O(10^{9})GeV$. We now check the compatibility of 
this bound on $M_1$ numerically by solving the Boltzmann 
equations~\cite{luty.92,plumacher.96,buch-bari-plum.02}.  

For demonstration purpose, we consider a model
~\cite{buc_gre_min.91} based on the gauge group 
$SU(2)\times U(1)_{Y}\times U(1)_{Y'}$, where $Y'$ is a linear 
combination of $Y$ and $B-L$. Since $B-L$ is a gauge symmetry, 
no primordial $L$ asymmetry exists. A net $L$-asymmetry is 
created dynamically after the $B-L$ gauge symmetry breaking 
phase transition. A part of this asymmetry is then transformed 
to $B$-asymmetry via the non-perturbative sphaleron processes.

In a diagonal basis the right handed Majorana neutrinos acquire 
masses $M_i=f_i\eta_{B-L}$, $\eta_{B-L}$ being the scale of $B-L$ 
symmetry breaking phase transition and $f$ being the Majorana 
Yukawa coupling matrix. Above the mass scale of $N_1$ all the 
interactions mediated by $N_1$ are fast enough to keep 
it in thermal equilibrium. This implies that there is no 
net $L$-asymmetry. However, as the temperature of thermal 
plasma falls and becomes comparable with the mass scale of 
$N_1$, an $L$-asymmetry is created through the $CP$-violating decays 
of $N_1$. However, a part of the created asymmetry is erased 
by the inverse decay and $L$-violating scatterings mediated 
by $N_1$. We study the dynamical generation of a net $L$-asymmetry 
by solving the relevant Boltzmann equations  
\begin{eqnarray}
\frac{dY_{N1}}{dZ} &=& -(D+S)\left(Y_{N1}-Y^{eq}_{N1}\right)
\label{Boltzmann-1}\\
\frac{dY_{B-L}}{dZ} &=& -\epsilon_{1} D\left(Y_{N1}-Y^{eq}_
{N1}\right)-W Y_{B-L},
\label{Boltzmann-2}
\end{eqnarray}
where $Y_a=(n_{a}/s)$ is the density of any species $`a'$ 
in a comoving volume and $s$ is the entropy density. Here 
$Z=M_1/T$ is a dimensionless parameter, where $T$ is related 
to the cosmic time $t$ through the time temperature relation    
\be
t=0.3g_*^{-1/2}\frac{M_{pl}}{T^2}.
\ee
A derivation of these equations is given in appendix A. 
The terms $D$, $S$ and $W$ occurring above are now discussed. 
\begin{enumerate}
 
\item $D=\frac{\Gamma_{D}}{HZ}$ accounts for the decay and 
inverse decay of lightest right handed neutrino into lepton 
and Higgs, $N_{1}\leftrightarrow \phi(\bar{\phi})+l(\bar{l})$, 
where
\begin{eqnarray}
\Gamma_{D} &=& \frac{1}{16\pi v^{2}} \tilde{m}_1 M_{1}^2
\frac{K_{1}(Z)}{K_{2}(Z)} \label{decay-1}\\
\Gamma_{ID} &=& \frac{n_{N1}^{eq}}{n_{l}^{eq}}\Gamma_{D}.
\end{eqnarray}
In the equation (\ref{decay-1}) the parameter $\tilde{m}_1$ is 
defined as 
\be 
\tilde{m}_1 = \frac{(m_{D}^{\dagger}m_{D})_{11}}{M_{1}},
\label{eff-neu-mass}
\ee
called the {\it effective mass} of the light neutrino
~\cite{plumacher.96}. $K_1$ and $K_2$ are modified Bessel 
functions whose ratio in equation (\ref{decay-1}) gives the 
time dilation factor. At a temperature above the mass scale 
of $N_1$ one can check that $\Gamma_{D}\simeq \Gamma_{ID}$. 
Below its mass scale the inverse decays are blocked. So the 
density of $N_1$ changes significantly due to the decays of 
$N_1$.                            

\item $S=\frac{\Gamma_{S}}{HZ}$ accounts for the lepton
number violating $\Delta L=1$ scatterings. The possible
reactions $N_{1}+l(\bar{l})\leftrightarrow \bar{t}(t)+q(\bar{q})$ 
via the exchange of $\phi$ in the s-channel and 
$N_{1}+t(\bar{t})\leftrightarrow \bar{l}(l)+q(\bar{q})$ through 
the exchange of $\phi$ in the t-channel, are shown in 
figure 2.2. The total rate of $\Delta L=1$ violating 
scatterings is given as 
\be
\Gamma_{S}=2\Gamma^{N_{1}}_{\phi,t}+4\Gamma^{N_{1}}_{\phi,s},
\ee
\begin{figure}
\begin{center}
\epsfig{file=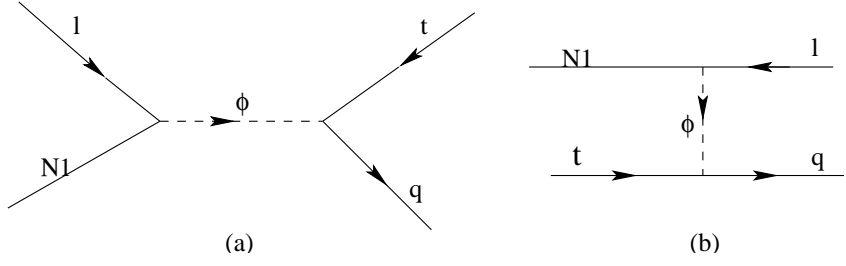, width=0.7\textwidth}
\caption{$\Delta L =1$ lepton number violating scatterings 
mediated by Standard Model Higgs through s or t-channel.}
\end{center}
\end{figure}     
where
\begin{eqnarray}
\Gamma^{N1}_{\phi, t} &=& \frac{m_t^2}{64 \pi^3
v^4} \frac{M_1^2 \tilde{m}_1}{K_2(Z)}\nonumber\\
&&\int_1^{\infty} dx \sqrt{x} K_1(Z\sqrt{x})\left[\frac{x-1}{x}
+\frac{1}{x}\ln \left(\frac{x-1+y'}{y'}\right)\right]\\
\Gamma^{N1}_{\phi, s} &=& \frac{m_t^2}{128 \pi^3
v^4} \frac{M_1^2 \tilde{m}_1}{K_2(Z)}
\int_1^{\infty} dx \sqrt{x} K_1(Z\sqrt{x})\left[
\frac{x-1}{x}\right]^2.
\end{eqnarray}
Here we have used $y'=(m^2_{\phi}/M_1^2)$ and the dimensionless
quantity $x=s'/M_{1}^{2}$, with $s'$ being the square of center 
of mass energy. Note that in the above scattering rates we 
have neglected the corrections due to second and third 
generation right handed heavy Majorana neutrinos.         

\item  $W=\frac{\Gamma_{W}}{HZ}$ constitute the wash out 
processes which compete with the decay term that actually produce 
the asymmetry. Here $\Gamma_{W}$ receives the contribution 
from the inverse decay ($\Gamma_{ID}$), $\Delta L=1$ scatterings 
($\Gamma^{N1}_{\phi,t}, \Gamma^{N1}_{\phi,s}$) and $\Delta L=2$ 
scatterings ($\Gamma^l_{N}, \Gamma^l_{N,t}$). The $\Delta L=2$ 
scattering processes, $\ell\phi \leftrightarrow \bar{\ell}\bar{\phi}$ 
via the exchange of $N_1$ and $\ell\ell \leftrightarrow 
\bar{\phi}\bar{\phi}$ mediated by $N_1$ in the $t$-channel, 
are shown in the figure 2.3. Combining all these processes we 
get the total scattering rate for the wash out processes to be
\begin{figure}[h]
\begin{center}
\epsfig{file=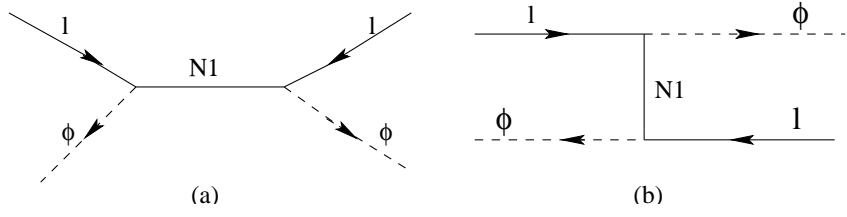, width=0.7\textwidth}
\caption{$\Delta L=2$, lepton number violating scatterings 
mediated by lightest right handed neutrino.}
\end{center}
\end{figure}
\end{enumerate}
\begin{equation}
\Gamma_{W}=\frac{1}{2}\Gamma_{ID}+2\left(\Gamma^{l}_{N1}+
2\Gamma^{l}_{N1,t}\right)+2\Gamma^{1}_{\phi,t}+
\frac{n_{N1}}{n_{N1}^{eq}}\Gamma_{\phi,s}.
\label{e24}
\end{equation}
Here 
\begin{eqnarray}
\Gamma^{l}_{N1} &=& \frac{Z^2}{256 \pi^3 v^4} M_1^3 \tilde{m}_1^2 
\int_0^{\infty} dx \sqrt{x}K_1(Z\sqrt{x}) \nonumber\\
&& \frac{1}{x}\{ x+\frac{x}{D_1(x)}+\frac{x^2}
{2D^2_1(x)}[1+(\frac{x+1}{D_1(x)})]\ln(1+x)\}\\
\Gamma^{l}_{N1,t} &=& \frac{Z^2}{256 \pi^3 v^4} 
M_1^3 \tilde{m}_1^2 \nonumber\\
&& \int_0^{\infty} dx \sqrt{x}K_1(Z\sqrt{x}) 
\{ \frac{x}{x+1} +(\frac{1}{x+2})\ln(1+x)\},
\end{eqnarray}
where 
\begin{equation}
\frac{1}{D_1(x)}=\frac{x-1}{(x-1)^2+c_1},~~\mathrm{with} ~~
c_1=\frac{\tilde{\Gamma}_D}{M_1}
\end{equation}
The quantities $\Gamma_{i}^{X}$ are thermally averaged reaction
rates per particle X. They are related to the reaction densities 
$\gamma_{i}$ as
\begin{equation}
\Gamma_i^X(Z)=\frac{\gamma_i(Z)}{n_X^{eq}}
\label{Gamma-gamma}
\end{equation}
The reaction densities are obtained from the reduced cross-sections 
$\hat{\sigma_i}(s'/M_1^2)$ as follows-
\begin{equation}
\gamma_{i}(Z)=\frac{M_{1}^{4}}{64\pi^{4}}\frac{1}{Z}\int^{\infty}_
{(m_{a}^{2}+m_{b}^{2})/M_{1}^{2}} dx~ \hat{\sigma}_{i}(x)\sqrt{x}
K_{1}(Z\sqrt{x}),
\end{equation}
where $m_{a}$ and $m_{b}$ are the masses of the two particles in
the initial state.  

Equations (\ref{Boltzmann-1}) and (\ref{Boltzmann-2}) have 
been solved numerically. We assume that far above its mass 
scale the species $N_1$ is in thermal equilibrium. So the 
initial abundance of $N_1$ is determined by its equilibrium 
distribution. Further at equilibrium, decays or lepton 
number violating scatterings will not produce any asymmetry. 
Therefore, we assume the following initial conditions
\begin{equation}
Y^{in}_{N1}=Y^{eq}_{N1}~~ {\mathrm and}~~ Y^{in}_{B-L}=0.
\label{in-assumption}
\end{equation}
\begin{figure}
\begin{center}
\epsfig{file=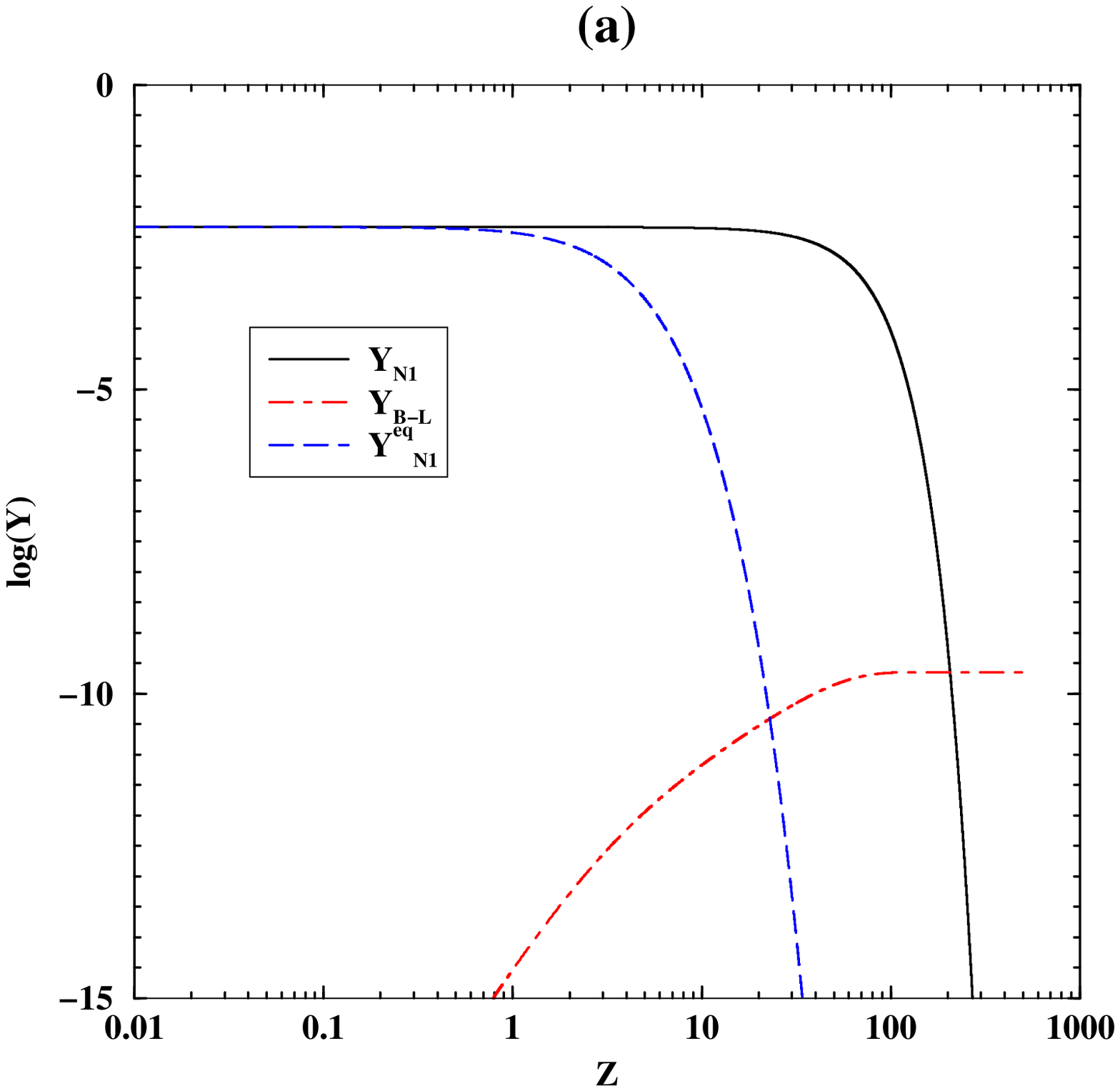, width=0.45\textwidth}
\epsfig{file=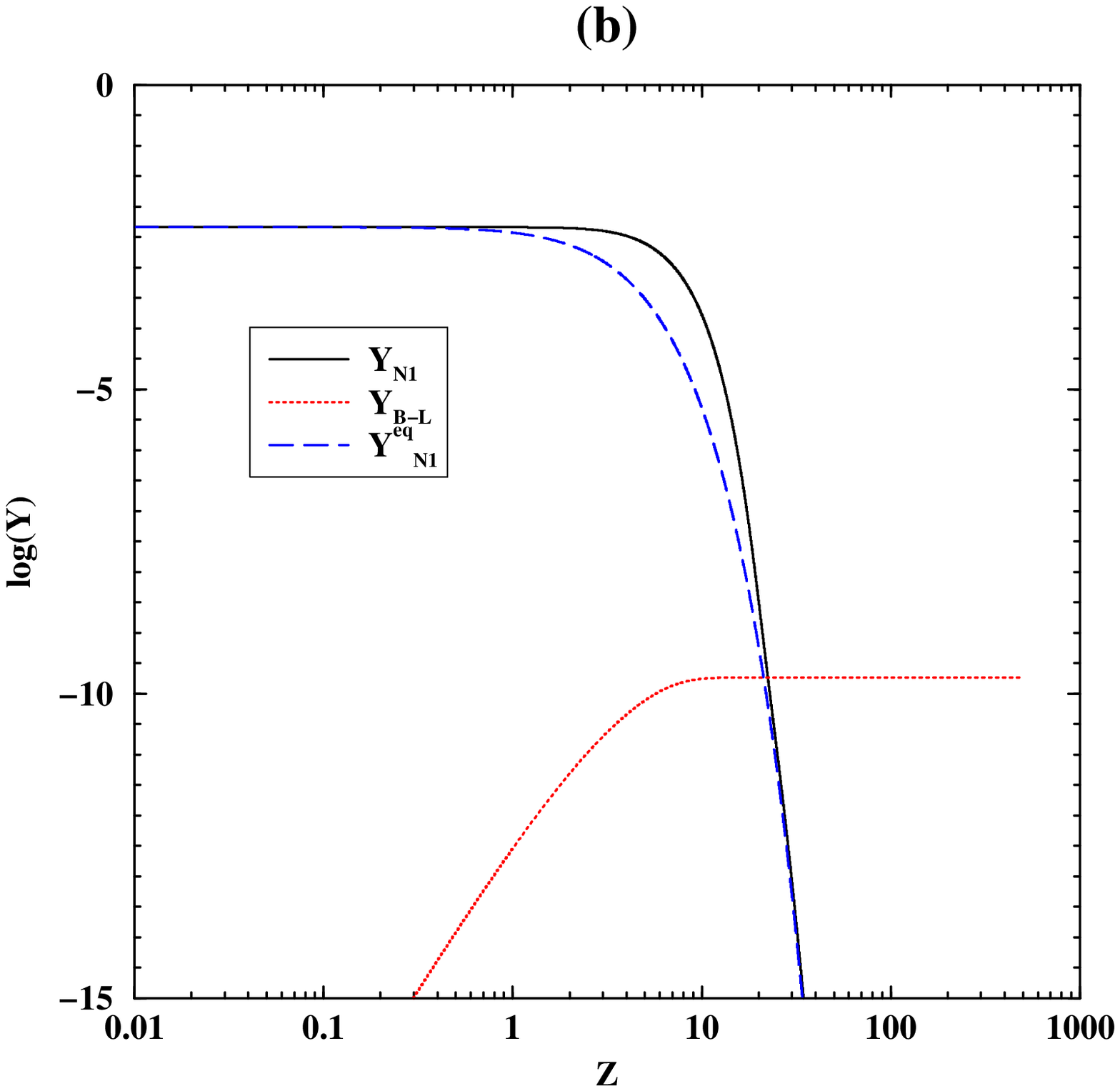, width=0.45\textwidth}
\caption{Dynamical evolution of neutrino number density
in a comoving volume and the produced $B-L$ asymmetry at 
$M_1=10^{10}GeV$ and $\epsilon_1=-5\times 10^{-8}$.}
\end{center}
\end{figure}
 
In fig. 2.4 (a) we have used $\tilde{m}_1=10^{-6}eV$. Far above 
the mass scale of $N_1$ (i.e. $Z\ll 1$) the $B-L$ asymmetry is 
zero. As the temperature falls (i.e. Z increases) the asymmetry 
builds up and finally it reaches a constant value $Y_{B-L}=2.6\times 
10^{-10}$ when the wash out effects become negligible. On the 
other hand, in fig. 2.4 (b) we have used $\tilde{m}_1=10^{-4}eV$ 
and we get a smaller asymmetry $1.8\times 10^{-10}$. This is 
because of larger effective neutrino mass. Note that the wash 
out effects not only depend on $M_1$ but also $\tilde{m}_1$. 
Therefore, for a larger $\tilde{m}_1$ the wash out effects 
is more and thus we get effectively a smaller asymmetry. 

We now introduce a new parameter $m_*$ to be called 
{\it equilibrium neutrino mass}. It is defined by~\cite{fglp.91} 
\be
m_*=\frac{\tilde{\Gamma}_D}{H},
\label{eqb-mass}
\ee
where 
\be
\tilde{\Gamma}_D=\frac{1}{16\pi v^2}\tilde{m}_1M_1^2 
\ee
is obtained from equation (\ref{decay-1}) in the limit 
$Z\rightarrow \infty$ and $H$ is the Hubble expansion parameter. 
At any epoch of temperature $T$, the Hubble expansion parameter 
is defined as 
\be
H=1.67 g_*^{1/2}\frac{T^2}{M_{pl}}.
\label{hubble}
\ee
Using (\ref{hubble}) in equation (\ref{eqb-mass}) we get the 
equilibrium mass at $T\sim M_1$ to be 
\be
m_*=16\pi g_*^{1/2}\frac{G_N^{1/2}}{\sqrt{2}G_F}\simeq 
2\times 10^{-3} eV, 
\label{eqb-mass-1}
\ee
where $G_N$ and $G_F$ are Newton and Fermi constants respectively 
and therefore $m_*$ may also be called {\it cosmological neutrino 
mass}~\cite{sahu&yajnik_prd.04}.

We now define a dimensionless parameter $K=\frac{\tilde{m}_1}{m_*}$, 
which determines whether the species $N_1$ is in thermal 
equilibrium. For $K>1$, the inverse decay processes are fast 
enough to ensure the species $N_1$ to be in equilibrium, 
irrespective of its initial abundance, in the epoch 
$Z\rightarrow 0$ ($T\rightarrow \infty$). In this regime, 
any pre-existing asymmetry gets erased by the rapid inverse 
decay processes. Therefore, it is called {\it strong wash out} 
regime~\cite{buch-bari-plum.02}. In this case the final lepton 
asymmetry doesn't depend on the initial conditions 
(\ref{in-assumption}). On the other hand, if $K\leq 1$, then the 
inverse decay processes are suppressed and the abundance of $N_1$ 
is not brought to equilibrium even for $Z\rightarrow 0$. 
In this case, any pre-existing asymmetry produced at the B-L 
symmetry breaking scale continue to be as it is until it gets 
some comparable contribution from the decays of $N_1$. So this 
regime is called {\it weak wash out regime}. In this 
regime the final lepton asymmetry strongly depends on the initial 
conditions. 

In the following we study the solution of Boltzmann equations 
(\ref{Boltzmann-1}) and (\ref{Boltzmann-2}) by taking the 
zero initial abundance of $N_1$ for two different values of 
$\tilde{m}_1$ in the case $K\leq 1$. For the numerical solution 
we assume that  
\begin{equation}
Y^{in}_{N1}=0~~ {\mathrm and}~~ Y^{in}_{B-L}=0
\end{equation}                                                        
\begin{figure}[h]
\begin{center}
\epsfig{file=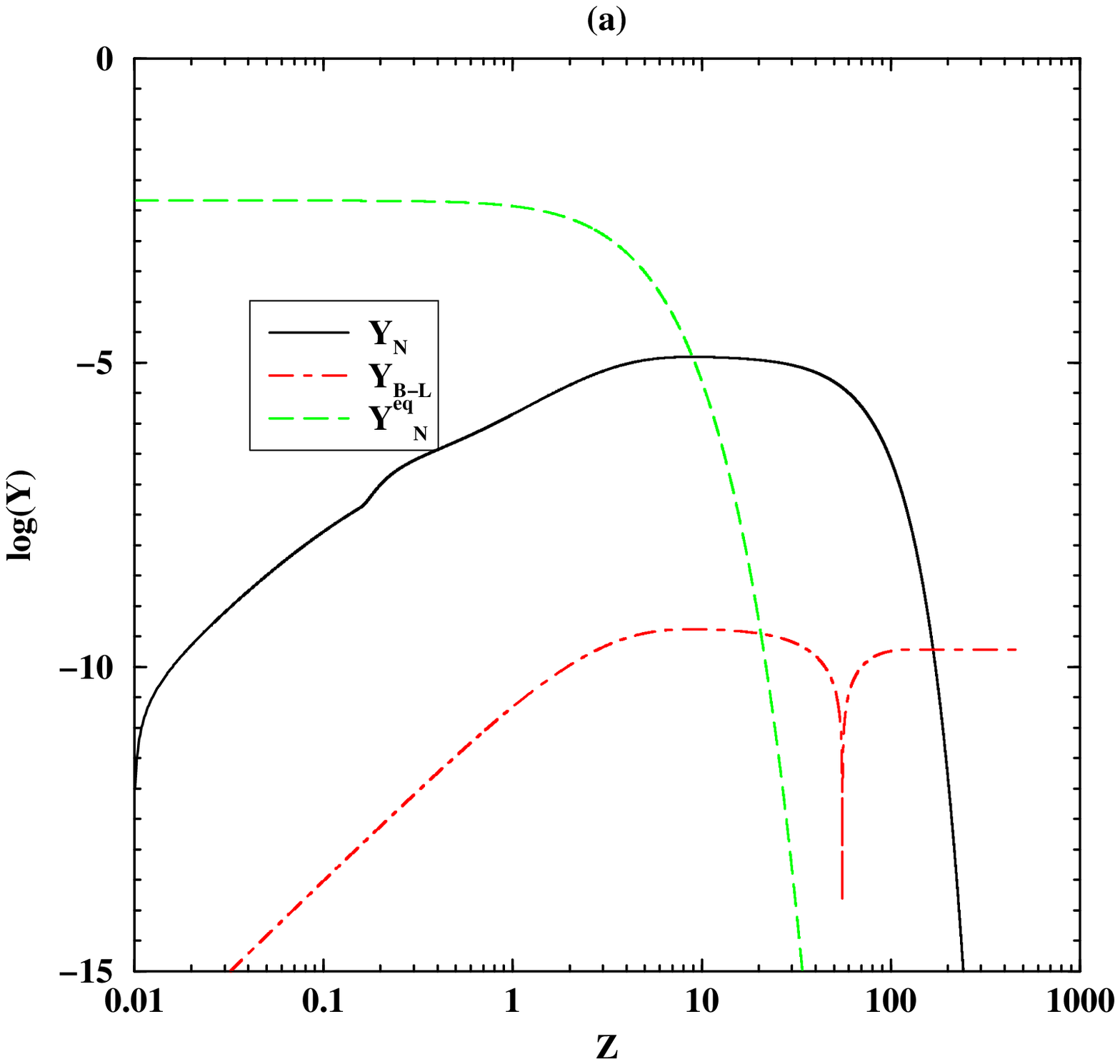, width=0.45\textwidth}
\epsfig{file=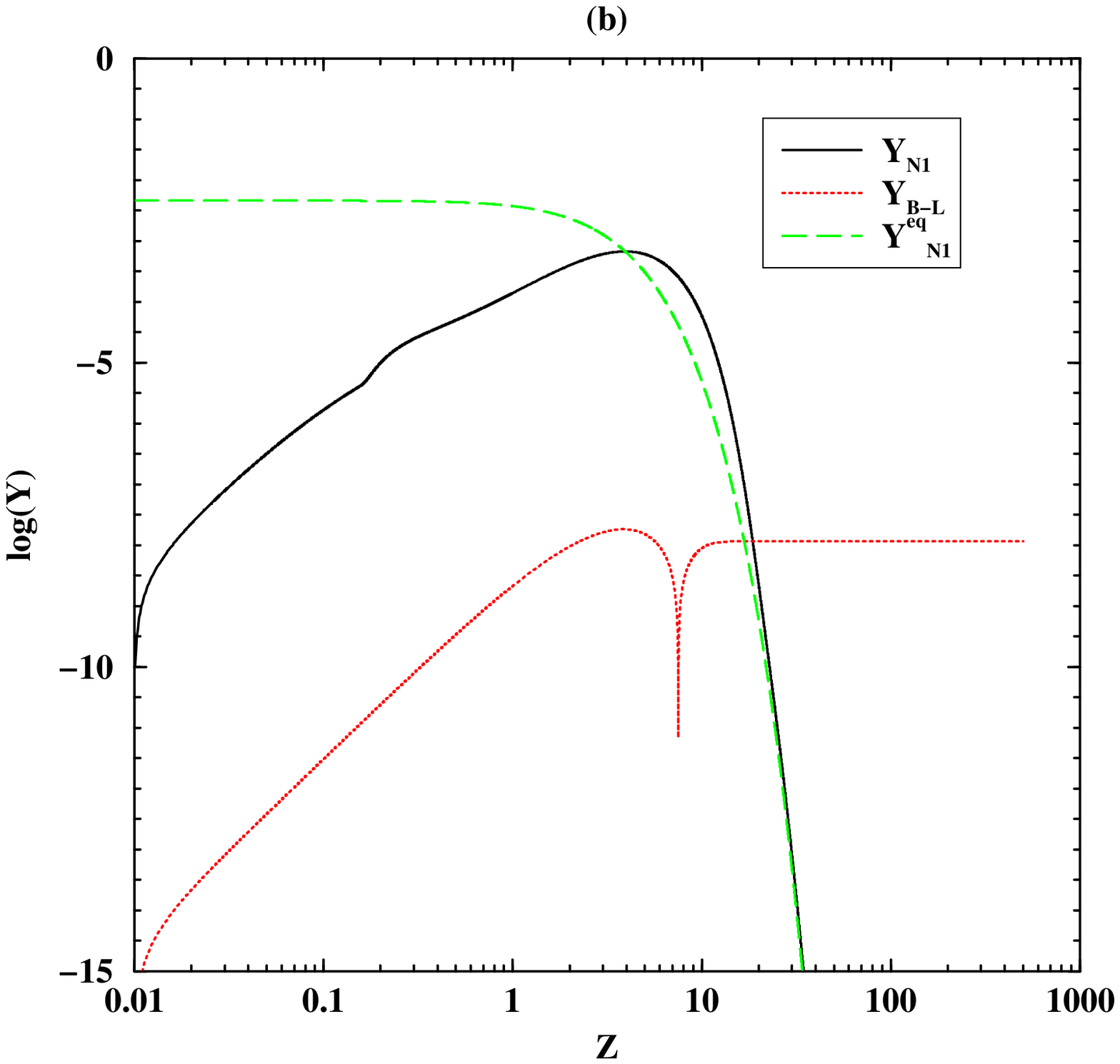, width=0.45\textwidth}
\caption{Dynamical evolution of neutrino number density
for $K\leq 1$ and $Y_{N_1}^{in}=0$ along with the produced 
$B-L$ asymmetry at $M_1=10^{10}GeV$ and $\epsilon_1=-5\times 
10^{-5}$.}
\end{center}
\end{figure}
In the figure 2.5(a) we have used $\tilde{m}_1=10^{-6}eV$, 
which is three orders of magnitude less than the equilibrium 
mass, $m_*\simeq O(10^{-3})eV$ of the light neutrino. In this 
scenario, the lightest right handed neutrino $N_1$ decays being out 
of equilibrium through out the evolution. Hence the erasure 
any pre-existing $L$-asymmetry is prevented. This remark is relevant 
to our study in chapter 4. On the other hand, in figure 2.5(b) we 
have used a larger $\tilde{m}_1=10^{-4}eV$, which is one order 
of magnitude less than $m_*$. Therefore, the neutrino abundance 
reaches the equilibrium value at an earlier time than the previous 
case. 

Similar calculations are done for the case 
$K>1$~\cite{buch-bari-plum.02}. It is shown that the species 
$N_1$ is brought to equilibrium quickly even if we start with 
the zero abundance of $N_1$, and hence erasing any pre-existing 
asymmetry, in the epoch $Z\rightarrow 0$.   
\chapter{Thermal leptogenesis in type-II seesaw models and 
bounds on neutrino masses}
\section{Introduction}
In the type-I seesaw models the upper left $3\times 3$ block 
of the $6\times 6$ neutrino mass matrix is zero; see e.g., 
section 2.2. This is because of the absence of $\nu_L-\nu_L$ 
interaction in the $SM$ Lagrangian as it violates the lepton 
number by two units. In contrast to it, in type-II seesaw 
models the presence of an additional scalar triplet $\Delta_L$ 
allows us to add a $\nu_L-\nu_L$ interaction to the $SM$ 
Lagrangian by compensating the two units of $B-L$ charge 
appearing in the $\nu_L-\nu_L$ interaction term. At a low 
scale the $\Delta_L$ acquires a $VEV$, thus providing an 
additional mass $M_L=f \langle \Delta_L\rangle$, $f$ being the 
Majorana Yukawa coupling, to the light neutrino mass eigenstate 
through the diagonalization of the $6\times 6$ neutrino mass 
matrix. As a result the light neutrino mass matrix takes the 
form $m_\nu=M_L-m_D^2/M_R$. The two terms are called type-II 
and type-I respectively. The class of models in which both 
type-I and type-II terms occurring in $m_\nu$ are called 
{\it type-II seesaw models}. 

\section{Type-II seesaw mechanism and neutrino masses}
In the minimal scenario, to achieve the light neutrino 
masses via the type-II seesaw mechanism, a scalar triplet 
$\Delta_L$ and a right handed Majorana neutrino per family 
are added to the $SM$. Thus the Lagrangian of this model 
reads 
\be
\mathcal{L}=\mathcal{L}_{SM}+\mathcal{L}_{new}\,
\ee
where $\mathcal{L}_{SM}$ is the $SM$ Lagrangian and 
$\mathcal{L}_{new}$ is the additional Lagrangian that 
contains the new interaction involving the right handed 
neutrinos and the triplet $\Delta_L$. The relevant terms of 
the Lagrangian are given to be  
\bea
-\mathcal{L}_{new} &\ni&\frac{1}{2}M_{Ri}N_{Ri}^T C N_{Ri}+
M_{\Delta}^2 Tr \Delta_L^{\dagger}\Delta_L+ h_{ij}\bar{\ell}_i
\phi \bar{N}_{Rj}\nonumber\\
&+& f_{ij}\ell^T C i\tau_2 \Delta \ell_j -\mu \phi^T i\tau_2 
\Delta_L \phi +H.C.\,
\eea
where $\ell^T=(\nu_{Li}, e_{Li})$ and $\phi^T=(\phi^0, \phi^-)$ 
and 
\be
\Delta_L=\begin{pmatrix}
\frac{1}{\sqrt{2}}\delta^+ & \delta^{++}\\
\delta^0 & -\frac{1}{\sqrt{2}}\delta^+
\end{pmatrix}.
\ee 
In the presence of these interactions, the neutral component 
of the triplet acquires a $VEV$, 
\be
\langle \Delta_L\rangle=\begin{pmatrix}
0&0\\
v_L & 0\end{pmatrix},
\ee
at a scale much below the electroweak symmetry breaking phase 
transition. Due to this $VEV$ there are now in general two sources 
of light neutrino masses
\bea
m_{\nu} &=& fv_{L}-v^2h M_R^{-1} h^T\nonumber\\
        &=&m_{\nu}^{II}+m_{\nu}^{I}.
\label{seesaw-II}
\eea
Note that $m_{\nu}^{II}$ that contribute to the neutrino mass 
matrix in the present case was absent in type-I models. 

We can diagonalize the light neutrino mass matrix $m_{\nu}$, 
through lepton flavor mixing matrix $U_{L}$~\cite{mns-matrix}. 
This gives us three light Majorana neutrinos of masses  
\be
U_L^{T} m_{\nu} U_L=dia(m_{1}, m_{2}, m_{3})\equiv D_m,
\label{dia-mass}
\ee
where the masses $m_1, m_2$ and $m_3$ can be chosen to be 
real and positive. 

\section{Thermal leptogenesis in type-II seesaw models}
In the type-II seesaw models the following decay modes:
\begin{displaymath}
\Delta\rightarrow \left\{ \begin{array}{lr}
\ell + \ell\\
\phi^\dagger + \phi^\dagger
\end{array}
\right.
\end{displaymath}
and 
\begin{displaymath}
N \rightarrow \left\{ \begin{array}{lr}
\bar{\ell} + \phi\\
\ell + \phi^\dagger
\end{array}
\right.
\end{displaymath}
violate lepton number by two units and hence produce the lepton 
asymmetry. In the above equations $\ell$ and $\phi$ are
$SM$ lepton and Higgs.

In what follows we assume a normal mass hierarchy in the heavy Majorana 
neutrino sector. We also assume that the quartic self coupling of the $SM$ 
Higgs, which is expected to be of order unity, is much larger 
than the Majorana Yukawa coupling of lightest right handed heavy 
Majorana neutrino $N_1$. In this case while the heavier right handed 
neutrinos, $N_2$ and $N_3$ and the triplet $\Delta_L$ decay, 
the lightest of heavy Majorana neutrinos is still in thermal 
equilibrium. Any asymmetry thus produced by the decay
of $N_2$, $N_3$ and $\Delta_L$ will be erased by the 
lepton number violating interactions mediated by $N_1$. 
Therefore, it is reasonable to assume that the final lepton 
asymmetry is given only by the $CP$-violating decays of $N_1$ to 
the $SM$ fields $\ell$ and $\phi$. 

\subsection{Upper bound on $CP$-asymmetry}
In comparison to the $CP$-asymmetry (\ref{cpasym1-1}) in 
type-I models there is an additional contribution
~\cite{donnell&sarkar.94, lazarides&shafi.98} in type-II 
seesaw models due to the one loop radiative correction through 
the virtual triplet $\Delta_L$ in the decays of lightest right 
handed Majorana neutrino. We assume that the masses of 
$\Delta_L$, $N_2$ and $N_3$ are much heavier than the 
the mass scale of $N_1$. In this scenario the total 
$CP$-asymmetry is given by
\be
\epsilon_1=\epsilon_1^I+\epsilon_1^{II},
\label{eff-epsilon}
\ee
where the contribution to $\epsilon_1^I$ comes from the
interference of tree level, self-energy correction and the
one loop radiative correction diagrams involving the heavier
Majorana neutrinos $N_2$ and $N_3$. This contribution is the
same as in type-I models~\cite{davidson&ibarra.02,
buch-bari-plum.02} and is given by
\be
\epsilon_1^{I}=\frac{3M_1}{16\pi v^2}\frac{\sum_{i,j}Im
\left[ (h^{\dagger})_{1i}(m_{\nu}^I)_{ij}(h^*)_{j1}\right]}
{(h^{\dagger}h)_{11}}.
\label{epsilon-I}
\ee 
\begin{figure}[h]
\begin{center}
\epsfig{file=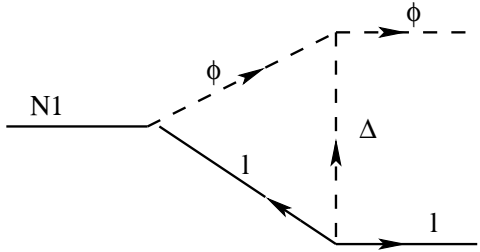, width=0.4\textwidth}
\caption{One loop radiative correction through the virtual triplet 
$\Delta_L$.}
\end{center}
\end{figure}  
On the other hand the contribution to $\epsilon_1^{II}$ in
equation (\ref{eff-epsilon}) comes from the interference of tree
level diagram and the one loop radiative correction diagram 
involving the triplet $\Delta_L$ as shown in fig. 3.1. It 
is given by~\cite{antusch.04,hamb-senj.03}
\be
\epsilon_1^{II}=\frac{3M_1}{16\pi v^2}\frac{\sum_{i,j}Im
\left[ (h^{\dagger})_{1i}(m_{\nu}^{II})_{ij}(h^*)_{j1}\right]}
{(h^{\dagger}h)_{11}}.
\label{epsilon-II}
\ee 
Substituting (\ref{epsilon-I}) and (\ref{epsilon-II}) in equation
(\ref{eff-epsilon}) and using (\ref{seesaw-II}), we get the total 
CP-asymmetry
\be
\epsilon_1 =\frac{3M_1}{16\pi v^2}\frac{Im(h^\dagger
m_\nu h^*)_{11}}{(h^\dagger h)_{11}}.
\label{cpasym-1}
\ee
Using (\ref{dia-mass}) in equation (\ref{cpasym-1}) we get 
\bea
\epsilon_1 &=& \frac{3M_1}{16\pi v^2} \frac{Im
(h^\dagger U_L^* D_m U_L^{\dagger} h^*)_{11}}
{(h^\dagger h)_{11}}\nonumber\\
&=&\frac{3M_1}{16\pi v^2} \frac{\sum_i m_i Im[(U_L^T h)_{i1}^*]^2}
{\sum_i |(U_L^Th)_{i1}|^2}.
\label{cpasym-2}
\eea
With an assumption of normal mass hierarchy for the light
Majorana neutrinos the upper bound on $CP$-asymmetry
(\ref{cpasym-2}) can be given by
\be
\epsilon_{1} \leq \frac{3 M_1}{16\pi v^2}m_3.
\label{type-II-cp}
\ee
Note that the above upper bound (\ref{type-II-cp}) for
$\epsilon_1$ as a function of $M_1$ and $m_3$ was first
obtained for the case of type-I seesaw
models~\cite{davidson&ibarra.02}. However, the same
relation holds in the case of type-II seesaw
models also~\cite{antusch.04} {\it independent of the relative
magnitudes of $m_{\nu}^{I}$ and $m_{\nu}^{II}$}. 

\subsection{Estimation of $L$-asymmetry and lower bound 
on the mass of lightest right handed neutrino}
Assuming $M_1\ll M_2$, $M_3$, $M_\Delta$, the final lepton 
asymmetry is given by the out of equilibrium decays of 
the lightest right handed Majorana neutrino $N_1$. A part of 
this asymmetry is then transformed to $B$-asymmetry by the 
thermally equilibrated sphaleron processes which violate 
$B+L$ quantum number of $SM$ fermions. In a comoving volume 
a net $B$-asymmetry can be written as  
\be
Y_B = \frac{n_B}{s}=0.55 \epsilon_1 Y_{N_1}\delta,
\label{Basym}
\ee
where the factor 0.55 in front~\cite{har&tur.90} is the 
fraction of $L$-asymmetry that is converted to $B$-asymmetry. 
Here $Y_{N_1}$ is the density of $N_1$ in a comoving volume 
which is given by $Y_{N_1}=n_{N_1}/s$, $s$ being the entropy 
density of the Universe at any epoch of temperature $T$ and 
$\delta$ is the wash out factor. 

We now recast equation (\ref{Basym}) in terms of a 
measurable quantity $(n_B/n_\gamma)$ which is given by  
\be
\left(\frac{n_{B}}{n_{\gamma}}\right) = 7 (Y_{B})
=3.85 (\epsilon_{1} Y_{N_1}\delta).
\label{baryon_asym1}
\ee
Substituting equation (\ref{type-II-cp}) in (\ref{baryon_asym1})
we get a bound on the baryon asymmetry to be 
\be
\left(\frac{n_{B}}{n_{\gamma}}\right) \leq 3.85\left(
\frac{3 M_1}{16\pi v^2} \right) m_3 Y_{N_1}\delta.
\label{baryon_asym2}
\ee
Using (\ref{atmo-mass}) in equation (\ref{baryon_asym2}) and 
comparing with the observed baryon asymmetry (\ref{b-asy-wmap}) 
we get a bound on the mass of $N_1$ to be 
\be
M_1\geq 0.84\times 10^{8} GeV \left( \frac{10^{-2}}{Y_{N_1}\delta}
\right)\left(\frac{0.05eV}{m_3}\right).
\ee
This bound was obtained in type-I seesaw models. However, in this 
case we revisit the same bound on the mass of $N_1$ irrespective 
of any assumption regarding the magnitude of type-I and type-II 
terms in the neutrino mass matrix (\ref{seesaw-II}).
\section{Spontaneous $CP$-violation and leptogenesis in 
Left-Right symmetric models}
In section 3.2 we demonstrated the type-II seesaw mechanism 
in a minimal scenario by adding a right handed Majorana 
neutrino per generation and a heavy triplet $\Delta_L$ to the $SM$. 
However, the light neutrino masses via type-II seesaw mechanism 
can be obtained naturally in Left-Right or $SO(10)$ models. In 
the following we consider the low energy left-right symmetric 
model in which we assume the case of spontaneous $CP$-violation 
($SCPV$). In this scenario we derive an upper bound on the 
$CP$-asymmetry. Moreover, we discuss the bounds on neutrino 
masses from the leptogenesis constraint.

\subsection{Left-Right symmetric model and $SCPV$}
In the low energy left-right symmetric model the right handed
charged lepton of each family, which was a singlet under the
$SM$ gauge group $SU(2)_L \otimes U(1)_Y$, gets a new partner
$\nu_R$. These two form a doublet under the $SU(2)_R$ of the
left-right symmetric gauge group $SU(2)_L \otimes SU(2)_R \otimes
U(1)_{B-L}$. Similarly, in the quark sector, the right handed up and
down quarks of each family, which were singlets under $SM$ gauge
group, combine to form a doublet under $SU(2)_R$. 
 
The Higgs sector of the model is dictated by two triplets
$\Delta_L$ and $\Delta_R$ and a bidoublet $\Phi$, which contains
two copies of $SM$ Higgs. Under $SU(2)_L\otimes SU(2)_R
\otimes U(1)_{B-L}$ the field content and the quantum numbers
of the Higgs fields are given as
\begin{eqnarray}
\Phi &=& \begin{pmatrix}
\phi_{1}^{0} & \phi_{1}^{+}\\
\phi_{2}^{-} & \phi_{2}^{0}
\end{pmatrix} \sim (1/2,1/2,0)\\
\Delta_{L} &=& \begin{pmatrix}
\delta_{L}^{+}/\sqrt{2} & \delta_{L}^{++}\\
\delta_{L}^{0} & -\delta_{L}^{+}/\sqrt{2}
\end{pmatrix}\sim (1,0,2)\\
\Delta_{R} &=& \begin{pmatrix}
\delta_{R}^{+}/\sqrt{2} & \delta_{R}^{++}\\
\delta_{R}^{0} & -\delta_{R}^{+}/\sqrt{2}
\end{pmatrix}\sim (0,1,2).
\end{eqnarray}  

To achieve the correct phenomenology, the various Higgs multiplets in
the model should have the following VEVs,
\be
\langle \Delta_{R}\rangle =\begin{pmatrix}
0 & 0\\
v_R e^{i\theta_R} & 0 \end{pmatrix},
\label{right_vev}
\ee
\be
\langle \Phi \rangle =\begin{pmatrix}
k_1 e^{i\alpha} & 0\\
0 & k_2 e^{i\beta}\end{pmatrix},
\label{dirac_vev}
\ee
and
\be
\langle \Delta_{L}\rangle =\begin{pmatrix}
0 & 0\\
v_L e^{i\theta_L} & 0 \end{pmatrix}.
\label{left_vev}
\ee
The electric charge of the fields is given by
\be
Q=T^3_L + T^3_R+ \frac{1}{2}(B-L).
\ee
In the above $v_L$, $v_R$, $k_1$ and $k_2$ are real parameters
and the electroweak symmetry breaking scale $v=174$ GeV is given by
$v^2=k_1^2+k_2^2$. Further we require that $v_L\ll v\ll v_R$. The
requirement of the spontaneous breakdown of parity gives rise to
\be
v_L v_R=\gamma (k_1^2+k_2^2)=\gamma v^2,
\ee
where $\gamma$ is parameter which is a function of the quartic couplings
in the Higgs potential.

The minimisation of the most general Higgs potential involving
$\Delta_L, \Delta_R$ and $\Phi$ was studied in refs.~\cite{scpv}. 
The relations between the various couplings, for which
the above set of VEVs are generated, were derived. In this
scenario, the gauge symmetry $SU(2)_L\times SU(2)_R \times U(1)_{B-L}$
is broken to $U(1)_{em}$ in a single step. Thus the $CP$-violating
phases come into existence at the same scale where the left-right
symmetry is broken. Since $v\ll v_R$, the $SM$ symmetry is present as
an approximate symmetry at the scale where symmetry breaking occurs. 

The fermions get their masses via Yukawa couplings. The
Lagrangian for one generation of quarks and leptons is
\bea
-\mathcal{L}_{yuk} &=& \tilde{h}_q\bar{q}_L\Phi q_R+
\tilde{g}_q\bar{q}_L\tilde{\Phi}q_R+ \tilde{h}_l\bar{\ell}_L\Phi l_R+
\tilde{g}_l\bar{\ell}_L\tilde{\Phi}l_R \nonumber \\
& &+if (\ell_L^T C\tau_2\Delta_L \ell_L
+\ell_R^T C\tau_{2}\Delta_{R}\ell_R) + H.c.\,
\label{yukawa}
\eea
where $q$ and $\ell$ are quark and lepton doublets,
$\tilde{\Phi}=\tau_2 \Phi^*\tau_2$ and $C$ is the Dirac charge
conjugation matrix. Further the Majorana Yukawa coupling
$f$ is the same for both left and right handed neutrinos to
maintain the discrete $L\leftrightarrow R$ symmetry.
 
Substituting the complex $VEV$s (\ref{right_vev}), (\ref{dirac_vev})
and (\ref{left_vev}) in (\ref{yukawa}) we obtain fermion mass 
terms to be
\bea
\mathcal{L}_{mass} &=& (\tilde{h}_q k_1 e^{i\alpha}+
\tilde{g}_q k_2 e^{i\beta})\bar{u}_L u_R+(\tilde{h}_q k_2
e^{i\beta}+\tilde{g}_q k_1e^{i\alpha})\bar{d}_L d_R \nonumber\\
& &+(\tilde{h}_l k_1e^{i\alpha}+\tilde{g}_l k_2 e^{i\beta})
\bar{\nu}_L \nu_R+(\tilde{h}_l k_2 e^{i\beta}+\tilde{g}_l
k_1e^{i\alpha})\bar{e}_L e_R \nonumber\\
& &+ f(\nu_L^T C v_L e^{i\theta_L}\nu_L +
\nu_R^T C v_R e^{i\theta_R}\nu_R)+H.C.
\label{eff-yukawa}
\eea
Generalizing the above equation (\ref{eff-yukawa}) for three
generation of matter fields we get the up and down quark mass
matrices to be
\be
(M_u)_{ij}=(\tilde{h}_q)_{ij} k_1 e^{i\alpha}+(\tilde{g}_q)_{ij}
k_2 e^{i\beta}~~\mathrm{and}~~
(M_d)_{ij}=(\tilde{h}_q)_{ij} k_2 e^{i\beta}+(\tilde{g}_q)_{ij}
k_1 e^{i\alpha}.
\label{quark-mass}
\ee 
We assume~\cite{scpv,ball.00} $k_1/k_2\sim m_t/m_b$.
In the seesaw mechanism, the Dirac mass matrix of the neutrinos
is assumed to be similar to the mass matrix of the charged leptons.
For $k_2 \ll k_1$, and further assuming $\tilde{h}_l\sim 
\tilde{g}_l$ in (\ref{eff-yukawa}), the Dirac mass matrix of 
the neutrinos to a good approximation becomes 
$\tilde{h}_l k_1 e^{i\alpha}$. Thus neglecting $k_2$ terms, 
the masses of three generations of neutrinos are given by
\be
\mathcal{L}_{\nu-mass}=\bar{\nu}_{L_i}k_1 e^{i\alpha}
(\tilde{h}_l)_{ij}\nu_{R_j}
+f_{ij}(v_L e^{i\theta_L} \nu_{L_i}^T C \nu_{L_j}
+v_R e^{i\theta_R}\nu_{R_i}^T C\nu_{R_j}) + H.C.
\ee
The Majorana mass matrix for the right handed neutrinos can
be diagonalized by making the following orthogonal transformation
on $\nu_R$
\be
N_R=O_R^T \nu_R.
\ee
In this basis, we have
\bea
O_R^T f O_R &=& f_{dia}\label{ortho_trans_f},\\
          h &=& \tilde{h} O_R\label{ortho_trans_h}.
\eea
In the transformed basis we get the mass matrix for the neutrinos
\be
\begin{pmatrix}
f v_L e^{i\theta_L} & k_1 e^{i\alpha}h\\
k_1e^{i\alpha} h^T & f_{dia}v_Re^{i\theta_R} \end{pmatrix}.
\ee
Diagonalizing the neutrino mass matrix into $3\times 3$ blocks
we get the light neutrino mass matrix to be
\be
m_{\nu} = f v_{L} e^{i\theta_L} - \frac{k_1^2}{v_R}
(h f_{dia}^{-1} h^T)e^{i(2\alpha-\theta_R)}
\label{see-saw}
\ee
 
Notice that the Lagrangian (\ref{yukawa}) is invariant under the
following unitary transformations of the fermion and Higgs fields,
\bea
\psi_L\longrightarrow U_L\psi_L~~ & \mathrm{and}~~ &
\psi_R\longrightarrow U_R\psi_R, \\
\Phi\longrightarrow U_L\Phi U_R^{\dagger}~~ & \mathrm{and}~~ &
\tilde{\Phi}\longrightarrow U_L\tilde{\Phi} U_R^{\dagger}\\
\Delta_L \longrightarrow U_L\Delta_L U_L^{\dagger}~~
& \mathrm{and}~~ & \Delta_R \longrightarrow U_R\Delta_R
U_R^{\dagger},
\eea
where $\psi_{L,R}$ is a doublet of quark or lepton fields.
The invariance under $U_L$ is the result of the remnant
global $U(1)$ symmetry which remains after the breaking of
the gauge symmetry $SU(2)_L$ and similarly for $U_R$.
The matrices $U_L$ and $U_R$ can be parametrized as
\be
U_L=\begin{pmatrix}
e^{i\gamma_L}&0\\
0& e^{-i\gamma_L}\end{pmatrix}~~ \mathrm{and}~~
U_R=\begin{pmatrix}
e^{i\gamma_R}&0\\
0& e^{-i\gamma_R}\end{pmatrix}.
\ee
By redefining the phases of the fermion fields we can rotate
away two of the phase degrees of freedom from the scalar
sector of the theory. Thus only two of the four phases of
Higgs $VEV$s have phenomenological consequences.
Under these unitary transformations, the $VEV$s (\ref{right_vev}),
(\ref{dirac_vev}) and (\ref{left_vev}) become
\be
\langle \Delta_{R}\rangle =\begin{pmatrix}
0 & 0\\
v_R e^{i(\theta_R-2\gamma_R)} & 0 \end{pmatrix},
\label{right_r_vev}
\ee
\be
\langle \Phi \rangle =\begin{pmatrix}
k_1 e^{i(\alpha+\gamma_L-\gamma_R)} & 0\\
0 & k_2 e^{i(\beta-\gamma_L+\gamma_R)}\end{pmatrix},
\label{dirac_r_vev}
\ee
and
\be
\langle \Delta_{L}\rangle =\begin{pmatrix}
0 & 0\\
v_L e^{i(\theta_L-2\gamma_L)} & 0 \end{pmatrix}.
\label{left_r_vev}
\ee
We choose $\gamma_R = \theta_R/2$ so that the masses of
the right handed neutrinos are real. The light neutrino mass
matrix (\ref{see-saw}) then becomes
\bea
m_{\nu} &=& f v_{L} e^{i(\theta_L-2\gamma_L)}-\frac{k_1^2}{v_R}
(h f_{dia}^{-1} h^T)e^{i(2\alpha+2\gamma_L-\theta_R)}
\label{rot-seesaw}\\
&=& m_{\nu}^{II}+m_{\nu}^{I}
\eea

Conventionally, in equation (\ref{rot-seesaw}), $\gamma_L$ was
chosen to be $-\alpha + \theta_R/2$ ~\cite{scpv,
mahanthapa.04}. This makes $m_\nu^I$ real leaving the imaginary
part purely in $m_\nu^{II}$.
We call this {\it type-II phase} choice. The
light neutrino mass matrix, with this phase choice, is
\be
m_\nu=f v_{L} e^{i\theta'_L}-\frac{k_1^2}{v_R}(h f_{dia}^{-1} h^T),
\label{con_seesaw}
\ee
where $\theta'_L=(\theta_L-\theta_R+2\alpha)$. On the other hand,
by choosing $\gamma_L = \theta_L/2$ in equation (\ref{rot-seesaw})
$m_\nu^{II}$ can be made real, with the phase occurring purely in
$m_\nu^I$. We call this {\it type-I phase} choice. Consequently
the light neutrino mass matrix (\ref{rot-seesaw}) becomes
\be
m_\nu = f v_{L}-\frac{k_1^2}{v_R}e^{i\theta'_R}
(h f_{dia}^{-1} h^T)\,
\label{chosen_seesaw}
\ee 
where $\theta'_R=(\theta_L-\theta_R+2\alpha)$. The $CP$-violating
parameter $\epsilon_1$ which gives rise to the lepton asymmetry
is independent of the phase choice. However, the theoretical
upper bound on $\epsilon_1$ is not a physical parameter of the
theory and can depend on the choice of phases as we see in
the next section. In numerical calculations, we take into account
the consistency of the bounds coming from the different phase
choices.
 
Using (\ref{dia-mass}) we can diagonalize the light neutrino 
mass matrix $m_{\nu}$. This gives us three eigenvalues, $m_1$, 
$m_2$ and $m_3$ which are chosen to be real.

\subsection{Upper bound on $CP$-asymmetry in Left-Right
symmetric models with $SCPV$}
Following the same convention in section 3.3.1 we can 
write the total $CP$-asymmetry in Left-Right symmetric model 
as 
\be
\epsilon_1=\frac{3M_1}{16\pi v^2}\frac{\sum_{i,j}Im
\left[ h^T_{1i}(m_{\nu}^I+m_{\nu}^{II})_{ij}h_{j1}\right]}
{(h^T h)_{11}}.
\label{total-epsilon}
\ee
From equation (\ref{total-epsilon}), we see that the physical
observable $\epsilon_1$ is not affected by the choice of phases.
In the following, we use bound on $\epsilon_1$ from the observed
baryon asymmetry to obtain bounds on right-handed neutrino
masses for the two different phase choices.

\subsubsection{A. The type-II choice of phases}
In this choice of phases the type-I mass term is
real. The only source of $CP$-violation in the light neutrino mass
matrix $m_\nu$ lies in the type-II mass term. Thus in this
case $\epsilon_1^{I}=0$ because of both h and $m_\nu^I$ are real.
The total $CP$-asymmetry in this choice of phases is therefore
given by
\bea
\epsilon_1 &=& \epsilon_1^{II}\nonumber\\
&=& \frac{3M_1v_L}{16\pi v^2} \frac{(h^T f h)_{11}}
{(h^T h)_{11}}Im (e^{i\theta'_L}).
\label{choice-2-cp}
\eea
Using (\ref{ortho_trans_f}) and (\ref{ortho_trans_h}) in
equation (\ref{choice-2-cp}) we get
\be
\epsilon_1=\frac{3M_1v_L}{16\pi v^2}\frac{
\sum_if_i(O_R^Th)_{i1}^2}{\sum_i(O_R^Th)_{i1}^2}sin\theta'_L,
\label{choice-2-cpasym}
\ee
where $f_i=(M_i/v_R)$. Up to a first order approximation it
is reasonable to assume that $\sum_if_i\approx 1$. In this
approximation the maximum value of the $CP$-asymmetry
(\ref{choice-2-cpasym}) is given by~\cite{davidson&ibarra.02,
buch-bari-plum.02,antusch.04,sahu&uma_prd.04}
\be
\epsilon_{1,max}=\frac{3M_1v_L}{16\pi v^2}.
\label{choice-2-max}
\ee
Thus, for type-II phase choice, a bound on $\epsilon_1$
leads to a bound on $M_1$.

\subsubsection{B. The type-I choice of phases}
In the type-I choice of phases the type-II mass term is real.
Hence the $CP$-violation comes through the type-I mass term only.
The total $CP$-asymmetry in this case is therefore given by
\bea
\epsilon_1 &=& \epsilon_1^I\nonumber\\
&=& \frac{3M_{1}k_1^2}{16\pi v^2 v_R} \frac{(h^Th f_{dia}^{-1}
h^Th)_{11}}{(h^T h)_{11}}Im (e^{-i\theta_R'}).
\label{cpasym2}
\eea
 
Let us consider the type-I term of the light neutrino mass matrix
\bea
m_\nu^I &=& m_{\nu}-m_{\nu}^{II}\nonumber\\
      &=& -\frac{k_1^2}{v_R'} h f_{dia}^{-1} h^T,
\label{fakemass}
\eea
where $v'_R=v_R e^{i\theta_R'}$. We can find a diagonalizing matrix
$U={\cal{O}}U_{phase}$ for $m_\nu^I$ such that
\be
U^T m_\nu^I U \equiv -D_{m_I}=-dia(m_{I_{1}}, m_{I_{2}}, m_{I_{3}})
\label{fakedia}
\ee
where $m_{I_{1}}, m_{I_{2}}$ and $m_{I_{3}}$ are made real by 
choosing $U_{phase}=e^{i\theta_R'/2}$. Therefore, from equation
(\ref{fakedia}) we have
\be
D_{m_{I}}=\frac{k_1^2}{v_R}{\cal{O}}^T \left(h f_{dia}^{-1}
h^T\right){\cal{O}}.
\label{mfdia}
\ee 
Using (\ref{mfdia}) in equation (\ref{cpasym2}) the $CP$-asymmetry
$\epsilon_1$ can be rewritten as
\bea
\epsilon_{1} &=& \frac{3M_{1}}{16\pi v^{2}}\frac{
\sum_{i}\left[(h^T{\cal{O}})_{1i} D_{m_{I_{ii}}} ({\cal{O}}^Th)_{i1}
\right]}{\sum_i \left[(h^T{\cal{O}})_{1i}({\cal{O}}^T h)_{i1}\right]}
Im (e^{-i\theta_R'})\nonumber\\
&=& \frac{3M_{1}}{16\pi v^{2}}\frac{
\sum_{i}m_{I_i}({\cal{O}}^Th)_{i1}^2}{\sum_i({\cal{O}}^Th)_{i1}^2)}
Im (e^{-i\theta_R'}).
\label{cpasym3}
\eea
In the above equation (\ref{cpasym3}) the maximum value of
$CP$-asymmetry is thus given by~\cite{davidson&ibarra.02,
buch-bari-plum.02}
\be
|\epsilon_{1, max}|=\frac{3M_{1}}{16\pi v^{2}}\sum_i m_{I_i}.
\label{cpasym-max}
\ee
In the equation (\ref{cpasym-max}) $m_{I}$s are the eigenvalues
of the matrix $m_\nu^I$ and {\it are not the physical light
neutrino masses}. It is desirable to express the $\epsilon_{1, max}$
in terms of physical parameters. In order to calculate the
$m_{I}$s we will assume a hierarchical texture of Majorana 
coupling
\be
f_{dia}= \frac{M_{1}}{v_{R}}\begin{pmatrix}1 & 0 & 0\\
0 & \alpha_{A} & 0\\
0 & 0 & \alpha_{B} \end{pmatrix},
\label{fdia-texture}
\ee 
where $1 \ll \alpha_{A}=(M_{2}/M_{1}) \ll \alpha_{B}=
(M_{3}/M_{1})$. We identify the neutrino Dirac Yukawa
coupling $h$ with that of charged leptons~\cite{gel-ram-sla}.
We assume $h$ to be of Fritzsch type~\cite{fritzsch.79}
\be
h=\frac{(m_{\tau}/v)}{1.054618}
\begin{pmatrix}
0 & a & 0\\
a & 0 & b\\
0 & b & c\end{pmatrix}.
\label{h-texture}
\ee
We make this assumption because Fritzsch mass matrices are
well motivated phenomenologically. By choosing the values
of $a$, $b$ and $c$ suitably one can get
the hierarchy for charged leptons and quarks. In particular
~\cite{fritzsch.79}
\begin{equation}
a=0.004,~~~ b=0.24 ~~~{\rm and}~~~ c=1
\label{abcvalues}
\end{equation}
can give the
mass hierarchy of charged leptons. For this set of values
the mass matrix $h$ is normalized with respect to the
$\tau$-lepton mass. The set of values of $a$, $b$ and $c$
are roughly in geometric progression. They can be expressed
in terms of the electro-weak gauge coupling $\alpha_{w}=
\frac{g^{2}}{4\pi}=\frac{\alpha}{sin^{2}\theta_{w}}$. In
particular $a=2.9 \alpha_{w}^{2}$, $b=6.5 \alpha_{w}$
and $c=1$. Here onwards we will use these set of values
for the parameters of $h$. Using equation (\ref{fdia-texture})
and (\ref{h-texture}) in equation (\ref{mfdia}), we now get
\bea
D_{mI} &=& \frac{v^{2}}{v_{R}}
\left(h f_{dia}^{-1}h^T\right)_{dia}\nonumber\\
 &\simeq & \frac{v^{2}}{M_{1}} \frac{(m_{\tau}/v)^{2}}
{(1.054618)^{2}}
\begin{pmatrix}
0& 0 & 0\\
0 & A & 0\\
0 & 0 & B\end{pmatrix},
\label{eigenvalue}
\eea
where the eigenvalues $A$ and $B$ are functions of $\alpha_{A}$
and $\alpha_{B}$ and their sum is given by
\be
A+B = \frac{1}{2}\left[a^{2}+\frac{1}{\alpha_{A}}(a^{2}+b^{2})
+\frac{1}{\alpha_{B}}(b^{2}+c^{2})\right].
\label{eigenvalue-sum}
\ee
Using equation (\ref{eigenvalue}) we can write
the maximum value of $CP$-asymmetry (\ref{cpasym-max})
\bea
\epsilon_{1, max} &=& \frac{3 M_{1}}{16\pi v^{2}}
(m_{I2}+m_{I3})\nonumber\\
&=& \frac{3}{16\pi}\frac{(m_{\tau}/v)^{2}}
{(1.054618)^{2}}(A+B).
\label{cpasym-max1}
\eea
Thus we see that, in type-I choice of phases, the leptogenesis
parameter $\epsilon_1$ constrains the hierarchy parameters
$\alpha_A$ and $\alpha_B$. In the following two sections, we
will obtain numerical bounds on $\alpha_A$ and $\alpha_B$
in a manner consistent with the bound $M_1$ coming from the
type-II phase choice.

\subsection{Estimation of $L$-asymmetry and bound on neutrino 
masses}
A net $B-L$ asymmetry is generated when left-right symmetry
breaks. A partial $B-L$ asymmetry is then gets converted to
$B$-asymmetry by the high temperature sphaleron transitions.
However these sphaleron fields conserve $B-L$. Therefore, the
produced $B-L$ asymmetry will not be washed out, rather they will
keep on changing it to $B$-asymmetry. Thus in a comoving volume
a net $B$-asymmetry is given by
\be
Y_B =\frac{n_B}{s}=\frac{28}{79} \epsilon_1 Y_{N1}\delta,
\label{B-asym}
\ee
where the factor $\frac{28}{79}$ in front~\cite{har&tur.90} is the
fraction of $B-L$ asymmetry that gets converted to $B$-asymmetry.
Here $\epsilon_{1}$ is given by equation (\ref{cpasym-max1}). 
The other symbols involved in equation (\ref{B-asym}) carry the 
usual meaning; see, e.g. section 3.3.2. However, the observed 
baryon asymmetry of the Universe is measured in terms of 
$n_B/n_\gamma$. Therefore, we rewrite equation (\ref{B-asym}) as 
\be
\left(\frac{n_{B}}{n_{\gamma}}\right) = 7 (Y_{B})
  = 2.48 (\epsilon_{1} Y_{N}\delta).
\label{baryon-asym1}
\ee
Substituting the type-II phase choice relation (\ref{choice-2-max})
in (\ref{baryon-asym1}) and comparing with the observed value
(\ref{b-asy-wmap}) of the baryon asymmetry we get the bound on 
the mass of lightest right handed neutrino to be 
\be
M_1\geq 1.25\times 10^{8} GeV \left( \frac{10^{-2}}{Y_{N_1}\delta}
\right)\left(\frac{0.1eV}{v_L}\right).
\label{M1-bound}
\ee 
On the other hand, substitution of $\epsilon_{1,max}$ from the
type-I phase choice (\ref{cpasym-max1}) in equation
(\ref{baryon-asym1}) and then comparison with the observed value
(\ref{b-asy-wmap}) gives the constraint
\be
A+B \geq 3.46\times 10^{-3}(10^{-2}/Y_{N}\delta)
\left(\frac{\left(n_{B}/n_{\gamma}\right)_{0}}
{6.1\times 10^{-10}} \right) \left(\frac{2GeV}{m_\tau}
\right) \left(\frac{v}{174GeV}\right)^{2},
\label{A+B-bound}
\ee
where the physical quantities are normalized with respect
to their observed values. The above equation, for the values
of $a$, $b$ and $c$ from (\ref{abcvalues}), gives only
one constraint on the two hierarchy parameters $\alpha_A$ and
$\alpha_B$. We will determine the individual parameters
$\alpha_A$ and $\alpha_B$ by demanding that their values
should reproduce the low energy neutrino parameters correctly,
while satisfying the inequalities $M_1 > O (10^8)$ GeV and
$\alpha_B > \alpha_A >> 1$.
Individual bounds on $\alpha_A$ and $\alpha_B$ can also be obtained
if we assume that the $\alpha_A$ term and the $\alpha_B$
term in the sum $A+B$ from equation (\ref{eigenvalue-sum}) are
roughly equal. We then get
\be
\alpha_{A}=(M_{2}/M_{1}) \leq 17 ~~{\rm and}~~\alpha_{B}=
(M_{3}/M_{1}) \leq 289.
\label{alpha-values-1}
\ee

\subsection{Checking the Consistency of $f$-matrix eigenvalues}
The solar and atmospheric evidences of neutrino
oscillations are nicely accommodated in the minimal
framework of the three-neutrino mixing, in which
the three neutrino flavors $\nu_{e}$, $\nu_{\mu}$, $\nu_{\tau}$
are unitary linear combinations of three neutrino mass eigenstates
$\nu_{1}$, $\nu_{2}$, $\nu_{3}$ with masses $m_{1}$,
$m_{2}$, $m_{3}$ respectively. The mixing among
these three neutrinos determines the structure
of the lepton mixing matrix~\cite{mns-matrix} which
can be parameterized as
\be
U_{L}=\begin{pmatrix}
c_{1}c_{3} & s_{1}c_{3} & s_{3}e^{i\delta}\\
-s_{1}c_{2}-c_{1}s_{2}s_{3}e^{i\delta} & c_{1}c_{2}-s_{1}s_{2}s_{3}
e^{i\delta} & s_{2}c_{3}\\
s_{1}s_{2}-c_{1}c_{2}s_{3} & -c_{1}s_{2}-s_{1}c_{2}s_{3}e^{i\delta} &
c_{2}c_{3}\end{pmatrix} dia(1, e^{i\alpha}, e^{i(\beta +\delta)}),
\label{mns-matrix}
\ee
where $c_{j}$ and $s_{j}$ stands for $\cos \theta_{j}$
and $\sin \theta_{j}$. The two physical phases
$\alpha$ and $\beta$ associated with the Majorana
character of neutrinos are not relevant for
neutrino oscillations~\cite{bilenkyetal.80} and will be
set to zero here onwards. While the Majorana phases
can be investigated in neutrinoless double beta decay
experiments~\cite{rodejohan_npb.01}, the CKM-phase
$\delta \in [-\pi, \pi]$ can be investigated in
long base line neutrino oscillation experiments. For
simplicity we set it to zero, since we are
interested only in the magnitudes of elements of $U_{L}$.
The best fit values of the neutrino masses and mixings from
a global three neutrino flavors oscillation analysis are~
\cite{gonzalez-garcia_prd.03}
\be
\theta_{1}=\theta_{\odot}\simeq 34^\circ, ~~\theta_{2}=\theta_{atm}
=45^\circ, ~~\theta_3 \leq 10^\circ,
\label{bestfit-theta}
\ee
and
\bea
\Delta m_{\odot}^{2}= m_2^2 - m_1^2 & \simeq & m_2^2 =
7.1\times 10^{-5} \ {\rm eV}^{2}\nonumber\\
\Delta m_{atm}^{2}= m_3^2 - m_2^2 & \simeq & m_3^2 =
2.6\times 10^{-3} \ {\rm eV}^{2}.
\eea
Using equation (\ref{see-saw}) we rewrite the $f$-matrix
\be
f=(\frac{eV}{v_{L}})\left[(m_{\nu}/eV)+\frac{4}{(1.054165)^{2}}
\frac{1}{(M_{1}/{\rm GeV})}
\begin{pmatrix}
\frac{a^{2}}{\alpha_{A}} & 0 & \frac{ab}{\alpha_{A}}\\
0 & a^{2}+\frac{b^{2}}{\alpha_{B}} & \frac{bc}{\alpha_{B}}\\
\frac{ab}{\alpha_{A}} & \frac{bc}{\alpha_{B}} & \frac{b^{2}}
{\alpha_{A}}+\frac{c^{2}}{\alpha_{B}}
\end{pmatrix} \right],
\label{fm1M1}
\ee
where the neutrino mass matrix $m_{\nu}$ is given by
equation(\ref{dia-mass}). The constrained eigenvalues
$\alpha_{A}$ and $\alpha_{B}$ are given by equation
(\ref{alpha-values-1}).

In the following, we choose $M_1$ to be larger than the bound
given by type-II phase choice (\ref{M1-bound}) and $m_1$ such
that $m_1^2 << \Delta_{sol}$. For such $m_1$ and $M_1$, we choose
suitable $\alpha_A$ and $\alpha_B$ that are compatible with the
low energy neutrino oscillation data. In particular here we
choose $m_{1}=1.0 \times 10^{-3}eV$, $M_{1}=1.0\times 10^{8}$ GeV,
$\alpha_{A}=17$, $\alpha_{B}=170$ and $\theta_{3}=6^\circ$. Then
we get
\be
f_{dia}=\frac{2.16\times 10^{-3}eV}{v_{L}}
\begin{pmatrix}
1 & 0 & 0\\
0 & 17.3 & 0\\
0 & 0 & 169.7 \end{pmatrix}.
\label{fdia-cal}
\ee
Thus, for the above values of $m_1$ and $M_1$, the assumed
hierarchy of right-handed neutrino masses is consistent with
global low energy neutrino data. Comparing equation (\ref{fdia-cal})
with (\ref{fdia-texture}) we get
\be
\frac{M_{1}}{v_{R}}=\frac{2.16\times 10^{-3}eV}{v_{L}}.
\ee
This implies that $v_{R}=O(10^{10})$ GeV for $v_{L}=0.1$ eV. These
values of $v_{L}$ and $v_{R}$ are compatible with genuine
seesaw $v_{L}v_{R}=\gamma v^{2}$ for a small value of
$\gamma\simeq O(10^{-4})$ \cite{nimai.04}.
On the other hand, if we choose the parameters $m_{1}=1.0
\times 10^{-3}$ eV, $M_{1}=1.0 \times 10^{9}$ GeV, $\alpha_{A}=17$,
$\alpha_{B}=65$ and $\theta_{3}=6^\circ$ we get
\be
f_{dia}=\frac{1.6 \times 10^{-3}eV}{v_{L}}
\begin{pmatrix}
1 & 0 & 0\\
0 & 16.76 & 0\\
0 & 0 & 64.68 \end{pmatrix}.
\label{fdia-cal-1}
\ee
Once again we have consistency between the assumed hierarchy
of right-handed neutrino masses and global low energy neutrino
data. Again comparing equation (\ref{fdia-cal-1}) with
(\ref{fdia-texture}) we get
\be
\frac{M_{1}}{v_{R}}=\frac{1.6 \times 10^{-3}eV}{v_{L}}.
\ee
Thus for $v_{L}=0.1$ eV one can get $v_{R}=O(10^{11}$ GeV). Again
these values are compatible with seesaw for $\gamma\simeq
O(10^{-3})$.
 
Here we demonstrated the consistency of our choice of the
matrix $f$ with neutrino data for two different choices of
$\alpha_A$ and $\alpha_B$. For other choices of these
parameters, to be consistent with $1<<\alpha_A<<\alpha_B$, one
can choose appropriate values of $m_1\leq 10^{-3}$ eV and
$M_1\geq 10^8$ GeV in equation (\ref{fm1M1}) which will
reproduce the correct eigenvalues of the matrix $f$.

\chapter{Gauged $B-L$ symmetry and upper bounds on neutrino masses}
\section{Introduction}
It has long been recognized that the existence of heavy Majorana
neutrinos has important consequences for the baryon asymmetry
of the Universe~\cite{fukugita.86}. With the discovery of the 
neutrino masses and
mixings, it becomes clear that only $B-L$ can be considered to
be a conserved global symmetry of the $SM$ and
not the individual quantum numbers $B-L_e$, $B-L_\mu$ and
$B-L_\tau$. Combined with the fact that the classical $B+L$
symmetry is anomalous~\cite{krs.86, arn_mac.88,aaps.91} it
becomes important to analyse the consequences of any $B-L$
violating interactions because the two effects combined can
result in the unwelcome conclusion of the net baryon
number of the Universe being zero.

At present two broad possibilities exist as viable explanations
of the observed baryon asymmetry of the Universe. One is the
baryogenesis through leptogenesis~\cite{fukugita.86}. This has
been analysed extensively in~\cite{luty.92, plumacher.96,
buch-bari-plum.02} and has provided very robust conclusions for
neutrino physics. Its typical scale of operation has to be high,
at least an intermediate scale of $10^9 GeV$. This has to do with
the intrinsic competition needed between the lepton number
violating processes and the expansion scale of the Universe.
While the mechanism does not prefer any particular unification
scheme, it has the virtue of working practically unchanged
upon inclusion of supersymmetry~\cite{plumacher.97}.

The alternative to this is provided by mechanisms which  work
at the TeV scale~\cite{susy_tev_group} and may rely on the new 
particle content implied in supersymmetric extensions of the $SM$.
It is worth investigating other possibilities
~\cite{sahu&yajnik_prd.04}, whether or not supersymmetry is 
essential to the mechanism. The starting point is the observation
~\cite{har&tur.90,fglp.91} that the heavy neutrinos participate 
in the erasure of any pre-existing asymmetry through scattering 
as well as decay and inverse decay processes. Estimates using 
general behavior of the thermal rates lead to a conclusion that 
there is an upper bound on the temperature $T_{\rm B-L}$ at which 
$B-L$ asymmetry could have been created. This bound is
$T_{B-L}\lsim 10^{13}$GeV$\times(1 eV/m_\nu)^2$, where $m_\nu$ is 
the typical light neutrino mass. This bound is too weak to be of 
accelerator physics interest. We extend this analysis by numerical 
solution of the Boltzmann equations and obtain regions of viability 
in the parameter space spanned by $\tilde{m}_1$-$M_1$, where 
$\tilde{m}_1$ is the effective light neutrino mass parameter as 
defined in eq. (\ref{eff-neu-mass}) and $M_1$ is the mass of the 
lightest of the heavy Majorana neutrinos. We find that our results 
are in consonance with~\cite{fglp.91} where it was argued that 
scattering processes provide a weaker constraint than the decay 
processes. If the scatterings become the main source of erasure 
of the primordial asymmetry then the constraint can be interpreted 
to imply  $\TBL<M_1$. Further, this temperature can be as low as 
$TeV$ range with $\tilde{m}_1$ within the range expected from 
neutrino observations. This is compatible with  see-saw mechanism if
the ``pivot" mass  scale is a factor of $10^2$ smaller than that 
of the charged leptons. 

Here we assume that a lepton asymmetry is produced when the 
$B-L$ gauge symmetry is broken without referring to any specific 
unification scheme. However, in~\cite{cynr.02} it was shown 
that the Left-Right 
symmetric model~\cite{leftright_group} presents just 
such a possibility. In this model $B-L$ appears as a gauged 
symmetry in a natural way. The phase transition is rendered 
first order so long as there is an approximate discrete symmetry 
$L\leftrightarrow R$, independent of details of other parameters. 
Spontaneously generated $CP$ phases then allow creation of lepton 
asymmetry. We check this scenario here against our numerical 
results and in the light of the discussion above. 

\section{Erasure constraints: An analytical estimation}
The presence of several heavy Majorana neutrino species ($N_i$)
gives rise to processes depleting the existing lepton
asymmetry in two ways. They are (i) scattering processes (S)
among the $SM$ fermions and (ii) Decay (D) and inverse decays (ID)
of the heavy neutrinos. We assume a normal mass hierarchy among the
right handed neutrinos such that only the lightest of the right
handed  neutrinos ($\non$) makes a significant contribution to
the above mentioned processes. At first we use a simpler picture, 
though the numerical results to follow are based on the complete
formalism. The dominant contributions to the two types of
processes are governed by the temperature dependent rates
\be
\label{rates}
\GD \sim \frac{h^2 M_1^2}{16\pi(4T^2 + M_1^2 )^{1/2}}
\hspace{1cm} {\rm and} \hspace{1cm}
\GS \sim \frac{h^4}{13\pi^3}\frac{T^3}{(9T^2 + M_1^2)},
\ee
where $h$ is typical Dirac Yukawa coupling of the neutrino.
 
Let us first consider the case  $M_1>\TBL$. For $T<\TBL$, the $\non$
states are not populated, nor is there sufficient free energy to
create them, rendering the D-ID processes unimportant.
We work in the scenario where the sphalerons are in equilibrium,
maintaining rough equality of $B$ and $L$ numbers. As the $B-L$
continues to be diluted we estimate the net baryon asymmetry
produced as~\cite{cynr.02} 
\be
10^{-d_{\sss B}}\equiv\exp\left(-\int_{t_{B-L}}^{t_{\sss EW}} 
\GS dt\right)=\exp\left(-\int_{T_{\sss EW}}^{\TBL} \frac{\GS}{H}\,
\frac{dT}{T}\right),
\ee
where $t_{B-L}$ is the time of the $(B-L)$-breaking phase transition,
$H$ is the Hubble parameter, and $t_{\sss EW}$  and $T_{\sss EW}$
corresponds to the electroweak scale after which the sphalerons
are ineffective. Evaluating the integral gives an estimate for
the exponent as
\be
\label{dsB}
d_{\sss B} \cong \frac{3\sqrt{10}}{13\pi^4\ln10\sqrt{g_*}}\,
h^4\frac{M_{Pl} \TBL}{M_1^2}.
\ee
The same result upto a numerical factor is obtained
in~\cite{buch_bari_plum.03}, the suppression factor 
$\omega^{(2)}$ therein. Eq.\ (\ref{dsB}) can be solved for the 
Yukawa coupling $h$ which gives the Dirac mass term for the 
neutrino
\be
h^4 \lsim 3200\, d_{\sss B}\left(\frac{M_1^2}
{\TBL M_{Pl}}\right)\,
\ee
where we have taken $g_* = 110$ for definiteness and $d_{\sss B}$ 
here stands for total depletion including from subdominant 
channels. This can be further transformed into an upper limit on 
the light neutrino masses using the canonical seesaw relation. 
The constraint (\ref{dsB}) can then be recast as
\be
\label{eq:mnudb}
\tilde{m}_1  \lsim \frac{180 v^2}{\sqrt{\TBL  M_{Pl}}}\,
\left(\frac{d_{\sss B}}{10}\right)^{1/2}.
\ee
This bound is useful for the case of large suppression. Consider 
$d_{\sss B}=10$.  If we seek $\TBL\sim 1$TeV and $M_1\sim10$TeV, 
this bound is saturated for $m_\nu\approx 50$keV, corresponding to 
$h\approx m_\tau/v$. This bound is academic in view of the WMAP 
bound $\sum m_{\nu_i}\approx 0.69 eV$~\cite{spergel.03} . On 
the other hand, for the phenomenologically interesting case 
$m_\nu \approx 10^{-2}$eV, with $h\approx 10^{-5}\approx m_e/v$ 
and with $M_1$ 
and $\TBL$ as above, eq. (\ref{eq:mnudb}) can be read to imply that 
in fact $d_{\sss B}$ is vanishingly small. This in turn demands, in 
view of the low scale we are seeking, a non-thermal mechanism for 
producing lepton asymmetry naturally in the range $10^{-10}$. Such 
a mechanism is discussed in sec. \ref{sec:lasymLR}.
 
In the opposite regime $M_1<\TBL$, both of the above types of
processes could freely occur. The condition that complete erasure
is prevented requires that the above processes are slower than
the expansion scale of the Universe for all $T>M_1$. It turns out
to be sufficient~\cite{fglp.91} to require $\GD<H$ which also
ensures that $\GS<H$. This leads to the requirement
\be 
\tilde{m}_1 < m_* \equiv 16\pi g_*^{1/2}\frac{G_N^{1/2}}
{\sqrt{2}G_F}\simeq 2\times10^{-3}eV
\label{FGLPbound}
\ee
where the parameter $m_*$~\cite{fglp.91} contains only universal
couplings and $g_*$, and may be called the \emph{cosmological 
neutrino mass}.
 
The constraint of equation (\ref{FGLPbound}) is compatible
with models of neutrino mass if we identify the neutrino
Dirac mass matrix $m_D$ as that of charged leptons. For a
texture of $m_D$ to be Fritzsch type (\ref{h-texture}), 
equation (\ref{eff-neu-mass}) gives
\be
\tilde{m_1} \simeq 2\times 10^{-3}eV \left(\frac{10^7 GeV}
{M_1}\right).
\ee 
Thus with this texture of masses, eq. (\ref{FGLPbound}) is 
satisfied for $M_1\gsim O(10^7) GeV$. If we seek $M_1$ mass 
within the TeV range, this formula suggests that the texture for 
the neutrinos should have the Dirac mass scale smaller by $10^{-3}$ 
relative to the charged leptons.
 
The bound (\ref{FGLPbound}) is meant to ensure that depletion 
effects remain unimportant and is rather strong. A more detailed 
estimate of the permitted values of $\tilde{m}_1$ and $M_1$ is 
obtained by solving the relevant Boltzmann equations in a scenario 
$\TBL>M_1$ and $\Gamma_D<H$. 

\section{Numerical solutions of Boltzmann equations}
The relevant Boltzmann equations~\cite{luty.92,plumacher.96,
buch-bari-plum.02,strumia.04} for our purpose are
\bea
\frac{dY_{N1}}{dZ} &=& -(D+S)\left(Y_{N1}-Y^{eq}_{N1}\right)
\label{boltzmann.1}\\
\frac{dY_{B-L}}{dZ} &=& -W Y_{B-L}.
\label{boltzmann.2},
\eea
The terminologies used in the above equations are explicitly 
elaborated in chapter 1. We will summarize later by considering 
their importance from the view of present context. The thermal 
corrections to the above processes as well as the processes 
involving the gauge bosons may have significance for final 
L-asymmetry~\cite{strumia.04}. However their importance is under 
debate~\cite{buch-bari-plum-ped.04}. Therefore, we limit 
ourselves to the same formalism as in~\cite{luty.92,plumacher.96,
buch-bari-plum.02}.

In equation (\ref{boltzmann.1}) $D=\Gamma_D/HZ$, where 
$\Gamma_D$ determines the decay rate of $N_1$, $S=\Gamma_S/HZ$, 
where $\Gamma_S$ determines the rate of $\Delta_{\rm L}=1$ lepton 
violating scatterings. In equation (\ref{boltzmann.2}) 
$W=\Gamma_W/HZ$, where $\Gamma_W$ incorporates the rate of 
depletion of the B-L number involving the lepton violating 
processes with $\Delta_{\rm L}=1$, $\Delta_{\rm L}=2$ as well as 
inverse decays creating $N_1$. The various $\Gamma$'s are related 
to the scattering densities~\cite{luty.92} $\gamma$s as 
given by equation (\ref{Gamma-gamma}). As the Universe expands 
these $\Gamma$'s compete with the Hubble expansion parameter 
(H). Therefore, for the $\Delta L=1$ lepton number violating 
processes in a comoving volume, we have 
\be
\left(\frac{\gamma_{D}}{sH(M_1)}\right), \left(\frac{
\gamma^{N1}_{\phi,s}}{sH(M_1)}\right), \left(\frac{
\gamma^{N1}_{\phi,t}}{sH(M_1)}\right) \propto k_1\tilde{m}_1.
\label{dilution}
\ee
On the other hand the dependence of the $\gamma$'s in
$\Delta_{\rm L}=2$ lepton number violating processes on
$\tilde{m}_1$ and $M_1$ are given by
\be
\left(\frac{\gamma^l_{N1}}{sH(M_1)}\right), \left(\frac{
\gamma^l_{N1,t}}{sH(M_1)}\right) \propto k_2 \tilde{m}_1^2 M_1.
\label{washout}
\ee
Finally there are also lepton conserving processes
where the dependence is given by
\be
\left(\frac{\gamma_{Z'}}{sH(M_1)}\right) \propto k_3 M_1^{-1}.
\label{l-conserve}
\ee
In the above equations (\ref{dilution}), (\ref{washout}), 
(\ref{l-conserve}), $k_i$, $i=1,2,3$ are dimensionful constants 
determined from other parameters. Since the lepton conserving 
processes are inversely proportional to the mass scale of $N_1$, 
they rapidly bring the species $N_1$ into thermal equilibrium 
for $M_1<T$. (\ref{washout}) are negligible because of their 
linear dependence on $M_1$. This is the regime in which we are
while solving the Boltzmann equations in the following.

\begin{figure}[ht]
\begin{center}
\epsfig{file=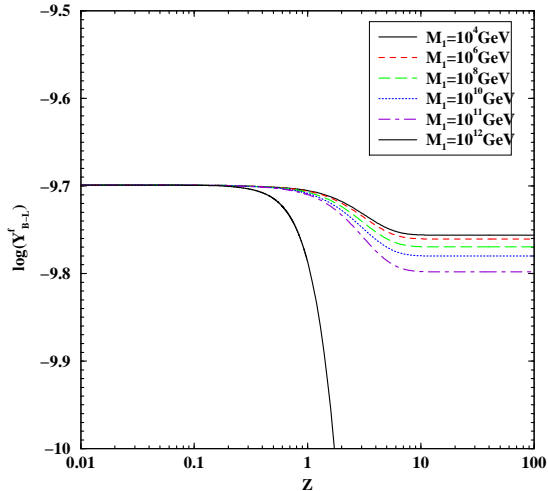, width=0.45\textwidth}
\caption{The evolution of B-L asymmetry for different values of
$M_1$ shown against $Z(=M_1/T)$ for $\tilde{m}_1=10^{-4}$eV
and $\eta^{raw}=2.0\times 10^{-10}$}
\label{figure-1}
\end{center}
\end{figure}
 
\begin{figure}[ht]
\begin{center}
\epsfig{file=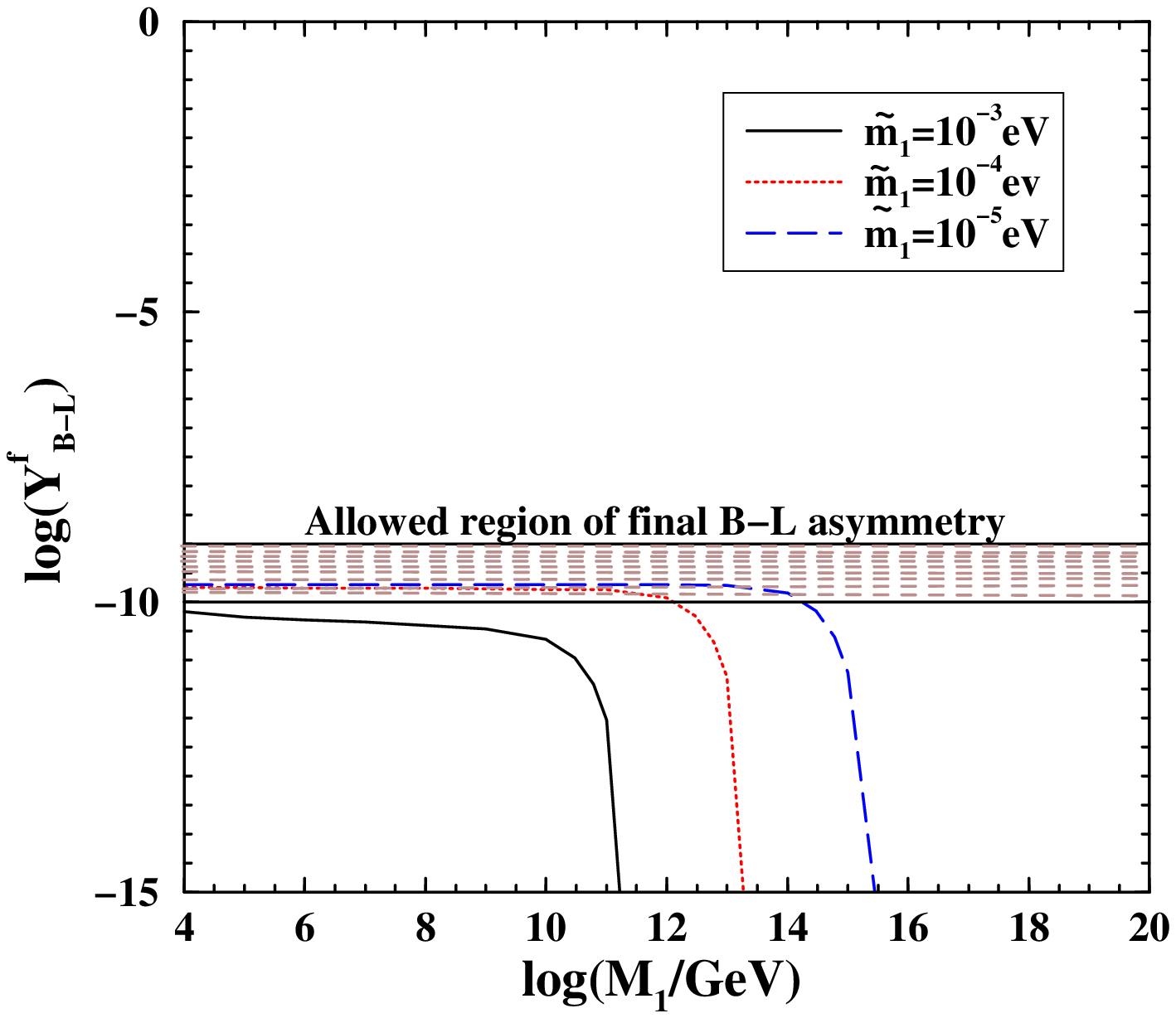, width=0.45\textwidth}
\caption{The allowed values of $M_1$ against the
required final asymmetry is shown for $\eta^{raw}=
2.0\times 10^{-10}$}
\label{figure-2}
\end{center}
\end{figure}
\begin{figure}[hbt]
\begin{center}
\epsfig{file=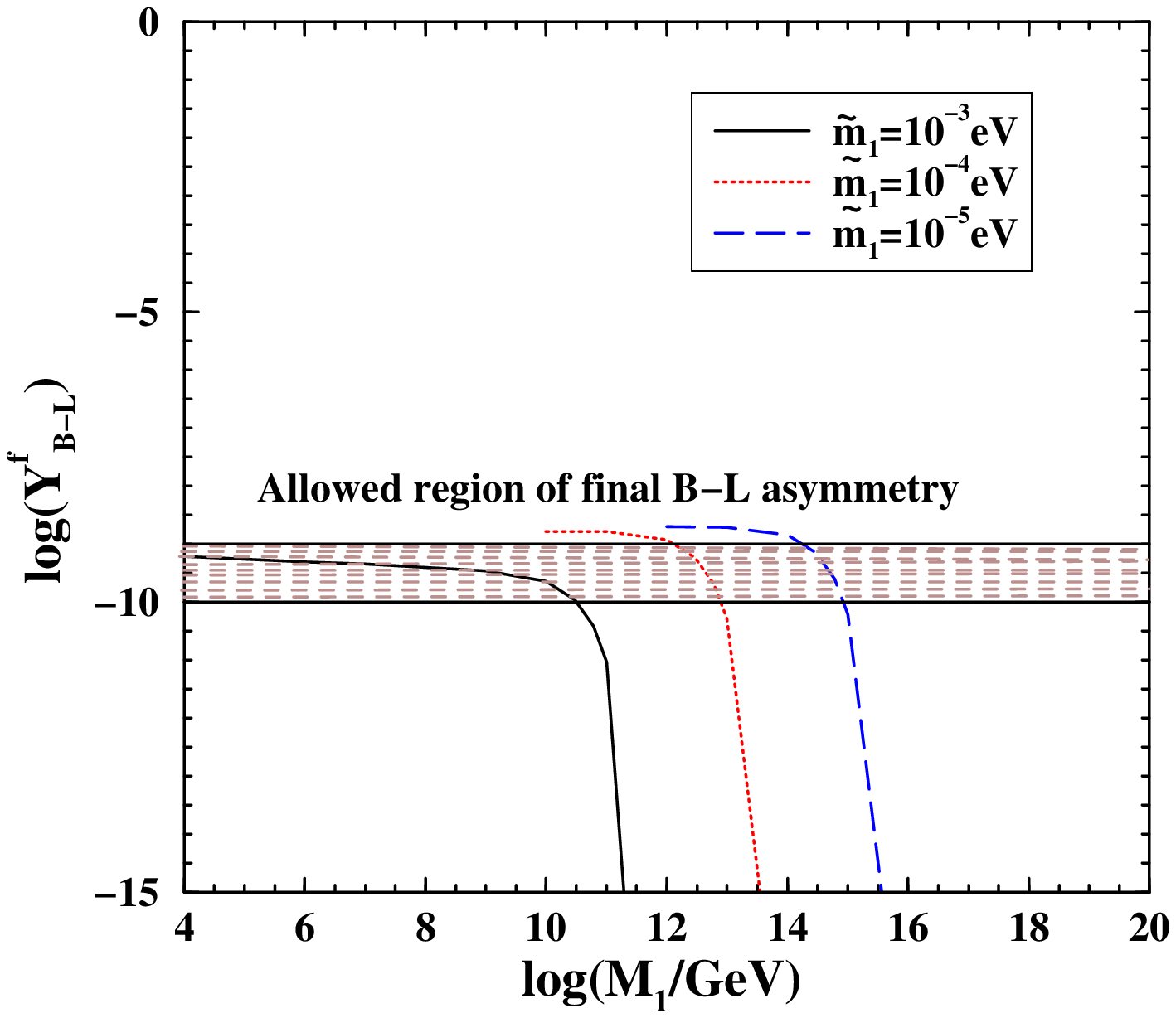, width=0.45\textwidth}
\caption{The allowed values of $M_1$ against the final
required asymmetry is shown for $\eta^{raw}=2.0\times 10^{-9}$}
\label{figure-3}
\end{center}
\end{figure} 

\begin{figure}[hbt]
\begin{center}
\psfig{file=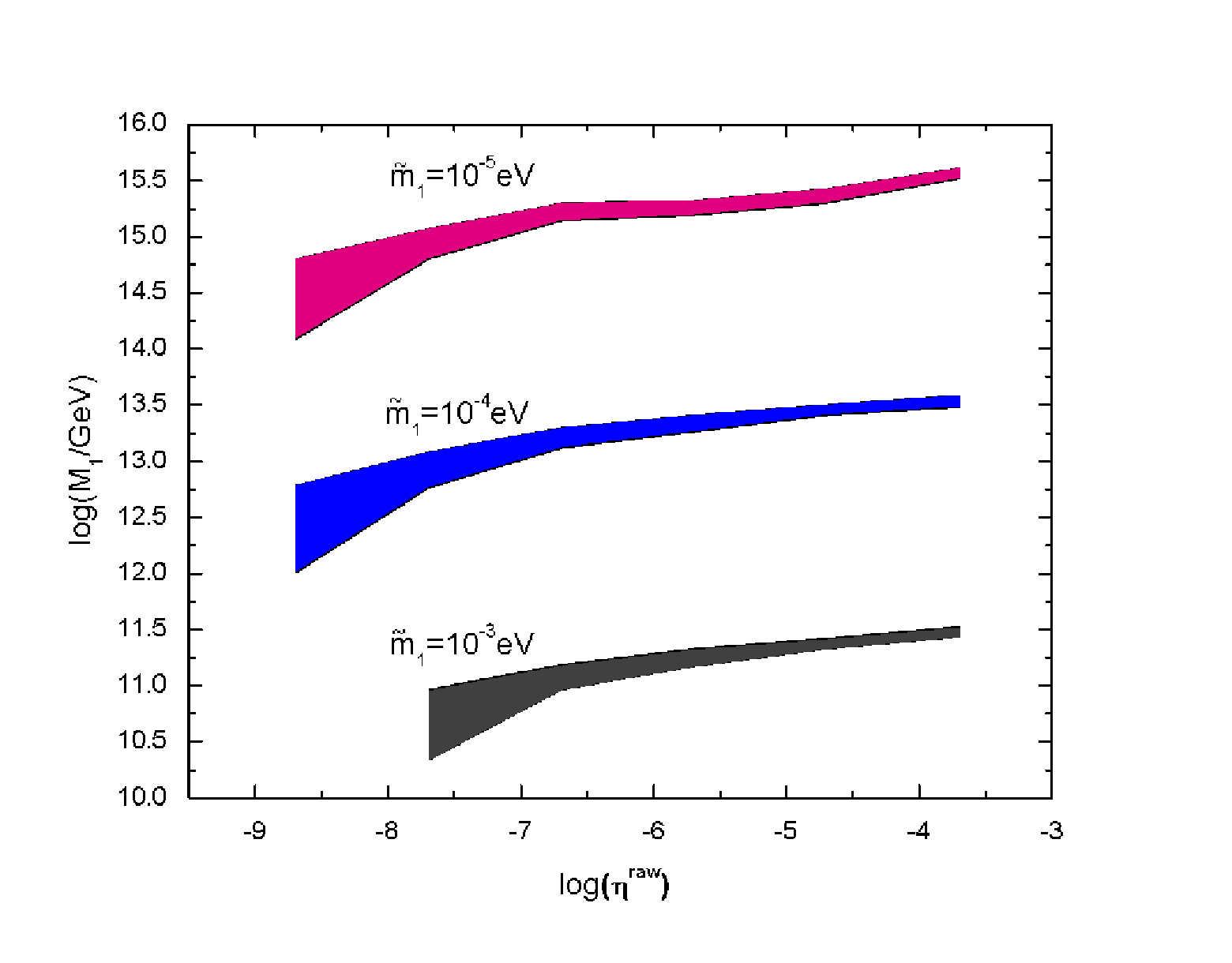, width=0.5\textwidth}
\caption{The allowed region of $M_1$ is shown for different
values of $\tilde{m}_1$ for large values of $\eta^{\sss raw}$}
\label{fig:regions}
\end{center}
\end{figure}

The equations (\ref{boltzmann.1}) and (\ref{boltzmann.2}) are solved
numerically. The initial $B-L$ asymmetry is the net raw
asymmetry produced during the B-L symmetry breaking phase
transition by any thermal or non-thermal process. As such we
impose the following initial conditions
\be
Y^{in}_{N1}=Y^{eq}_{N1}~~{\mathrm and}~~ Y^{in}_{B-L}=
\eta^{raw}_{B-L}.
\label{initial-cond}
\ee
 
At temperature $T\geq M_1$, wash out effects involving $N_1$ are
kept under check due to the $\tilde{m}_1^2$ dependence in (\ref{washout})
for small values of $\tilde{m}_1$. As a result a given
raw asymmetry suffers limited erasure. As the temperature
falls below the mass scale of $M_1$ the wash out processes
become negligible leaving behind a final lepton asymmetry.
Fig. \ref{figure-1} shows the result of solving the Boltzmann 
equations for different values of $M_1$.
 
If we demand that the initial raw asymmetry is of the order
of $\alpha \times 10^{-10}$, with $\alpha\sim O(1)$, then
in order to preserve the final asymmetry of the same order
as the initial one it is necessary that the neutrino mass
parameter $\tilde{m}_1$ should be less than $10^{-3}eV$.
This can be seen from fig. \ref{figure-2}. For 
$\tilde{m}_1= 10^{-3}eV$ we can not find any value of $M_1$ 
to preserve the final asymmetry,
$\alpha \times 10^{-10}$ in the allowed region. This is because
of the large wash out effects as inferred from the equation
(\ref{washout}). However, for $\tilde{m_1}= 10^{-4}eV$ we get
a {\it lowest threshold} on the lightest right handed neutrino
of the order $10^{12}GeV$. For any value of $M_1\leq 10^{12}GeV$ the  final asymmetry lies in the allowed region. This bound increases by
two order of magnitude for further one order suppression of
the neutrino parameter $\tilde{m}_1$. The important point
being that $M_1=10 TeV$ is within the acceptable range.
 
We now consider the raw asymmetry one order more than the
previous case {\it i.e.} $\eta^{\sss raw}_{\sss B-L}= \alpha
\times 10^{-9}$. From fig. \ref{figure-3} we see that
for $\tilde{m}_1=10^{-3}eV$ there is only an upper bound
$M_1= 10^{10.5}GeV$, such that the final asymmetry lies in the
allowed region for all smaller values of $M_1$. Thus for the case
of raw asymmetry an order of magnitude smaller, the upper bound
on $M_1$ decrease by two orders of magnitude (e.g. compare previous
paragraph). However, the choice of smaller values of $\tilde{m}_1$
leads to a small window for values of $M_1$ for which we end up
with the final required asymmetry. In particular for
$\tilde{m}_1= 10^{-4}eV$ the allowed range for $M_1$
is ($10^{12}~ - ~10^{13})GeV$, while for $\tilde{m}_1=
10^{-5}eV$ the allowed range shifts to ($10^{14}~ -
~10^{15})GeV$. The window effect can be understood as
follows. Increasing the value of $M_1$ tends to lift the
suppression imposed by the $\tilde{m}_1^2$ dependence
of the wash out effects, thus improving efficiency of
the latter. However, further increase in $M_1$ makes the
effects too efficient, erasing the raw asymmetry to
insignificant levels.

The windowing effect emerges clearly as we consider the cases
of large raw asymmetries. This is shown in fig. \ref{fig:regions}.
It is seen that as the raw asymmetry increases the allowed regions
become progressively narrower and lie in the range 
$(10^{10}~-~10^{15}) GeV$. Thus a given raw lepton asymmetry 
determines a corresponding
small range of the heavy Majorana neutrino masses for which we can
obtain the final asymmetry of the required order $\alpha\times
10^{-10}$. Again smaller is the effective neutrino mass $\tilde{m}_1$
larger is the mean value of the allowed mass of the heavy Majorana
neutrino and this is a consequence of normal see-saw.

Finally, in the following, we give an example for
non-thermal creation of L-asymmetry in the context of
left-right symmetric model.

\section{Lepton asymmetry in left-right symmetric model}
\label{sec:lasymLR}
We discuss qualitatively the possibility of lepton
asymmetry during the left-right symmetry breaking phase
transition~\cite{cynr.02}. In the following we recapitulate
the important aspects of left-right symmetric model
for the present purpose and the possible non-thermal mechanism 
of producing raw lepton asymmetry. This asymmetry which gets
converted to baryon asymmetry, can be naturally small if
the quartic couplings of the theory are small. Smallness
of zero-temperature $CP$ phase is not essential for this
mechanism to provide small raw $L$ asymmetry. 

\subsection{Left-Right symmetric model and transient
domain walls}
The important features of the left-right symmetric model 
based on the gauge group $SU(2)_L\times SU(2)_R\times U(1)_{B-L}$ 
are elucidated in chapter 3. The Higgs potential of the theory 
naturally entails a vacuum structure wherein at the first stage 
of symmetry breaking, either one of $\Delta_L$ or $\Delta_R$ 
acquires a vacuum expectation value the left-right symmetry, 
$SU(2)_L\leftrightarrow SU(2)_R$, breaks. It is required that 
$\Delta_R$ acquires a VEV first, resulting in $SU(2)_R \otimes 
U(1)_{B-L}$ $\rightarrow U(1)_Y$. Finally $Q=T^3_L + T^3_R+ 
\frac{1}{2}(B-L)$, survives after the bidoublet $\Phi$ and the 
$\Delta_L$ acquire VEVs.

If the left-right symmetry were exact, the first stage of breaking
gives rise to stable domain walls~\cite{ywmmc,lazar,lew-rio}
interpolating between the L and R-like regions. By L-like we 
mean regions favored by the observed phenomenology, while in the 
R-like regions the vacuum expectation value of $\Delta_R$ is zero. 
Unless some non-trivial mechanism prevents this domain structure, 
the existence of R-like domains would disagree with low energy 
phenomenology. Furthermore, the domain walls would quickly come 
to dominate the energy density of the Universe. Thus in this
theory a departure from exact symmetry in such a way as to 
eliminate the R-like regions is essential.
 
The domain walls formed can be transient if there exists a
slight deviation from exact discrete symmetry. As a result the
thermal perturbative corrections to the Higgs field free energy
will not be symmetric and the domain walls will be unstable.
This is possible if the low energy ($\sim10^4$GeV-$10^9$GeV)
left-right symmetric theory is descended from a Grand Unified 
Theory (GUT) and the effect is small, suppressed by the GUT 
mass scale.  In the process of cooling the Universe would first 
pass through the phase transition where this approximate 
symmetry breaks. The slight difference in free energy between 
the two types of regions causes a pressure difference across 
the walls,  converting all the R-like regions to L-like regions. 
Details of this dynamics can be found in ref.~\cite{cynr.02}.

\subsection{Leptogenesis mechanism}
In order to produce adequate lepton asymmetry the
following criteria of Skharov's have to be satisfied~
\cite{sakharov.67}. $C$ and $CP$-violation and finally
all the thermal processes have to be out of equilibrium.
The first two properties can be realized if there is
a $CP$-violating condensate exists in the domain wall.
Finally the out-of-equilibrium condition can be realized
from the directional motion of the domain walls to R-like
regions and thus making all the domains L-like.
 
We now consider the interaction of neutrinos from the
L-R wall, which is moving towards the energetically
disfavored phase, the R-like region. The left-handed
neutrino, $\nu_L$s, are massive in this domain as they
couple with their $CP$-conjugate states, where as they are
massless in the phase behind the wall as 
$\langle \Delta_L\rangle =0$.
This can be seen from the Yukawa coupling 
\be
\mathcal{L}_{yuk}= f\Delta_L \bar{\nu_L^c} \nu_L+ h.c.\, 
\ee
and since $\Delta_L$ has a wall like profile, the mass of 
$\Delta_L$ is zero behind the wall and is $O(v_R)$ in front of 
it. 

To get a net lepton-asymmetry one needs the asymmetry between
the reflection and transmission coefficients from the wall 
between $\nu_L$ and its $CP$-conjugate state $(\nu^c_L)$.
If it favors the transmission of $\nu_L$ to the L-like region
then the excess of antineutrinos ($\nu^c_L$) reflected in front
of the wall will equilibrate with the $\nu_L$ due to helicity
flipping scatterings. However the transmitted excess of $\nu_L$
remain as it is, since it does not couple with $\nu^c_L$ behind
the wall.

At least two of the Higgs expectation values in L-R model
are generically complex, thus providing natural $CP$-violation~
\cite{scpv} permitting all parameters in the 
Higgs potential to be real. We choose the following profiles
for the Higgses
\be
\langle \phi \rangle = \begin{pmatrix}
k_1 e^{i\alpha} & 0\\
0 & k_2 \end{pmatrix}, ~~ \langle \Delta_L\rangle =\begin{pmatrix}
0 & 0\\
v_L e^{i\theta} & 0 \end{pmatrix}, ~~\langle \Delta_R \rangle=
\begin{pmatrix}
0 & 0\\
v_R & 0\end{pmatrix}.
\ee
In the classical approximation where the wall width is assumed
to be large the $CP$ violating phase inside the domain wall
becomes position dependent. Under these circumstances a formalism
exists~\cite{jpt,clijokai,clikai}, wherein the chemical potential
$\mu_\Ls$ created for the Lepton number can be computed as
a solution of the diffusion equation 
\be
\label{eq:diffeq}
-D_\nu \mu_\Ls'' - v_w \mu_\Ls'
+ \theta(x)\, \Gamma_{\rm hf}\,\mu_\Ls = S(x).
\ee
Here $D_\nu$ is the neutrino diffusion coefficient, $v_w$
is the velocity of the wall, taken to be moving in the $+x$
direction, $\Gamma_{\rm hf}$ is the rate of helicity
flipping interactions taking place in front of the wall (hence
the step function $\theta(x)$), and $S$ is the source term
which contains derivatives of the position dependent complex
Dirac mass.

After integration of the above equation and using inputs
from the numerical solutions we find the raw Lepton
asymmetry \cite{cynr.02}
\be
\eta^{\rm raw}_{\sss L} \cong 0.01\,  v_w \frac{1}{g_*}\,
        \frac{M_1^4}{T^5\Delta_w}\,
\label{eq:ans2}
\ee
where $\eta^{\rm raw}_{\sss L}$ is the ratio of $n_L$
to the entropy density, $s=(2\pi^2/45)g_* T^3$. In the right hand
side $M_1$ stands for the Majorana neutrino mass,
$\Delta_w$ is the wall width and $g_*$ the effective
thermodynamic degrees of freedom at the epoch with
temperature $T$. However, the high temperature sphalerons
are efficiently converting the $L$ asymmetry into $B-L$
asymmetry. The standard chemical equilibrium calculation~
\cite{har&tur.90} gives $\eta^{\rm raw}_{\sss B-L} =
\frac{79}{51} \eta^{\rm raw}_{\sss L}$. Using $M_1=f 
\Delta_{\sss T}$, with $\Delta_{\sss T}$ is the temperature
dependent VEV acquired by the $\Delta_{\sss R}$ in the phase
of interest, and $\Delta_w^{-1} = \sqrt{\lambda_{eff}}
\Delta_{\sss T}$ in equation (\ref{eq:ans2}) we get
\be
\eta^{\rm raw}_{\sss B-L} \cong 10^{-4} v_w \left(
\frac{\Delta_{\sss T}}{T} \right)^5 f^4 \sqrt{\lambda_{eff}}.
\ee
Here we have used $g_*=110$. Depending on the Majorana
Yukawa coupling the raw asymmetry can take a range a values
of $O(10^{-4}~-~10^{-10})$.

\section{Results and discussions}
We have assumed a non-thermal production of raw lepton 
asymmetry during the $B-L$ breaking phase transition 
in a generic theory incorporating $B-L$ as a gauge 
symmetry. If this asymmetry passes without much 
dilution to be the currently observed baryon asymmetry 
consistent with WMAP and Big Bang nucleosynthesis, then the 
effective neutrino mass parameter $\tilde{m}_1$ must be less 
than $10^{-3}eV$. Solution of the relevant Boltzmann equations 
shows that for $\tilde{m}_1=10^{-4}eV$ the  mass of lightest 
right handed neutrino $N_1$ has to be smaller than $10^{12}GeV$ 
and can be  as low as $10$ TeV. In a more restrictive scenario 
where the neutrino Dirac mass matrix is identified with that of
the charged leptons it is necessary that $M_1>10^7 GeV$ in order 
to satisfy $\tilde{m}_1 < 10^{-3}eV$. Therefore in the more 
restricted scenario all values $M_1$, $10^8 GeV < M_1 < 10^{12}GeV$ 
can successfully create the required asymmetry. If the Dirac mass 
scale of neutrinos is less restricted, much lower values of 
$M_1$ are allowed. In particular, a right handed neutrino as low 
as 10 TeV is admissible.
 
If the raw asymmetry is large, the numerical solutions show a 
small window for $M_1$ to get the final asymmetry of the required 
order. The allowed range gets smaller as the raw asymmetry gets 
larger. This is true for all allowed values of the neutrino mass 
parameter $\tilde{m}_1$.   

In summary, if the $B-L$ symmetry is gauged, we start with a clean 
slate for $B-L$ number and an asymmetry in it can be generated
by a non-perturbative mechanism at the scale where it breaks.
The presence of heavy right handed neutrinos still permits
sufficient asymmetry to be left over in the form of baryons
for a large range of values of the $B-L$ breaking scales.
While other mechanisms of leptogenesis become unnatural
below $10^7$ GeV (see for example chapter 1 and chapter 2) this 
mechanism even tolerates TeV scale. A specific mechanism of this 
kind is possible in the context of Left-Right symmetric model, 
presumably embedded in the larger unifying group $SO(10)$. 
Here we conjecture that upon incorporation of supersymmetry, the 
qualitative picture will remain unaltered and the present gravitino 
bound of $10^9$ GeV for reheating temperature after inflation 
can be easily accommodated.  
\newpage
~~
\vskip 7\baselineskip
\begin{LARGE}
\pagestyle{empty}
\begin{center}
{\bf Part-II}
\vskip1cm
{\bf Topological Defects}
\vskip2mm
{\bf Cosmic Strings}
\vskip2mm
{\bf and}
\vskip2mm
{\bf Baryogenesis via Leptogenesis}
\end{center}
\end{LARGE}
\addcontentsline{toc}{chapter}{Part II - Topological 
defects, cosmic strings and baryogenesis via leptogenesis}

\chapter{Soliton fermion systems in Quantum Field Theory}
\section{Introduction}
``Solitary waves" and ``solitons" emerge as the non perturbative 
solutions of non-linear wave equations in classical field theory. 
These are non-dispersive localized packets of energy moving
uniformly in space. Although they have many common features, the 
former class of solutions do not keep their shape intact during
a collision among themselves, whereas the latter do in the 
asymptotic time domain. Thus the latter class of solutions is 
a special subset of the former class of solutions, but the 
converse is not necessarily true. Having  differentiated 
solitons from solitary waves, we should mention (without 
embarking into details) that the distinction between them 
in field theory is completely blurred. So here onwards,
what merely called as ``solitary wave", we call it
``soliton" as it is more crisp and appealing. A prototype 
example in field theory is the well known massive Thirring 
model where the solution is exactly solitonic. 

It was believed that elementary particles in nature can be 
thought of as localized packets of energy, the prime among 
such theories is the famous Skyrme model. But till date no
satisfactory formalism has been developed to view elementary 
particles as solitons as the present effective theory takes 
particles as point like, nor is there any experimental evidence 
at latest attainable energies to prove that particles have 
indeed extended features. We bypass this issue here but
concentrate on defects occurring in gauge theories. Of course 
field theories describing elementary particles are quantum 
theories, whereas solitons are solutions of classical field 
theories. Although it is a classical solution, the quantization 
of bosonic fluctuation in the background of this non trivial 
vacuum, the so called ``semiclassical treatment", leads to a 
finite correction~\cite{rajaraman.82}. 

A topological solution can be obtained as a solution of 
a differential equation with a difficulty that the 
boundary values are known only at infinity. In the context 
of field theories a prototype example is the `kink', which emerge 
as the solution of $\phi^{4}$-theory in 1+1 dimensions. For 
demonstration purpose, in section 5.2 we consider a toy model in 
1+1 dimensions. 

A more curious phenomenon happens in Quantum Field Theory (QFT) 
when a fermion interacts with a kink. This was first pointed out 
by Jackiw and Rebbi~\cite{jackiw&rebbi.76} in 1976. It was shown 
that the resulting fermion number becomes fractionalized. 
Although it is surprising, the consequences of this phenomena 
have already been studied extensively in literature in the 
context of condensed matter systems, polyacetylene being the 
standard example~\cite{sushreefer&hegger.80, rice.79,
rajaraman.01,jackiw.99}, and in high energy physics, cosmic 
string being the example~\cite{jackiw.81, witten.85}. 

\section{Solitary waves in 1+1 dimensions}
Soliton or solitary wave is a static, localized and 
finite energy solution of non-linear wave equation in 
classical field theory. For the present purpose we 
consider only bosonic fields $\phi(x, t)$ in one
space and one time dimensions. The dynamics of the fields
is governed by the Lagrangian 
\be
\mathcal{L}=\frac{1}{2}\partial_{\mu}\phi\partial^{\mu}
\phi-V(\phi)\\,
\label{lagrangian}
\ee
where $V(\phi)$ is the required potential. In $\lambda
\phi^{4}$-theory it is given by
\be
V(\phi)=\frac{\lambda}{2}\left(\phi^{2}- \frac{\mu^{2}}
{\lambda}\right)^{2}.
\label{potential}
\ee
The Hamiltonian corresponding to the Lagrangian (\ref{lagrangian}) 
is
\be
\mathcal{H}=\frac{1}{2}\partial_{\mu}\phi \partial^{\mu}
\phi + V(\phi).
\label{hamiltonian}
\ee 
In order to get solitonic solutions we demand that $\phi$
is independent of time. Thus the equation of motion
will be
\be
\frac{\partial^{2}\phi}{\partial x^{2}}=
\frac{\partial V(\phi)}{\partial \phi}.
\label{eqn-motion}
\ee
Again the finiteness of the energy requires that
the energy functional
\be
E[\phi]=\int_{-\infty}^{+\infty}dx\left[\frac{1}{2}\left(
\frac{d\phi}{dx}\right)^{2}+V(\phi)\right]< \infty.
\label{eng-fun}
\ee
If the absolute minima of $V(\phi)$, which are also
its zeros, occur at n points i.e
\begin{equation}
V(\phi)=0\hspace{1cm}
for\hspace{5mm} \phi=g^{i}, i=1,2\cdots n
\label{minima}
\end{equation}
then the energy functional $E[\phi]$ can be minimized
when the field $\phi(x, t)$ is constant in space-time
and takes any one of these values. That is $E[\phi]
=0$ if and only if $\phi(x, t)=g^{i}$. Further to
make $E[\phi]$ finite at spatial infinity
$(x\rightarrow \pm\infty)$, it requires that $V(\phi)=0$
at $x\rightarrow \pm\infty$, which is the absolute
minimum of $V(\phi)$. In addition to that
$\frac{d\phi}{dx}=0$. This implies that $\phi$=constant.
 
We now solve Eqn.(\ref{eqn-motion}). One of the easiest
way to solve this equation is that, if we replace
\bea
\phi &\leftrightarrow & y \nonumber\\
x &\leftrightarrow & t \nonumber\\
V(\phi)&\leftrightarrow & -V(\phi)\nonumber
\eea
in equation (\ref{eqn-motion}) then it will resemble with newtons
second law of motion of a particle of unit mass moving
in a negative potential. The solution $\phi(x)$
represents this particle's motion of this analogue particle. 
So the total energy of this motion is conserved as x, the time, 
varies  and is given by
\be
W=\frac{1}{2}\left(\frac{d\phi}{dx}\right)^{2}-V(\phi).
\ee
Note that this energy is different from the energy
of the soliton. Now using the above boundary conditions
(used for soliton), we have $W=0$. This gives rise to
\be
\int_{0}^{x}dx = \pm \int_{\phi(0)}^{\phi(x)}
\frac{d\phi}{\sqrt{2V(\phi)}}.
\label{int-equation}
\ee
Solving this equation for the particular potential
defined in Eqn.(\ref{potential}) we get
\be
\phi_{\pm}(x) = \pm \frac{\mu}{\sqrt{\lambda}}\tanh(\mu x).
\label{sol-solution}
\ee
The solution corresponding to $\phi_{+}(x)$ is called `kink'
and that of $\phi_{-}(x)$ is called `antikink'. Note that
the coupling constant $\lambda$ appear in the denominator
of the above Eqn.(\ref{sol-solution}). If $\lambda\rightarrow 0$
then the corresponding solution $\phi\rightarrow \infty$.
Thus the phenomenon is completely non-perturbative one.
The energy corresponding to these solutions is given by
\be
E=\int_{-\infty}^{+\infty}\epsilon  dx,
\label{energy}
\ee
where
\be
\epsilon = \frac{\mu^4}{\lambda}{\mathrm sech}^4(\mu x).
\label{sol-mass}
\ee
Sometimes this energy is called classical `kink mass'
$M_{cl} = E = \frac{4\mu^{3}}{3\lambda}$. At
this point it is worth to point out the symmetric
properties between kink and antikink. It is clear
from equation (\ref{sol-solution}) that
\bea
\phi_{+}(x) &=& -\phi_{-}(x)\nonumber \\
             &=& \phi_{-}(-x).
\eea
Thus a kink starts its motion from a trivial vacuum
$\phi=-\frac{\mu}{\sqrt{\lambda}}$ at $x\rightarrow
-\infty$ to reach other trivial vacuum $\phi=\frac{\mu}
{\sqrt{\lambda}}$ at $x\rightarrow +\infty$ through
zero. Although these are extended objects still its
energy is concentrated within narrow region of
space of $O(1/\mu)$. So these solutions is expected to
behave as particles in high energy physics. Fortunately these
solutions admit a Lorentz boost with a definite
velocity ($v$). For example, the energy of the soliton
transforms as 
\be
E\rightarrow E^{'}= \frac{E}{\sqrt{1-v^{2}}}.
\ee
 
It is worth to point out that kink is merely a
solitary wave and not a soliton. That means they
do not survive collisions. Since in our current
discussion (i.e. in context of $\phi^{4}$-theory)
there is only a pair of kink and antikink solution
exist, so it is difficult to check their shape
retaining property in a collision. Of course, a
numerical calculation is needed to check it. But
it is beyond the scope of this manuscript.

\section{Topological charge}
Often we need to distinguish the topological solutions of a 
system of equations. So we define a `topological index', which 
is conserved in time. It is very similar to `quantum number' in 
QFT, but its origin is completely different.
Now we will illustrate it in context of $\lambda\phi^{4}$
-theory. The potential has two degenerate minima, at
$\phi=\pm \frac{\mu}{\sqrt{\lambda}}$. Consequently solutions
of this system, whether static or time dependent, fall
into four topological sectors. These are characterized
by the pair of indices $(-\frac{\mu}{\sqrt{\lambda}}, \frac{\mu}
{\sqrt{\lambda}})$, $(\frac{\mu}{\sqrt{\lambda}},
-\frac{\mu}{\sqrt{\lambda}})$, $(-\frac{\mu}
{\sqrt{\lambda}}, -\frac{\mu}{\sqrt{\lambda}})$,
$(\frac{\mu}{\sqrt{\lambda}}, \frac{\mu}{\sqrt{\lambda}})$
respectively. Where the quantities inside the parenthesis
represents the value of $\phi_{\pm}(x)$ at their spatial
infinity. Now we choose a particular sector $(-\frac{\mu}
{\sqrt{\lambda}}, -\frac{\mu}{\sqrt{\lambda}})$, which
is nothing but the value of kink at $x\rightarrow -\infty$
and antikink at $x\rightarrow \infty$. Even though we
may not be able to calculate easily what happens after
they collide, but we can be sure that the resulting
field configuration will lie in the $(-\frac{\mu}
{\sqrt{\lambda}}, -\frac{\mu}{\sqrt{\lambda}})$ sector.
At this juncture it is worth to designate such sectors
by a definite quantity called `topological charge', defined
by 
\be
Q = \frac{\sqrt{\lambda}}{\mu}\left[\phi(x=+\infty)-
\phi(x=-\infty)\right],
\label{top-charge}
\ee
associated with a conserved current
\be
k^{\alpha} = \frac{\sqrt{\lambda}}{\mu}\epsilon^{\alpha\beta}
\partial_{\beta}\phi\,
\label{top-current}
\ee
where $\alpha$, $\beta$= 0, 1. From equation (\ref{top-current}) 
we get the topological charge
\be 
Q= \int_{-\infty}^{+\infty} k_{0} dx .
\label{}
\ee
Thus we need both $\phi(\infty)$ and $\phi(-\infty)$
to identify a topological sector. If $Q=0$, then the
object is a non topological object. Thus topological charges are 
nothing but the boundary conditions of the problem in contrast to 
usual Noether charge in QFT which comes from the continuous 
symmetry associated with the theory.

\section{Fractional fermion charge in QFT}
The phenomenon of fractional charge was first discovered by 
Jackiw and Rebbi~\cite{jackiw&rebbi.76} in their pioneering work 
in 1976. We begin by reviewing their work. Note that the charges 
we are talking about are not the wellknown and widely accepted 
fractional charge of quarks. But it is the charge which appear 
as the eigen value of number operator in the QFT. For our 
illustration purpose, we consider the 1+1-dimensional theory.
 
As an extension of equation (\ref{lagrangian}) to include 
fermion field $\psi$, the additional terms are given by the 
Lagrangian
\be
\mathcal{L}_{fermion}=\bar{\psi}\left(i\gamma^{\nu}
\partial_{\nu}-g\phi(x,t) \right)\psi.
\label{}
\ee
For the potential defined in equation (\ref{potential}), the
trivial vacuua are $\phi=\pm \frac{\mu}{\sqrt{\lambda}}$.
It is well known that the presence of two degenerate
classical solutions indicate spontaneous breaking
of the $\phi\rightarrow -\phi$ symmetry of the
Lagrangian (\ref{lagrangian}). Around each of these classical
vacuua a whole tower of Fock states can be built and hence 
to be called {\it vacuum sectors}. In addition to that the 
system has also two other static solutions. Those are the 
solitonic solutions, the so called kink and its reflected 
partner the anti-kink, given respectively by
\be
\phi_{s}=\pm \frac{\mu}{\sqrt{\lambda}}\tanh(\mu x).
\label{kink-antikink}
\ee
As per the general theory of semi classical quantization
of quantum fields, one can build two other separate towers
of states, one around each of these solitonic solutions. 
In short, we have four sectors of states for this 
interacting system: two are the vacuum sectors built around  
$\phi=\pm \frac{\mu}{\sqrt{\lambda}}$ and others two are 
the soliton sectors $\pm(\mu/\sqrt{\lambda})\tanh(\mu x)$.

\subsection{The vacuum sector}
In the vacuum sector, $\phi=+\frac{\mu}{\sqrt{\lambda}}$, the 
Lagrangian for the fermion field $\psi$ is given by  
\be
\mathcal{L} =\bar{\psi}\left(i\gamma^{\nu}\partial_{\nu} - m_{F} 
\right)\psi\\, 
\label{e3.3}
\ee
where $m_{F}=g(\mu/\sqrt{\lambda})$ is the mass of the fermion and 
$\nu$ stands for 0,1. Since we have only one space dimension, we 
use the representation of the Dirac matrices as $\gamma^{1}=
\beta \alpha=i\sigma^{3}$ and $\gamma^{0}=\beta=\sigma^{1}$. Let 
us denote by $u_{k}(x)$ and $v_{k}(x)$ the positive and negative 
energy spinorial solutions of the Dirac equations
\begin{eqnarray}
(-i\alpha \partial_{x} + \beta m_{F})u_{k}(x) &=&
E_{k}u_{k}(x)\label{dirac_eqn}\\
(-i\alpha \partial_{x} + \beta m_{F})v_{k}(x) &=&
-E_{k}v_{k}(x),
\label{e3.4}
\end{eqnarray}
where $E_{k} = +\sqrt{k^{2}+m_{F}^{2}}$ and spinor indices
have been suppressed. The Dirac field can be expanded in
terms of these solutions and the destruction operators $b_{k}$
and $d_{k}$, obeying the usual anticommutation rules, as
\begin{equation}
\psi(x, t)=\sum_{k}[b_{k}u_{k}e^{-iE_{k}t} + d_{k}^
{\dagger}v_{k}e^{iE_{k}t}].
\label{e3.5}
\end{equation}
The vacuum state in the $\phi=(+\mu/\sqrt{\lambda})$ sector is given
by the conditions
\begin{equation} 
b_k \state{0}=d_k \state{0}=0
\label{e3.6}
\end{equation}
with all the bosonic oscillators being in the ground state.
 
Note that the third Pauli matrix $\sigma^{3}$ acts as the
charge conjugation matrix. It anticommutes with the
Dirac Hamiltonian in Eq. (\ref{e3.4}) and generates for
every positive energy solution $u_{k}(x)$ of energy
$E_{k}$ the corresponding negative energy solution
$v_{k}(x)$ of energy $-E_{k}$, i.e.
\begin{equation}
\sigma^{3}u_{k}(x)=v_{k}(x).
\label{e3.7}
\end{equation}
Hence all modes of the expansion (\ref{e3.5}) come in pairs
with equal positive and negative energies. There are no zero
energy solutions for $\psi$ in the vacuum sector that allowed 
by the Dirac equations (\ref{dirac_eqn}) and (\ref{e3.4}).
 
Finally we consider the number density operator
\begin{equation}
\rho(x, t)=\frac{1}{2}\left[\psi^{\dagger}(x, t),
\psi(x, t)\right].
\label{e3.8}
\end{equation}
The form of $\rho$ is designed so as to be odd under
charge conjugation. Inserting the mode expansion
(\ref{e3.5}) and using the orthonormality conditions of the
Dirac spinors, the total charge becomes
\begin{eqnarray}
Q &=& \int dx \rho(x, t)\nonumber \\
  &=& \frac{1}{2}\sum_{k}\left([b^{\dagger}_{k}, b_{k}]
+ [d_{k}, d^{\dagger}_{k}]\right)\nonumber\\
  &=& \sum_{k}\left([b^{\dagger}_{k}b_{k}-1/2]
- [d^{\dagger}_{k}d_{k} - 1/2]\right)\nonumber \\
  &=& \sum_{k}\left(b^{\dagger}_{k}b_{k} -
d^{\dagger}_{k}d_{k}\right).
\label{e3.9}
\end{eqnarray}
Notice that the half-integers cancel term by term
because of the existence of paired positive and negative energy
modes. Hence the familiar result in the vacuum sector
that the charge operator has only integer eigenvalues.

\subsection{The soliton Sector}
In this section, we repeat the same procedure as we did 
in the trivial vacuum sector. We replace the $\phi$ by its 
solitonic value $\phi_s$ from Eq. (\ref{kink-antikink}). The 
corresponding Dirac equation becomes
\begin{equation}
\left(-i\alpha \partial_{x} + \beta
m_{F} \tanh(\mu x)\right)\psi(x)= E\psi(x).
\label{e3.10}
\end{equation}
This equation has a set of positive energy solutions 
$\psi_{k}(x)$ associated with energy $E_{k}$. However, the 
charge conjugation matrix $\sigma^{3}$ again anticommutes with 
the Dirac Hamiltonian, $C^{-1}H C=-H$. Therefore, for 
every positive energy solution $\psi_{k}(x)$ there is 
a negative energy solution $\tilde{\psi_{k}}(x)$ with 
energy $-E_{k}$. 

For the two components $\psi_{1, 2}$ of the Dirac spinor, 
equation (\ref{e3.10}) yields the coupled equations
\begin{eqnarray}
\left(-\partial_{x} + m_{F} \tanh (\mu x)\right)
\psi_{2} &=& E\psi_{1}\\
\left(\partial_{x} + m_{F} \tanh (\mu x)\right)
\psi_{1} &=& E\psi_{2}.
\label{e3.11}
\end{eqnarray}
Solving these coupled equations we get an unpaired zero-energy 
solution
\begin{equation}
\psi_{0}=\begin{pmatrix}
N exp \left( -m_{F} \int_{0}^{x} dx' \tanh (\mu x')\right)\\
0\end{pmatrix}.
\label{e3.12}
\end{equation}
In contrast to the normal vacuum, in the soliton sector 
we have a zero energy solution (\ref{e3.12}). This is 
because of the soliton function $(\mu/\sqrt{\lambda}) 
\tanh(\mu x)$ which forms the background potential for the Dirac 
spinor tends to opposing limits $\pm(\mu / \sqrt {\lambda})$ as 
$x \rightarrow \pm\infty$. In infinite spatial volume this 
solution has no partner. It is self charge conjugate i.e. 
$\sigma^{3}\psi_{0}=\psi_{0}$. The mode expansion of the Dirac 
field operator now becomes
\begin{equation}
\psi(x, t)= \sum_{k\neq 0}\left[b_{k}\eta_{k}(x)e^{-iE_{k}t}
+ d^\dagger_{k}\tilde{\eta_{k}}(x)e^{iE_{k}t}\right]
+a\psi_{0}(x)
\label{e3.13}
\end{equation}
where `a' is the destruction operator for the zero mode $\psi_0$.
 
Unlike the vacuum sector built around  $(\mu/\sqrt{\lambda})$
which had a unique ground state, in the solitonic sector
there exists two degenerate ground states because of the
existence of a zero energy solution of $\psi$. They are 
$\state{sol}_{-}$, being the unfilled state, and 
$\state{sol}_{+}$, being the filled state, obeying
\begin{equation}
a_k\state{sol}_{-}=b_{k}\state{sol}_{-}=d_{k}\state{sol}_{-}=0
\label{e3.14}
\end{equation}
and
\begin{eqnarray}
\state{sol}_{+} &=& a^\dagger\state{sol}_{-}\nonumber\\
a \state{sol}_{+} &=& \state{sol}_{-}.
\label{e3.15}
\end{eqnarray}
These are the two basic quantum soliton states of this
system. They are energetically degenerate, but are
distinguishable by their charge
\begin{eqnarray}
Q &=& \frac{1}{2}\int dx \left[\psi^{\dagger}(x, t) ,
\psi(x, t) \right]\nonumber \\
  &=& \frac{1}{2}\sum_{k}\left( [b^{\dagger}_{k} , b_{k}]  
+ [d_{k} , d^{\dagger}_{k}] \right) + (1/2)[a^{\dagger}, a]
\nonumber \\
  &=& \sum_{k}\left( (b^{\dagger}_{k}b_{k}-1/2) -
 (d^{\dagger}_{k}d_{k} - 1/2) \right) + (a^{\dagger}a - 1/2)
\nonumber\\
  &=& \sum_{k}\left( b^{\dagger}_{k}b_{k} - d^{\dagger}_{k}d_{k}
\right) + (a^{\dagger}a - 1/2)
\label{e3.16}
\end{eqnarray}
Notice that the piece (-1/2) coming from the zero mode
commutator remains uncanceled because it does not have
a charge conjugate partner. Thus it is obvious that the
total charge (the number operator) Q has half integral eigen
values. The two degenerate soliton states have eigen
values $\pm 1/2$ respectively for the total number
operator Q:
\begin{eqnarray}
Q\state{sol}_{-} &=& -(1/2)\state{sol}_{-}\nonumber\\
Q\state{sol}_{+} &=&  (1/2)\state{sol}_{+}
\label{e3.17}
\end{eqnarray}
Thus the solitonic states are superselected from the 
normal vacuum due to the unavailability of fractional 
states in the translational invariant vacuum. Note that the 
fractionalization of the fermion number in the background of 
a nontrivial vacuum is one of the manifestation of Dirac 
negative energy sea.

\section{Implication of fermion zero modes and superselection 
rules}
By the name `superselection' we mean that there are
certain restrictions on nature and scope of possible
measurements. It is shown that~\cite{wicketal.52} there
exist superselection rules for spinor fields which can
be proved in context of the assignment of parity to the
quantum mechanical states of a system.
 
The usual assumption in quantum mechanics is that it is
possible to carry out a complete set of measurements; the
result of which determines the state of any vector completely
except for the usual phase factor. If there exist sectors
$A$, $B$, $C$, etc. in the Hilbert space such that vectors
in each subspace can be independently rescaled by phases,
such sectors must be understood to be completely independent,
and superposition of states from such unrelated sectors
should be forbidden. Specifically, let state vectors
$\Psi_{A}$, $\Psi_{B}$,$\Psi_{C}$ etc. belong to such
independent sectors, so that it is assumed that no
physical measurements distinguish between the state
\begin{equation}
\Psi_{A} + \Psi_{B} + \Psi_{C} + \ldots
\end{equation}
and
\begin{equation}
e^{i\alpha}\Psi_{A} + e^{i\beta}\Psi_{B}
+ e^{i\gamma}\Psi_{C} + \ldots
\end{equation}
where $\alpha$, $\beta$, $\gamma$, $\ldots$ are arbitrary
phases. Then the expectation value of any operator possessing
matrix elements connecting subspaces $A$ and $B$, or $A$
and $C$, etc.  must be completely undefined. Hence such an
operator will not correspond to a measurable quantity.
 
It is customary to say that a selection rule operates
between subspace of the total Hilbert space if the state 
vectors of each subspace remain orthogonal to all state
vectors of other subspaces as long as the system is isolated.
There is, for instance, a selection rule which prevents any
state of an isolated system from changing its total linear
momentum. Similarly, the  state vectors of the subspace
containing all states with total angular momentum J will
remain, in a closed systems, orthogonal to all states with
any other total angular momentum. So we shall say that
a superselection rule operates between subspaces if
there does not exist any selection rule between them and
if, in addition to this, there are no measurable quantities
with finite matrix elements between their state vectors.
For example if we superpose states  of integral spin (bosonic)
with half integral spin (fermionic) then the probability
of finding of each states are equal and 1/2 each. So
it is impossible to compare directly a spin half particle
with a spin one particle. It is conjectured that~\cite{wicketal.52} 
superselection rule not only work between the states of 
different intrinsic parity but also it works between the 
states of different total charge. For instance if we 
construct a state, by superposing the states $\state{1/2}$ 
and $\state{1}$, as
\begin{equation}
\psi_{s} = \frac{1}{2}[\state{1/2}+\state{1}]
\label{state}
\end{equation}
then it is expected that under $P^{4}$ operation it
will come back to it's original state up to a phase.
To verify this we can assign a parity $\pm 1$ to
a Dirac neutrino and $\pm i$ to a Majorana neutrino
in the trivial vacuum sector. But in the vortex sector
the corresponding state is doubly degenerate and thus
the parity assignment is half of their trivial vacuum
values. Now the  $P^{4}$ operation on (\ref{state})
for a Majorana neutrino will give
\begin{equation}
P^{4}\psi_{s} = \frac{1}{2}[-\state{1/2}+\state{1}].
\end{equation}
This state is another state orthogonal to (\ref{state})
and hence implies the impossibility of superposition of
such states. We discuss more about it in context of 
the decay of cosmic strings in chapter 6.

\chapter{Metastable topological defects and fermion zero modes}
\section{Introduction}
In the process of cooling down from the highest possible energy
scale till the present epoch, early Universe has been passed 
through several phase transitions. A phase transition occurs 
through the spontaneous breaking of a larger gauge symmetry 
group $G$ to any of its subgroup $H$. If the corresponding 
vacuum manifold 
${\cal M}=G/H$ is non trivial then the topological defects are 
formed. Depending on the geometry of the manifold the possible 
defects are domain walls, cosmic strings, monopoles and textures. 
Domain walls form if ${\cal M}$ has disconnected components, 
strings can form if ${\cal M}$ has unshrinkable loops and 
monopoles form when ${\cal M}$ contains unshrinkable surfaces. 
The relevant properties of the manifold ${\cal M}$ are most 
conveniently studied using homotopy theory; the $n^{th}$ 
homotopy group $\pi_n({\cal M})$ classifies qualitatively 
distinct mappings from the n-dimensional sphere $S^n$ into the 
manifold ${\cal M}$. If $\pi_0({\cal M})\neq I$ then the formed 
defects are called domain walls. On the other hand, if 
$\pi_1({\cal M})\neq I$ then cosmic strings are formed and 
if $\pi_2({\cal M})\neq I$ then monopoles are formed. These are 
extended objects, cosmic strings being 1+1-dimensional, domain 
walls the 2+1-dimensional and finally monopoles the 
3+1-dimensional but not point like. 

The topological defects emerge as the solitonic solutions of 
non linear wave equations in gauge theories. Therefore, in the 
asymptotic time domain these objects are very stable and keep 
their original configurations in tact. However, they can be unwound 
by overcoming finite energy barriers. The dynamics of these 
objects may have several consequences that may link to the physics 
of early Universe with the recent cosmology. In particular, we 
discuss baryogenesis via leptogenesis from the decay products 
of cosmic strings in chapter 7. 

\section{Metastable topological defects}
If there are several stages of symmetry breaking, the 
topological stability of these extended objects depend on 
the structure of the vacuum manifold at every stage. Under 
certain conditions defects stable at the first stage of 
symmetry breaking are rendered unstable at lower temperatures. 
Likewise objects that seem to enjoy topological stability 
in the low energy effective theory are actually unstable in 
the complete description of the theory. The instability however 
tends to energetically exorbitant thus rendering the object 
metastable~\cite{pres&vil.92}.
 
There are several reasons that may stabilize objects
that are unstable on topological grounds. For instance
superconducting strings lead to small loops called vortons,
stabilized  by the electric current flowing through them.
Here we consider the occurrence of fermionic zero-energy
modes. As a rule of thumb, if the number of trapped
zero-energy modes is $n$, the ground state of the defect
carries fermion number $n/2$. Thus if $n$ is odd one
obtains curious occurrence of fractionalization of fermion
number, proven to be integer in translation invariant
field theory. In general there are several internal
charges the fermions may carry and the fractionalization
rule affects all of them. It has been observed by several
authors~\cite{devega.78,stern_prl.83,stern&yaj.86} that the 
unavailability of final states of matching charge should 
forbid the decay of such objects in isolation. 

Here we consider the sequential breakdown of the gauge
group $G\rightarrow H_{1} \rightarrow H_{2}$. If the
vacuum manifold ${\cal M}=G/H_{1}$ is non-trivial then
topological defects are formed. However, these defects
in their corresponding low energy theories loss their
stability after the breaking of second phase of symmetry
if the manifold $G/H_{2}$ is simply connected and thus 
making the defect metastable. As an example we study the 
metastable cosmic strings in sec. 6.2.1. 

\subsection{Metastable cosmic strings and fermion zero modes}
Here we construct two examples in which the cosmic strings 
of a low energy theory are metastable due to the embedding of 
the low energy symmetry group in a larger symmetry group at 
higher energy. Examples of this kind were considered 
in~\cite{pres&vil.92}. Borrowing the strategies for bosonic sector 
from there, we include appropriate fermionic content to ensure 
the zero-modes.

\subsection*{\textbf{A.} Dirac fermions}
Consider first a model with two stages of symmetry breaking 
similar to~\cite{pres&vil.92}, but with local $SU(2)$ gauge 
invariance. The two scalars $\vec \Sigma$ and $\sigma$ are 
respectively real triplet and complex doublet. The Lagrangian 
is taken to be
\begin{eqnarray}
\mathcal{L} &=&
-\frac{1}{4} F^{\mu\nu a}F^a_{\mu\nu}
+  \frac{1}{2} D_\mu {\vec \Sigma} \cdot D^\mu \vec\Sigma
+ D_\mu\sigma^\dag D^\mu\sigma   \nonumber \\
&-&\lambda_1(\vec\Sigma\cdot\vec\Sigma - \eta_1^2)^2
-\lambda_2(\sigma^\dag\sigma - \eta_2^2)^2  \nonumber \\
&+&\lambda_{12}\eta_1 \vec\Sigma\cdot\sigma^\dag{\vec \tau}\sigma
\label{case_one}
\end{eqnarray}
where isovector notation is used and $\tau$ are the Pauli
matrices. It is assumed that $\eta_1\gg\eta_2$ and that the
coupling $\lambda_{12}>0$ satisfies $\lambda_{12}\eta_1^2\ll$
$\lambda_2\eta_2^2$ 

The vacuum expectation value (VEV)
$\vec\Sigma = (0\  0\ \eta_1)^T$ breaks the $SU(2)$ to the 
$U(1)$ generated by $\exp(i\tau^3 \alpha/2)$. The effective 
theory of the $\sigma$ can be rewritten as the theory of two 
complex scalar $\sigma_u$ and $\sigma_d$ for the up  and the 
down components respectively. The potential of the effective 
theory favors the minimum
\be
\sigma_u=\eta_2, \qquad \sigma_d=0
\label{eq:vev}
\ee
In the effective theory $\sigma_u$ enjoys a $U(1)$ invariance
$\sigma_u \rightarrow e^{i\alpha}\sigma_u$ which is broken by the
above VEV to $\mathbb{Z}\equiv\{e^{2n\pi i}\}$, $n=1,2,\ldots$.
This makes possible vortex solutions with an ansatz in the 
lowest winding number sector
\be
\sigma_u(r, \phi) = \eta_2 f(r) e^{-i\phi}
\label{eq:str_ansatz} 
\ee
where $r, \phi$ are planar coordinates with vortex aligned along 
the $z$ axis. The vortex configuration is a local minimum, 
however it can decay by spontaneous formation of a 
monopole-antimonopole pair~\cite{pres&vil.92}. These monopoles 
are permitted by the first breaking $SO(3)\rightarrow SO(2)$ 
in the $\Sigma$ sector. Paraphrasing the discussion of~
\cite{pres&vil.92}, we have $SU(2)\rightarrow$$U(1)\rightarrow 
\mathbb{Z}$. The vortices are stable in the low energy 
theory because $\pi_1(U(1)/\mathbb{Z})$ is nontrivial. But 
in the $SU(2)$, the $\mathbb{Z}$ lifts to $\{e^{4n\pi i
\tau^3/2}\}=I$ making it possible to unwind the vortex by 
crossing an energy barrier.
 
Consider now the introduction of a doublet of fermion species
$\psi_L^T\equiv (N_L, E_L)$ assumed to be left handed and
a singlet right handed species $N_R$. The Yukawa coupling of 
these to the $\sigma_u$ is given by $h{\overline{N_R}}
\sigma^\dag\psi_L$, which in the vortex sector reads
\be
\mathcal{L}_{\sigma-\psi} \sim h \eta_2 f(r)(e^{-i\phi}
{\overline{N_R}} N_L + h.c. )
\label{fermion_case_one}
\ee
The lowest energy bound states resulting from this coupling
are characterized by a topological index,~\cite{weinberg.81}
\( \mathcal{I} \equiv n_L - n_R\) where $n_L$ and $n_R$ are 
the zero modes of the left handed and the right handed fermions 
respectively. This index can be computed using the 
formula~\cite{weinberg.81,ganoulis.88}
\be
\mathcal{I} = \frac{1}{2\pi i}(\ln \textrm{det}M)
\vert^{2\pi}_{\phi=0}
\label{eq:index}
\ee
where  $M$ is the position dependent effective mass matrix for 
the fermions. In the present case this gives rise to a single 
zero-energy mode for the  fermions of species $N$. According 
to well known reasoning~\cite{jackiw&rebbi.76} to be recapitulated
below, this  requires the assignment of either of the values
$\pm1/2$ to the fermion number of this configuration.
 
\subsection*{\textbf{B.} Majorana fermions}
The example above can be extended to the case where the $N_R$ 
is a Majorana fermion. Being a singlet $N_R$ admits a mass term
$M_{\ssM} {\overline{N_R^{\ssC}}}N_R$, $M_{\ssM}$ signifying 
Majorana mass. This could also be a spontaneously generated 
mass due to the presence of a neutral scalar $\chi$ with 
coupling terms $h_{\ssM}\chi{\overline{N_R^{\ssC}}}N_R + 
h.c.$. If this $\chi$ acquires a VEV at energies higher than 
the $\Sigma$, the $N_R$ particles possess a Majorana mass and
fermion number is not a conserved observable.
 
Finally we present the case where Majorana mass is spontaneously
generated at the same scale at which the vortex forms.
Consider a theory with local $SU(3)$ symmetry broken
to $U(1)$ by two scalars, $\Phi$ an octet acquiring a VEV 
$\eta_1\lambda_3$ ($\lambda_3$ here being the third 
Gell-Mann matrix) and $\phi$, a ${\bar 3}$, acquiring the VEV 
$\langle\phi^k\rangle=\eta_2\delta^{k 2}$, with 
$\eta_2\ll \eta_1$. Thus
\be
SU(3) {\buildrel 8 \over{\longrightarrow}}
U(1)_3\otimes U(1)_8 {\buildrel \overline{3} 
\over{\longrightarrow}} U(1)_+
\ee
Here $U(1)_3$ and $U(1)_8$ are  generated by $\lambda_3$
and $\lambda_8$ respectively, and $U(1)_+$ is generated by
\( (\sqrt{3}\lambda_8 + \lambda_3)/2 \) and likewise $U(1)_-$
to be used below.
It can be checked that this pattern of VEVs can be generically
obtained from the quartic scalar potential of the above Higgses.
The effective theory at the second breaking $U(1)_-\rightarrow 
\mathbb{Z}$ gives rise to cosmic strings. However the 
$\mathbb{Z}$ lifts to identity in the $SU(3)$ so that the string 
can break with the formation of monopole-antimonopole pair.
 
Now add a multiplet of left-handed fermions belonging to 
$\overline{15}$. Its mass terms arise from the following coupling 
to the $\overline{3}$\be \sL_{\textrm{Majorana}} = h_{\ssM} 
\overline{\psi^{\ssC}}^{\{ij\}}_{k}\psi^{\{lm\}}_{n}\phi^{r}
(\epsilon_{ilr}\delta^{n}_{j}\delta^{k}_{m})
\ee
The indices symmetric under exchange have been indicated by 
curly brackets. No mass terms result from the $8$ because it 
cannot provide a singlet from tensor product with $\overline{15}
\otimes \overline{15}$~\cite{slansky_rep}. After substituting the 
$\phi$ VEV a systematic enumeration shows that all but the two 
components $\psi^{\{22\}}_{1}$ and $\psi^{\{22\}}_{3}$ acquire   
Majorana masses at the second stage of the breaking. Specifically
we find the Majorana mass matrix to be indeed rank $13$. Thus, using
either of the results~\cite{jackiw.81} or~\cite{ganoulis.88} 
i.e.,\  eq.(\ref{eq:index}) we can see that there will be 
$13$ zero modes present in the lowest winding sector of 
the cosmic string. Thus the induced fermion number differs 
from that of the vacuum by half-integer as required. 

\subsection*{\textbf{C.} Final state zero-modes}
\begin{figure}[htb]
{\par\centering \resizebox*{0.3\textwidth}{!}
{\rotatebox{0}{\includegraphics{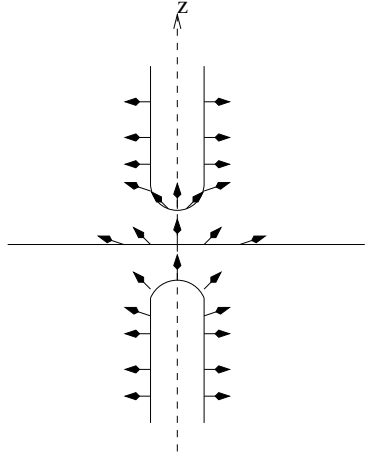}}} \par}\vspace{0.3cm}
\caption{Schematic configuration of isospin vectors after the
rupture of a string. Internal orientations are mapped to external
space. They are shown just outside the core of the two resulting
pieces and on the  mid-plane symmetrically separating the two.}
\label{fig:string}
\end{figure}

The stability argument being advanced is in jeopardy if the 
final state after rupture of the topological object also 
possesses half-integral fermionic charge. To see that this 
is not the case it is necessary to study the zero-modes on the 
two semi-infinite strings shown in fig. \ref{fig:string}. 
Generically we expect each of the halves to support the same 
number of zero-modes, making the total fermion number of the 
putative final state integer valued, as required for the validity 
of our argument. 

Consider the ansatz for the lower piece ($l$) with origin at the
corresponding monopole and coordinates \((r_l, \theta_l, \phi)\).
For the domain $z<0$ let the ansatz for the field $\sigma$ be
\be
U_{l}^\infty(\theta_l,\phi)\begin{pmatrix}
0 \\ 
\eta_2\end{pmatrix} f_l(r_l)
\equiv  \exp\{\frac{i}{2}\theta_l {\vec \tau}\cdot \hat{\phi} \}
\begin{pmatrix}
0 \\ 
\eta_2\end{pmatrix} f_l(r_l)
\ee
so that \( \langle \sigma \rangle \) has the behavior
\begin{displaymath}
\langle\sigma\rangle = \left\{ 
\begin{array}{lr}
\begin{pmatrix}0 \\ \eta_2 \end{pmatrix}f_l(r_l) & \textrm{for}\ 
\theta_l\approx 0\\
& \\
\begin{pmatrix}\eta_2e^{-i\phi} \\ 0\end{pmatrix}f_l(r_l)  & 
\textrm{for}\ \theta_l\approx \pi
\end{array}
\right.
\end{displaymath}
which agrees with the ansatz (\ref{eq:str_ansatz}) for the 
cosmic string at the South pole. Likewise for the domain 
$z>0$, i.e. the upper piece ($u$), we choose
\be
U_{u}^\infty(\theta_u,\phi) = \exp\{\frac{i}{2}(\pi-\theta_u)
{\vec \tau}\cdot \hat{\phi} \}
\ee
resulting in the behavior
\begin{displaymath}
\langle\sigma\rangle = \left\{ \begin{array}{lr}
\begin{pmatrix}\eta_2e^{-i\phi} \\ 0 \end{pmatrix} f_u(r_u) & 
\textrm{for}\  \theta_u\approx 0\\
 &  \\
\begin{pmatrix}0 \\ \eta_2\end{pmatrix}f_u(r_u) & \textrm{for}\  
\theta_u\approx \pi
\end{array}
\right.
\end{displaymath}
thus matching correctly with the cosmic string at the North pole.
The ansatz for the heavier scalar $\Sigma$ needs to be 
appropriately set up, \( U^\infty {\vec\Sigma} \cdot{\vec\tau} 
{U^\infty}^\dag \) in both $l$ and $u$ domains. This scalar 
however does not contribute to fermion mass matrix.
 
The two maps match at the mid-plane where $\theta_l=\pi - \theta_u$
and \( \sigma_u \sim e^{-i\phi} \) at \( \theta_l=\theta_u =\pi/2\)
so that we have ensured that the combined map is within the same
homotopy class as the string we began with. Finally, as the two
pieces move far away, each can be seen to have the same number of
zero-modes. To see this we can choose~\cite{DHN_74,nohl.75} fermion
ansatz for the zero-modes compatible with the scalar field ansatz,
in each of the patches $l$ and $u$. In (isospin)$\otimes$(two 
component spinor) notation for $\psi_L$ and for the two component 
fermion $N_R$,
\be
\psi_L = U^\infty(\theta,\phi) \begin{pmatrix}0 \\ 1\end{pmatrix}
\otimes \begin{pmatrix}\varphi_1(r) \\ \chi_1(r)\end{pmatrix}
\quad
N_R = \begin{pmatrix}\varphi_2(r) \\ \chi_2(r)\end{pmatrix}
\ee
where the labels $l$, $u$ have been dropped. To analyse the 
asymptotic radial dependence choose $\gamma^{\ r}=\sigma^2$ 
the Pauli matrix. In each patch one finds the pair 
\( \varphi_1(r), \chi_2(r) \sim e^{-h\eta_2 r}\) to constitute 
the zero-mode while for the other pair, \( \varphi_2(r), \chi_1(r)
\sim e^{+h\eta_2 r}\) which are therefore not normalizable.
In any case, since each of the pieces acquires  the same number
of zero-modes, the total fermion number of the putative
final state has been proved to be integer as required.
 
\section{Assignment of fermion number}\label{sec:ferno}
We now recapitulate the reasoning behind the assignment of
fractional fermion number. We focus on the Majorana fermion
case, which is more nettlesome, while the treatment of the
Dirac case is standard~\cite{jackiw&rebbi.76,jackiw.77}. 
In the prime example~\cite{jackiw.81} in $3+1$ dimensions of a single left-handed 
fermion species $\Psi_L$ coupled to an abelian Higgs model 
according to 
\be
\mathcal{L}_\psi = i\overline{\Psi_L}\gamma^\mu D_\mu \Psi_L
-\frac{1}{2}( h \sigma \overline{\Psi_L^C}\Psi_L + h.c.)
\label{fermion_case_two}
\ee
the following result has been obtained. For a vortex oriented 
along the $z$-axis, and in the winding 
number sector $n$, the fermion zero-modes are of the form
\be
\Psi_{L0}({\bf x})=\begin{pmatrix}1\\ 0 \end{pmatrix}
\left[U(r)e^{il\phi}+V^*(r)e^{i(n-1-l)\phi}\right]g_l(z+t)
\ee
In the presence of the vortex, $\tau^3$ (here representing
Lorentz transformations on spinors) acts as the
matrix which exchanges solutions of positive frequency
with those of negative frequency. It is therefore identified
as the ``particle conjugation'' operator. In the above ansatz, 
the $\Psi_L$ in the zero-frequency sector are charge self-conjugates, 
$\tau^3\Psi_{L0}=\Psi_{L0}$, and have an associated left 
moving zero mode along the vortex. The functions satisfying 
$\tau^3\Psi_{L0}=-\Psi_{L0}$ are not normalizable. The 
situation is reversed when the winding sense of the scalar 
field is reversed, ie, for $\sigma_u\sim$ 
$e^{-in\phi}$. In the winding number sector $n$, regular 
normalizable solutions~\cite{jackiw.81} exist for for $0\leq l
\leq n-1$. The lowest energy sector of the vortex is
now \(2^n\)-fold degenerate, and each zero-energy mode needs to be
interpreted as contributing a value $\pm 1/2$ to the total
fermion number of the individual states~\cite{jackiw.81}. This 
conclusion is difficult to circumvent if the particle spectrum 
is to reflect the charge conjugation symmetry of the theory~
\cite{sud&yaj.86}. The lowest possible value of the induced number 
in this sector is $-n/2$. Any general state of the system is 
built from one of these states by additional integer number 
of fermions. All the states in the system therefore possess 
half-integral values for the fermion number if $n$ is odd. 

One puzzle immediately arises, what is the meaning of
negative values for the fermion number operator for
\textit{Majorana} fermions? In the trivial vacuum, we can
identify the Majorana  basis as
\be
\psi\ =\ \frac{1}{2}(\Psi_L +  \Psi_L^C).
\label{majdef}
\ee
This leads to the Majorana condition which results in
identification of particles with anti-particles according to
\be
\sC \psi \dsC\  =\   \psi
\label{majcond}
\ee
making negative values for the number meaningless. Here 
$\sC$ is the charge conjugation operator. We shall first verify 
that in the zero-mode sector we must indeed assign negative 
values to the number operator. It is sufficient to treat
the case of a single zero-mode, which generalizes easily
to any larger number of zero-modes. The number operator 
possesses the properties
\be
[ N, \psi ]\ =\ -\psi\qquad{\rm and}
\qquad [ N, \psi^\dagger ]\ =\ \psi^\dagger
\label{psiN}
\ee
\be
\sC N \dsC\ =\ N
\label{ccN}
\ee        
Had it been the Dirac case, there should be a minus sign
on the right hand side of eq. (\ref{ccN}). This is absent due
to the Majorana condition. The fermion field operator for the
lowest winding sector is now expanded as
\be
\psi\ =\ c\psi_0 + \left\{ \sum_{{\bf \kappa},s}
a_{{\bf \kappa},s}\chi_{{\bf \kappa},s}(x)\ \\
+\ \sum_{{\bf k},s} b_{{\bf k},s}u_{{\bf k},s}(x)\
+\ h.c. \right\}
\label{eq:expansion}
\ee
where the first summation is over all the possible bound states of
non-zero frequency with real space-dependence of the form $\sim
e^{-{\bf \kappa\cdot x_\perp}}$ in the transverse space directions
${\bf x}_\perp$, and the second summation is over all
unbound states, which are asymptotically plane waves.
These summations are suggestive and their exact connection to
the Weyl basis mode functions~\cite{Fuk&Suz_book} are not essential
for the present purpose. Note however that no "$h.c.$" is needed 
for the zero energy mode which is self-conjugate. Then the Majorana 
condition (\ref{majcond}) requires that we demand
\be
\sC\ c\ \dsC\ =\   c \qquad{\rm and}
\qquad \sC\  c^\dagger\  \dsC\ =\  c^\dagger
\label{ccc}
\ee 
Unlike the Dirac case, the $c$ and $c^\dagger$ are not
exchanged under charge conjugation. The only non-trivial 
irreducible realization of this algebra is to require the 
existence of a doubly degenerate ground state with states 
$\state{-}$ and $\state{+}$ satisfying
\be
c\state{-}\ =\ \state{+}\qquad {\rm and}\qquad
c^\dagger\state{+}\ =\ \state{-}
\label{cstates}
\ee
with the simplest choice of phases. Now we find
\begin{eqnarray}
\sC\  c\  \dsC \sC \state{-}\ &=\ \sC\state{+}\\ \vspace{3mm}
\Rightarrow \hspace{2mm}  c (\sC \state{-})\ &=\ (\sC\state{+})
\label{Ctransform}
\end{eqnarray}
This relation has the simplest non-trivial solution
\be
\sC\state{-}\ =\ \eta^-_{\ssC} \state{-}\qquad
{\rm and}\qquad \sC\state{+}\ =\ \eta^+_{\ssC} \state{+}
\label{Cproperty}
\ee
where, for the consistency of  (\ref{cstates}) and (\ref{Ctransform})
$\eta^-_{\ssC}$ and $\eta^+_{\ssC}$ must satisfy
\be
(\eta^-_{\ssC})^{-1}\eta^+_{\ssC}\ =\ 1
\ee
Finally we verify that we indeed get values $\pm 1/2$ for $N$.
The standard fermion number operator which in the Weyl basis is
\be
N_F = \frac{1}{2}[ \Psi_L^\dag \Psi_L - \Psi_L \Psi_L^\dag ]
\ee
acting on these two states gives,
\be
\frac{1}{2}(c\ c^\dagger\  -\  c^\dagger\  c)\ \state{\pm}\ =\
\pm\frac{1}{2}\state{\pm}
\ee
The number operator indeed lifts the degeneracy of the
two states. For $s$ number of zero modes, the ground
state becomes $2^s$-fold degenerate, and the fermion number
takes values in integer steps ranging from $-s/2$ to $+s/2$.
For $s$ odd the values are therefore half-integral.
Although uncanny, these conclusions accord with some known
facts. They can be understood as spontaneous symmetry breaking
for fermions~\cite{wilczek.84}. The negative values of the number 
thus implied occur only in the zero-energy sector and do not
continue indefinitely to $-\infty$. Instead of an unfathomable
\textit{Dirac sea} we have a small \textit{Majorana pond}
at the threshold.

\section{Quantum mechanical stability}\label{sec:indsta}
The theory of eq. (\ref{case_one}) possesses a gauge symmetry which
is  reflected in the effective theory (\ref{fermion_case_one}) as
$N_L\rightarrow$ $ e^{i \alpha} N_L$, $N_R\rightarrow$$ e^{i \alpha} 
N_R$ giving rise to the usual conserved number for Dirac fermions.
The lowest winding vortex sector results in half-integer values for 
this number. Quantum Mechanical stability of this sector follows 
from well known arguments~\cite{wicketal.52,Sweinberg_book} 
which can now be understood as either following from 
distinctness of sectors of 
different values of $(-1)^{N_F}$, or as a consequence of a 
residual subgroup of the gauge symmetry. For the Majorana case we 
shall now carry out this kind of argument explicitly.
 
It is known that Majorana fermions can be assigned a unique
parity~\cite{Sweinberg_book},  either of the values $\pm i$.
Accordingly let us choose $i$ to be the parity of the free 
single fermion states in the trivial vacuum.

As a step towards deriving our superselection rule, we 
determine the parities of the zero-energy states. The fermion 
spectrum should look the same as trivial vacuum far away from 
the vortex. In turn the parities of the latter states should 
be taken to be the same as those of the trivial ground state. 
Next, any of these asymptotic free fermions is capable of being 
absorbed by the vortex (see for instance~\cite{Davis:1999ec}). 
In the zero energy sector this absorption would cause a 
transition from \( \state{-1/2} \) to \( \state{1/2} \)
and cause a change in parity by $i$.
Thus the level carrying fermion number \(+1/2\) should be
assigned a parity $e^{i\pi/2}$ relative to the \(-1/2\) state.
Symmetry between the two states suggests that we assign parity
$e^{i\pi/4}$ to the $N_F=1/2$ and $e^{-i\pi/4}$ to the $N_F=-1/2$
states. 

Similar reasoning applies to a residual discrete symmetry belonging
to the original $U(1)$ gauge group of Lagrangian 
(\ref{fermion_case_two}). According to eq. (\ref{majdef}), under  
gauge transformation,
\be
\psi \rightarrow \psi_{[\alpha]} \equiv
\frac{1}{2}(e^{i\alpha}\Psi_L + e^{-i\alpha}\Psi_L^C)
\ee
Thus $\alpha=\pi$ preserves the choice of the Majorana  basis
upto a sign. After  symmetry breakdown and  Higgs mechanism, the
Yukawa coupling takes the form $\sim (m+{\tilde \phi})
\overline{\psi} \psi$, which is invariant under the residual 
$\mathbb{Z}_2$ symmetry $\psi$  $\rightarrow -\psi$. We can use
this as a discrete symmetry distinguishing states of even and odd
Majorana fermions. Since single Majorana fermions can
be absorbed by the vortex~\cite{Davis:1999ec}, the ground states 
$|\pm \rangle$ are distinguished from each other by a relative 
negative sign. To be symmetric we can assign the value $\pm i$ 
to these states under this discrete symmetry with sign same as 
in the value of the number operator. It is possible to prove the 
superselection rule using this conserved quantity. However we also 
see that this discrete symmetry can be used to change our 
convention of the parity for free Majorana particles from $+i$ 
to $-i$. Thus the two are intimately related and in what follows 
we shall use the parity with convention as in the preceding 
paragraph.

We now show the inappropriateness of superposing states of
half-integer valued fermion number and integer valued fermion
number \cite{wicketal.52}. The operation $\sP^{\displaystyle4}$, 
parity transformation performed four times  must return the system
to the original  state, upto a phase. Consider forming the
state $\psi_{\ssS}=$$\frac{1}{\sqrt{2}}(\state{1/2}+\state{1})$
from states of half-integer and integer value for the fermion number.
But
\be
\sP^4\ \psi_{\ssS}\ =\ \frac{1}{\sqrt{2}}(-\state{1/2}+\state{1})\
\ee
Thus this operation identifies a state with another orthogonal
to it. Similarly, application of $\sP^{\displaystyle2}$ which
should also leave the physical content of a state unchanged
results in yet another linearly independent state,
$\frac{1}{\sqrt{2}}(i\state{1/2}-\state{1})$.
Thus the space of superposed states collapses to a trivial
vector space. The conclusion therefore is that it is not
possible to superpose such sectors. In turn there can be
no meaningful operator  possessing non-trivial matrix
elements between the two spaces. This completes our proof
of the theorem.   

\chapter{B-L cosmic strings and baryogenesis via leptogenesis}
\section{Introduction}
Cosmic strings, which are one dimensional topological defects, 
are expected to be formed during the early Universe
phase transitions. When a larger symmetry group $G$ breaks to $H$ 
and if the corresponding vacuum manifold ${\cal M}=G/H$ is not 
simply connected (for definition see Appendix C) then the 
formation of cosmic strings is assured. 

The presence of cosmic string in the real physical space can be 
ensured by encircling it with a closed path (for definition see 
Appendix C) or loop f(t) as shown in the figure~
\ref{fig:cosmicstring}
\begin{figure}[h]
\begin{center}
\epsfig{file=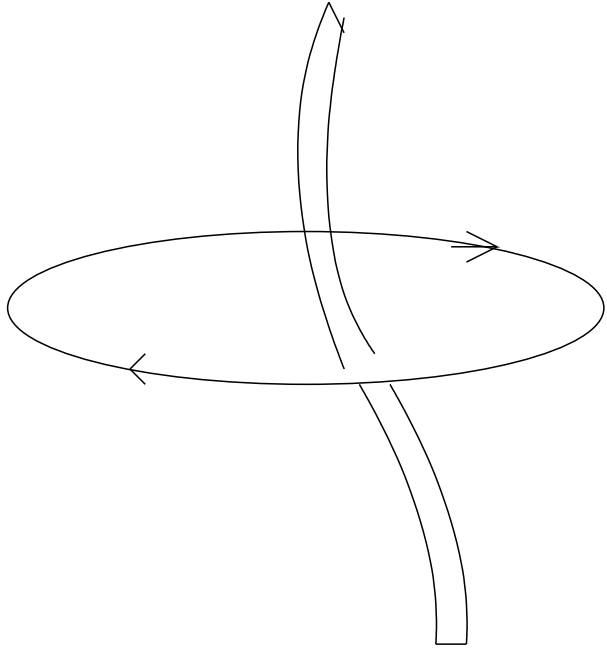, width=0.5\textwidth}
\caption{A non-zero winding in the phase ensures a
cosmic string within.}
\end{center}
\label{fig:cosmicstring}
\end{figure} 
The field values $\phi(x, t)$ at points along the loop, $f(t)$, 
take values in ${\cal M}$, and so they map points from
a physical space $R^{3}$ into the vacuum manifold ${\cal M}$. 
Hence $\phi(x, t)$ completes a mapping from $S^{1}$ onto a
path g(t) in ${\cal M}$, $g(t)=\phi\left(f(t)\right)$. If 
the path $g(t)$ possesses a non-trivial winding $n$ in ${\cal M}$, 
then a string must be present in the physical space. But this 
is exactly the criterion for $g(t)$ to belong to the homotopy 
class $n$. This correspondence is one to one since each type 
of string can be identified with a unique element of 
$\pi_{1}(S^{1})\equiv Z$. One might suppose, therefore, that 
string solutions in general can be classified by the elements 
of $\pi_{1}({\cal M})$. Thus in general if $\pi_{1}({\cal M})
\neq I$ then the Higgs field $\phi(x)$ will necessarily wind 
around ${\cal M}$ in a nontrivial way and a network of cosmic 
strings is formed. 

If $G$ is a local gauge group then the strings formed after the 
breaking of $G$ harbor quanta of massive gauge bosons as 
well as zero-modes of right handed Majorana fermions
~\cite{jackiw.81}. Usually the same Higgs which forms the 
string also gives mass to the Majorana fermions. Presence 
of right handed Majorana zero modes on the string leads to 
a curious phenomenon of fermion number fractionalization. 
However, the existence of the number of zero modes on a string 
depends on its winding number~\cite{weinberg.81}. If the 
winding number of the string is odd then the induced fermion number
on the string is half integral. Hence the decay of a cosmic 
string to fundamental particles is forbidden because of the 
unavailability of quanta of particles carrying fractional fermion 
number in the translational invariant vacuum. In this chapter 
we shall study the decay of strings by forming loops since 
the string loops may not carry any fractional fermion number.

In the context of type-I models, we study the formation, evolution 
and the decay of $B-L$ cosmic strings. When the string loop 
decays it emits those harbored particles, which are collectively 
called $X$-particles. The delayed decay of these massive 
$X$-particles provides a link between the early Universe and 
recent cosmology. In particular, we consider the case of 
baryogenesis via the route of leptogenesis. Assuming the 
hierarchical mass spectrum in the right handed neutrino 
sector we discuss the leptogenesis constraint on the mass 
scale of lightest right handed Majorana neutrino as well as 
the scale of $B-L$ symmetry breaking phase transition. 
 
\section{$U(1)_{\rm B-L}$ cosmic strings and neutrino zero modes}
An inevitable feature of $U(1)$ gauge symmetry breaking
phase transition is the formation of cosmic strings.
In the vacuum manifold, ${\cal M}=\left(U(1)/I\right)$,
since $\pi_1({\cal M})\neq I$, a loop can not be trivially
shrunk to a point and thus the existence of strings is
assured. There are several realistic particle physics models 
where a gauged $U(1)_{\rm B-L}$ symmetry exists and breaks at a 
certain scale. Since $SO(10)$ minimally incorporates 
$U(1)_{\rm B-L}$ gauge symmetry we consider the models 
embedded in $SO(10)$. The 
following breaking schemes can potentially accommodate 
cosmic strings. One of the breaking schemes, motivated 
by supersymmetric $SO(10)$~\cite{jeannerot&davis.95,
jeannerot_prd.96}, involves the intermediate left-right 
symmetric model:
\bea
SO(10) &\underrightarrow{54+45}& SU(3)_{C}\otimes SU(2)_{L}\otimes
SU(2)_{R}\otimes U(1)_{\rm B-L}\nonumber\\
&\underrightarrow{126+\overline{126}}& SU(3)_{C}\otimes
SU(2)_{L}\otimes U(1)_{Y} \otimes Z_{2}\nonumber\\
&\underrightarrow{10+10^{'}}& SU(3)_{C}\otimes U(1)_{Q}\otimes
Z_{2}\,.
\eea  
During the first phase of symmetry breaking, presumably at a
GUT scale of $\sim10^{16}\gev$, monopoles are formed. However, 
during the second and third phases of symmetry breaking cosmic 
strings are formed since $\pi_{1}(\frac{3221}{321})=
\pi_{1}(\frac{321}{31})=Z_{2}$, where the numbers inside the 
parentheses symbolize the group structures. The monopole problem 
in this model can be solved by using a hybrid inflation ending at 
the left-right symmetric phase of the Universe~\cite{jeannerot_prd.96}
thus inflating away the monopoles. The formation of cosmic
strings in the later phases is of great interest since
these ``light'' (i.e., lighter than GUT scale) cosmic strings
do not conflict with any cosmological observations. The $Z_{2}$
strings of the low energy theory was investigated in an
earlier work~\cite{ywmmc}.

Another scheme is to break supersymmetric $SO(10)$
directly to $SU(3)_{C}\otimes SU(2)_{L}\otimes U(1)_{R}\otimes
U(1)_{\rm B-L}$ with the inclusion of extra $54+45$. The
first $45$ acquires a vacuum expectation value along
the direction of $B-L$. However, the latter $45$ acquires
a vacuum expectation value along the direction of $T_{3R}$:
\bea
SO(10) &\underrightarrow{54+45+54^{'}+45^{'}}& SU(3)_{C}\otimes
SU(2)_{L}\otimes U(1)_{R}\otimes U(1)_{\rm B-L}\nonumber\\
&\underrightarrow{126+\overline{126}}& SU(3)_{C}\otimes
SU(2)_{L}\otimes U(1)_{Y} \otimes Z_{2}\nonumber\\
&\underrightarrow{10+10^{'}}& SU(3)_{C}\otimes U(1)_{Q}\otimes Z_{2}
\eea
 
For the demonstration purpose it is sufficient to consider a
model based on the gauge group $SM \otimes U(1)_{\rm B-L}$ which is
spontaneously broken to $SM$. Existence of cosmic strings and the 
related neutrino zero-modes in this model can be established as 
follows. Let the gauge field corresponding to the $U(1)_{\rm B-L}$ 
symmetry be denoted by $C_{\mu}$, and the symmetry be broken by 
a $SM$ singlet $\chi$. Let $\langle\chi\rangle$ be $\eta_{\rm B-L}$ 
below the critical temperature $T_{\rm B-L}$. In a suitable gauge a 
long cosmic string oriented along the $z$-axis can be represented 
(in cylindrical polar coordinates) by the ansatz~\cite{vilen&shell}
\begin{eqnarray}
\chi&=&\eta_{\rm B-L} f(r)e^{in{\theta}} \label{string_higgs}\,,\\
C_{\mu}&=&\frac{ng(r)}{\alpha r}\delta^{\theta}_{\mu}\,,
\label{string_gauge}
\end{eqnarray}
where $n$ is an integer giving the winding number of the phase 
of the complex Higgs field $\chi$, and $\alpha$ is the gauge 
coupling constant for the group $U(1)_{\rm B-L}$. In order for the 
solution to be regular at the origin we set $f(0)=g(0)=0$. Also 
requiring the finiteness of energy of the solution, we set 
$f(r)=g(r)=1$ as $r\rightarrow\infty$. It turns out that both 
$f(r)$ and $g(r)$ take their asymptotic values everywhere outside 
a small region of the order of $\eta_{\rm B-L}^{-1}$ around the 
string. Thus away from the string $\langle\chi\rangle=
\eta_{\rm B-L}$ up to a phase $\theta$, and $C_{\mu}$ is a pure 
gauge. The arbitrary phase $\theta= \theta(x)$ can vary in
different regions of space. For $\chi$ to be single valued  
$\theta$ must change by an integer multiple of $2\pi$ 
around a closed loop. When the loop is shrunk to a point, 
$\theta$ becomes undefined, so that there exists point where 
$\langle \chi \rangle=0$ and the symmetry is unbroken i.e. 
thin tubes of false vacuum get trapped some where inside 
the loop. It is in this sense that the object is referred 
to as a `defect', a region of unbroken symmetry (false vacuum) 
surrounded by broken symmetry region (true vacuum). Such 
strings are either infinitely long or closed loops. While 
an infinite straight string is topologically stable due to the non 
trivial winding of the Higgs, a closed string loop can decay by 
emitting gravitational radiation. The strings trap 
excess energy density associated with gauge and Higgs fields, 
and, in addition, fermion zero modes which make the strings 
massive. The mass scale of the string is fixed 
by the energy scale of the symmetry breaking phase transition
$\eta_{\rm B-L}$ at which the strings are formed. Then the mass
per unit length of a cosmic string, $\mu$, is of order
$\eta^{2}_{\rm B-L}\sim T^{2}_{\rm B-L}$.
 
The Lagrangian for the right-handed neutrino is
\be
\mathcal{L}_{\nu_R}= i\overline{\nu_{R}}\sigma^{\mu}
D_{\mu}\nu_{R} -{1\over 2}
[if\overline{\nu_{R}}\chi\nu_{R}^{c}+H.C]\,,
\label{lagrangian_nu_R}
\ee
where $h$ is the Yukawa coupling constant, $\sigma^{\mu}=(-I,
\sigma^{i})$, and $\nu_{R}^{c}=i \sigma^{2}\nu_{R}^{*}$ defines
the Dirac charge conjugation operation. The resulting equations
of motion have been shown~\cite{jackiw.81} to possess $|n|$
normalisable zero-modes in winding number sector $n$. The field
equations in the  $U(1)$ example are~\cite{sdavis.00}
\be
\begin{pmatrix} -e^{i\theta}[\partial_{r}+{i\over r}
\partial_{\theta}+{ng(r)\over 2r}] & \partial_{z}+\partial_{t}\\
\partial_{z}-\partial_{t} & e^{-i\theta}[\partial_{r}-{i\over r}
\partial_{\theta}-{ng(r)\over 2r}]\end{pmatrix}\nu_{R}-M_{R}
e^{in\theta}\nu_{R}^{*}=0\,,
\label{field_equations_nu_R}
\ee
where the expressions (\ref{string_higgs}) and (\ref{string_gauge})
have been substituted for $\chi$ and $C_{\mu}$, and
$M_{R}=f\eta_{\rm B-L}$. In the winding number sector $n$ the
normalizable zero-modes obey $\sigma^{3}\nu_R=\nu_R$ and are
of the form
\be
\nu_{R}(r, \theta)=\begin{pmatrix}1\\
0\end{pmatrix}\left(U(r)e^{il\theta}+V^{*}(r)e^{i(n-1-l)
\theta}\right)\,,
\label{zeromode}
\ee
where $U(r)$ and $V(r)$ are well behaved functions at the
origin and have the asymptotic behavior $\sim \exp(-M_{R}r)/
\sqrt{r}$. When nontrivial $z$ and $t$ dependences are included,
these modes have solutions that depend on $z+t$ and are Right movers.
For $n<0$, normalizable solutions obey $\sigma^{3}\nu_R=-\nu_R$,
and form the zero-energy set of a Left moving spectrum. On a straight
string these modes are massless. However on wiggly strings
they are expected to acquire effective masses proportional
to  the inverse radius of the string curvature.

\section{Evolution of cosmic strings: Formation and evolution
of closed loops and production of massive particles}
\subsection{Scaling solution and closed loop formation}
The evolution of cosmic strings in the expanding Universe has
been studied extensively, both analytically as well as
numerically; for a text-book review, see the
monograph~\cite{vilen&shell}. Here we briefly summarize
only those aspects of cosmic string evolution that are
relevant for the present purpose, namely the
formation and subsequent evolution of closed loops
of strings and production of massive particles from them.
This closely follows the discussion in section 6.4 of
Ref.~\cite{pijushreport}.

Immediately after their formation at a phase transition, the
strings would in general be in a random tangled configuration.
One can characterize the string configuration in terms of a
coarse-grained length scale  $\xi_s$ such that the overall
string energy density $\rho_s$ is given by $\rho_s=\mu/\xi_s^2$.
Initially, the strings move under a strong damping force due
to friction with the ambient thermal plasma. In the friction
dominated epoch a curved string segment of radius of curvature
$r$ acquires a terminal velocity $\propto 1/r$. As a result the
strings tend to straighten out so that the total length of
the strings decreases. Thus the overall energy density in the form
of strings decreases as the Universe expands. This in turn means
that the length scale $\xi_{s}$ increases. Eventually, $\xi_s$
reaches the causal horizon scale $\sim t$. After the damping
regime ends (when the background plasma density falls to a
sufficiently low level as the Universe expands), the strings
start to move relativistically. However, causality prevents the
length scale $\xi_{s}$ from exceeding the horizon size $\sim
t$. Analytical studies supported by extensive numerical simulations
show that the subsequent evolution of the system is such that
the string configuration reaches a ``scaling regime'' in which
the ratio $\frac{\xi_{s}}{t}\equiv x$ remains a constant.
Numerical simulations generally find the number $x$ to lie
approximately in the range $\sim$ 0.4--0.7. This is called
the scaling regime because then the energy density in the form
of strings scales as, and remains a constant fraction of, the
energy density of radiation in the radiation dominated epoch or
the energy density of matter in the matter dominated epoch both
of which scale as $t^{-2}$.

The fundamental physical process that maintains the string
network in the scaling configuration is the formation of
{\it closed loops} which are pinched off from the network 
whenever a string segment curves over into a loop, intersecting 
itself. It is difficult to calculate ab initio the length distribution
of the closed loops so formed, but numerical simulations
find that loops are formed typically on the scale of the smallest scale
structure allowed by the spatial resolution of the simulation. It is
expected~\cite{vilen&shell} that the smallest scale structure on
the string at any time time $t$ would be $\sim\Gamma G\mu t$, which is
determined by gravitational radiation from the small scale structure 
and its back-reaction on the string. Here $\Gamma\sim 100$ is a 
geometric factor~\cite{vilen&shell}. Thus, the closed loops at birth 
can be assumed to have a typical length~\cite{vilen&shell}
\be
L_b=K\Gamma G\mu t\,,
\label{birth_length}
\ee
and they are formed at a rate (per unit volume per unit time) 
which, in the radiation dominated epoch, is given by
\be
\frac{dn_{b}}{dt}=\frac{1}{x^{2}}\left(\Gamma G \mu\right)
^{-1} K^{-1}t^{-4}\,,
\label{birth_rate}
\ee
where $\Gamma\sim 100$ is a geometrical factor that determines the
average loop length, and $K$ is a numerical factor of order unity.

The whole string network consisting of closed loops as well as
long strands of strings stretched across the horizon gives rise
to density fluctuations in the early Universe which could
potentially contribute to the process of formation of structures
in the Universe. More importantly, they would produce specific
anisotropy signatures in the cosmic microwave background (CMB).
Using a large-scale cosmic string network simulation and comparing
the resulting prediction of CMB anisotropies with observations,
a recent analysis~\cite{landriau-shellard} puts an upper limit
on the fundamental cosmic string parameter $\mu$, giving
$G\mu\lsim 0.7\times 10^{-6}$. This translates to an upper limit,
$\eta_{\rm B-L}\lsim 1.0\times10^{16}\gev$, on the symmetry-breaking
energy scale of the cosmic string-forming phase transition. This
probably rules out cosmic string formation at a typical GUT
scale $\sim10^{16}\gev$. However, lighter cosmic strings arising
from symmetry breaking at lower scales, such as the $B-L$ cosmic
strings in the case of the $SO(10)$ model discussed in the
previous section, are not ruled out by CMB anisotropy constraints. 
For detailed discussions of constraints on topological defects in 
general and cosmic strings in particular from CMB anisotropies see, 
for example~\cite{td_cmb}.

It should be noted here that, in the standard scenario of cosmic
string evolution described above, the loops are formed on a 
length scale that is a constant fraction of the horizon length, 
as given by equation (\ref{birth_length}). Thus, the average 
size of the newly formed loops increases with time. At the 
relevant times of interest, these loops, although small in 
comparison to the horizon scale, would still be of macroscopic 
size in the sense that they are much larger than the microscopic 
string width scale $w\sim\eta_{\rm B-L}^{-1}\sim\mu^{-1/2}$.
 
In contrast, results of certain Abelian Higgs (AH) model
simulations of cosmic string evolution~\cite{vah} seem to
indicate that scaling configuration of the string network is 
maintained primarily by loops formed at the smallest fixed 
length scale in the problem, namely, on the scale of the 
width $w\sim\eta_{\rm B-L}^{-1}\sim\mu^{-1/2}$ of the string. These 
microscopic ``loops'' quickly decay into massive particles 
(quanta of gauge bosons, Higgs bosons, heavy fermions etc.) 
that ``constitute'' the string. In other words, in this scenario, 
there is essentially no macroscopic loop formation at all; 
instead, the scaling of the string network is maintained
essentially by massive particle radiation. In order for the 
scaling configuration of the string network to be maintained by 
this process, the microscopic loops must be formed at a rate
\be
\left(\frac{dn_{b}}{dt}\right)_{\rm AH}=\frac{1}{x^{2}}
\mu^{1/2}t^{-3}\,.
\label{birth_rate_AH}
\ee
 
The above scenario of cosmic string evolution in which massive 
particle radiation rather than gravitational radiation plays the 
dominant role is, however, currently a subject of 
debate~\cite{moore-shell}. One of the major problems hindering a 
resolution of the issues involved is the insufficient dynamic 
range possible in the currently available AH model simulations 
and the consequent need for extrapolation of the simulation 
results to the relevant cosmological scales, which is not 
straightforward. In the present case, we shall primarily restrict 
ourselves to consideration of the ``standard'' macroscopic loop 
formation scenario described by equations (\ref{birth_length}) 
and (\ref{birth_rate}) above, although we shall have occasions to 
refer to the massive particle radiation scenario below (see, in 
particular, section 7.3.2.B).

\subsection{Fate of the closed loops and massive particle 
production}
The behavior of the closed loops after their formation may be
broadly categorized into following two classes:

\subsubsection{\textbf{A.}~Slow death}
Any closed loop of length $L$ in its center of momentum frame
has an oscillation period $L/2$~\cite{kibble&turok-82}. However,
a loop may be either in a self-intersecting or non-selfintersecting
configuration. In general, a closed loop configuration can be
represented as a superposition of waves consisting of various
harmonics of $\sin$'s and $\cos$'s. Some explicit low harmonic
number analytical solutions of the equations of motion of
closed loops representing non-selfintersecting loops are known
in literature~\cite{kibble&turok-82, turok-84, chendicarlo&hotes-88,
delaneyetal-90}, and it is possible that there exists a large
class of such non-selfintersecting solutions. Indeed, numerical
simulations, while limited by spatial resolution, do seem to
indicate that a large fraction of closed loops are born in
non-selfintersecting configurations. 

A non-selfintersecting loop oscillates freely. As it oscillates,
it loses energy by emitting gravitational radiation, and
thereby shrinks. When the radius of the loop becomes of the
order of its width $w\sim\eta_{\rm B-L}^{-1}\sim\mu^{-1/2}$, the
loop decays into massive particles. Among these particles will be the
massive gauge bosons, Higgs bosons, and in the case of the
$B-L$ strings, massive right-handed neutrinos ($\nu_R$) which
were trapped in the string as fermion zero modes. We shall
hereafter collectively refer to all these particles as $X$
particles. We are, of course, interested here only in the 
$\nu_R$'s. In addition to those directly released from the 
loop's final decay, there will also be some $\nu_R$'s coming 
from the decays of the gauge and Higgs bosons released in the 
final loop decay. It is difficult to calculate exactly the 
total number of $\nu_R$'s so obtained from each loop, but we 
may expect that it would be a number of order unity. For the 
present purpose we shall assume that each final
demise of a loop yields a number $N_N\sim O(1)$ of heavy right 
handed Majorana neutrinos; we shall keep this number $N_N$ as 
a free parameter. 

The rate of release of $\nu_R$'s at any time $t$ by the above
process can be calculated as follows. The lifetime of a loop of
length $L$ due to energy loss through gravitational wave 
radiation is
\be
\tau_{\rm GW}\sim \left(\Gamma G \mu\right)^{-1}L\,.
\label{gw_lifetime}
\ee
Equations (\ref{birth_length}) and (\ref{gw_lifetime}) thus 
show that loops born at time $t$ have a lifetime $\sim 
Kt\gsim H^{-1}(t)$, where $H^{-1}(t)\sim t$ is the Hubble 
expansion time scale. It is thus a slow process. From the 
above, we see that the loops that are disappearing at any 
time $t$ are the ones that were formed at the time $(K+1)^{-1}t$. 
Taking into account the dilution of the number density of loops 
due to expansion of the Universe between the times of their 
birth and final demise, equation (\ref{birth_rate}) gives the 
number of loops disappearing due to this ``slow death'' (SD) 
process per unit time per unit volume at any time $t$ (in the 
radiation dominated epoch) as 
\be
\frac{dn_{\rm SD}}{dt}=f_{\rm SD}\frac{1}{x^{2}}\left(\Gamma 
G \mu\right)^{-1} \frac{(K+1)^{3/2}}{K}t^{-4}=
f_{\rm SD}(K+1)^{3/2}\frac{dn_b}{dt}\,,
\label{loop_SD_rate}
\ee
where $f_{\rm SD}$ is the fraction of newly born loops which
die through the SD process.
 
The rate of release of the heavy right-handed neutrinos (we shall 
hereafter denote it by $N$; see section 7.4 below) due to SD 
process can then be written as 
\be
\left(\frac{dn_N}{dt}\right)_{\rm SD}=N_N \frac{dn_{\rm SD}}{dt}
=N_N f_{\rm SD}\frac{1}{x^{2}}\left(\Gamma G \mu\right)^{-1}
\frac{(K+1)^{3/2}}{K}t^{-4}\,.
\label{N_SD_rate}
\ee  
In a comoving volume the above injection rate (\ref{N_SD_rate}) 
of massive Majorana neutrinos is given by
\be
\left(\frac{dY^{st}_{N}}{dZ}\right)_{SD}= \frac{1.57\times 
10^{-17}}{Z^4}N_N f_{\rm SD}\left(\frac{M_N}{\eta_{\rm B-L}}\right)^2
\left(\frac{M_N}{GeV}\right),
\label{slowdeath-rate}
\ee
where $Y^{st}_{N}=n_N/s$ is the density of massive Majorana 
neutrino in a comoving volume with $s=43.86 g_* T^3$ being 
the entropy density and $Z=M_N/T$ is the dimensionless variable 
with respect to which the evolution of the various quantities 
is studied. Here we have used the numerical values for the 
constants $\Gamma=100$, $x=0.5$, $g_{*}=100$ and $K=1$. We 
shall use this equation while solving the Boltzmann equations
numerically in section 7.5. For our numerical purpose we shall 
be interested in the lightest right handed neutrino $N_1$ whose 
mass is given by $M_1$. Since $M_1$ relates to the scale of symmetry 
breaking phase transition as $M_1=f_1\eta_{\rm B-L}$ hence for all 
values of $f_1\leq 1 $, $M_1$ in general satisfies the constraint 
$M_1\leq \eta_{B-L}$.

\subsubsection{\textbf{B.}~Quick death}
Some fraction of the loops may be born in configurations with
waves of high harmonic number. Such string loops have been
shown~\cite{siemens&kibble.95} to have a high probability of
self-intersecting. Ref.~\cite{siemens&kibble.95} gives the
self-intersecting probability of a loop as
\be
P_{SI}=1-e^{-\alpha-n\beta}\,,
\label{self_int_prob}
\ee
where $\alpha=0.4$, $\beta=0.2$, and n is the harmonics number.
 
A self-intersecting loop would break up into two or more
smaller loops. The process of self-intersection leaves behind
``kinks'' on the loops, which themselves represent high harmonic
configurations. So, the daughter loops would also further
split into smaller loops. If a loop does self-intersect, it
must do so within its one oscillation period, since the
motion of a loop is periodic. Under this circumstance, since
smaller loops have smaller oscillation periods, it
can be seen that a single initially large loop of length $L$
can break up into a debris of tiny loops of size $\eta_{\rm B-L}^{-1}$
(at which point they turn into the constituent massive particles)
on a time-scale $\sim L$. Equation (\ref{birth_length}) then
implies that a loop born at the time $t$ in a high harmonic
configuration decays, due to repeated self-intersection, into
massive particles on a time scale $\tau_{QD}\sim K\Gamma G \mu t
\ll H^{-1}(t)$. It is thus a ``quick death'' (QD) process ---
the loops die essentially instantaneously (compared to
cosmological time scale) as soon as they are formed.
Equation (\ref{birth_rate}), therefore, directly gives the
rate at which  loops die through this quick death process:
\be
\frac{dn_{\rm QD}}{dt}=f_{\rm QD}\frac{dn_b}{dt}\,,
\label{loop_QD_rate}
\ee   
where $f_{\rm QD}$ is the fraction of newly born loops that
undergo QD.
 
Note that, since these loops at each stage self-intersect
and break up into smaller loops before completing one oscillation,
they would lose only a negligible amount of energy in
gravitational radiation. Thus, almost the entire original
energy of these loops would eventually come out in the
form of massive particles.
 
Assuming again, as we did in the SD case, that each segment of 
length $\sim w \sim \mu^{-1/2}$ of the loop yields a number 
$N_N\sim O(1)$ of heavy right-handed Majorana neutrinos, we can 
write, using equations (\ref{loop_QD_rate}), (\ref{birth_rate}) 
and (\ref{birth_length}), the rate of release of the $N$'s due 
to QD process as
\be
\left(\frac{dn_N}{dt}\right)_{\rm QD}=N_N\, f_{\rm QD}\frac{1}{x^2}
\mu^{1/2} t^{-3}\,.
\label{N_QD_rate}
\ee
In a comoving volume the above rate (\ref{N_QD_rate}) can be 
rewritten as
\be
\left(\frac{dY^{st}_{N}}{dZ}\right)_{QD}\simeq \frac{1.36\times
10^{-36}}{Z^2} f_{\rm QD} N_N \left(\frac{\eta_{\rm B-L}}{GeV}\right)
\left(\frac{M_N}{GeV}\right).
\label{quickdeath-rate}
\ee
We shall use this equation while solving Boltzmann equations 
numerically in section 7.5. However, for the leptogenesis 
purpose we shall be interested in the lightest right handed 
neutrino $N_1$ and hence we replace $M_N$ by $N_1$ and correspondingly 
$Z=M_1/T$. 

It is interesting to note here that if all loops were to die 
through this QD process, i.e., if we take $f_{\rm QD}=1$ in 
equations (\ref{loop_QD_rate}) and (\ref{N_QD_rate}), then the 
situation is in effect exactly equivalent to the microscopic 
loop formation scenario described by equation (\ref{birth_rate_AH}), 
although the primary loops themselves are formed with macroscopic 
size given by equation (\ref{birth_length}). 

While the important issue of whether or not massive particle 
radiation plays a dominant role in cosmic string evolution remains 
to be settled, the standard model may, of course, still allow a 
small but finite fraction, $f_{\rm QD}\ll 1$, of quickly dying 
loops. There already exist, however, rather stringent 
astrophysical constraints~\cite{pijush&rana,pijushreport} on
$f_{\rm QD}$ from the observed flux of ultrahigh energy cosmic rays
(UHECR) above $10^{11}\gev$~\cite{uhecr_obs} and the cosmic
diffuse gamma ray background in the energy region 10 MeV -- 100
GeV measured by the EGRET experiment~\cite{sreekumar}. This comes
about in the following way: 

The massive $X$ particles released from the string loops would
decay to $SM$ quarks and leptons. The hadronization of the
quarks gives rise to nucleons and pions with energy up to
$\sim M_X$, the mass of the relevant $X$ particle. The neutral
pions decay to photons. These extremely energetic nucleons and 
photons, after propagating through the cosmic radiation background, 
can survive as ultrahigh energy particles. The observed flux of 
UHECR, therefore, puts constraints on the rate of release of the
massive $X$ particles, thereby constraining $f_{\rm QD}$.
The most stringent constraint on $f_{\rm QD}$, however, comes
from the fact that the electromagnetic component (consisting of
photons and electrons/positrons) of the total energy injected
in the Universe from the decay of the $X$ particles initiates
an electromagnetic cascade process due to interaction of the
high energy electrons/positrons and photons with the
photons of the various cosmic background radiation fields (such as
the radio, the microwave and the infrared/optical backgrounds);
see, e.g., Ref.~\cite{pijushreport} for a review. As a result,
a significant part of the total injected energy cascades down
to lower energies. The measured flux of the cosmic gamma ray
background in the 10 MeV -- 100 GeV energy region~\cite{sreekumar}
then puts the constraint~\cite{pijushreport}
\be
f_{\rm QD}\eta_{16}^{2}\leq 9.6\times 10^{-6}\,,
\label{f_QD_constraint}
\ee
where $\eta_{16}\equiv(\eta_{\rm B-L}/10^{16}\gev)$. For GUT
scale cosmic strings with $\eta_{16}=1$, for example, the above
constraint implies that $f_{\rm QD}\leq 10^{-5}$, so that most loops
should be in non-selfintersecting configurations, consistent with the
standard scenario of cosmic string evolution.
Note, however, that $f_{\rm QD}$ is not constrained by the above
considerations for cosmic strings formed at a scale $\eta_{\rm B-L}
\lsim 3.1\times 10^{13}\gev$.
 
In this context, it is interesting to note that there is no
equivalent constraint on the corresponding parameter
$f_{\rm SD}$ for the slow death case from gamma ray background
consideration. The reason is that, unlike in the QD case where
the entire initial energy of a large loop goes into $X$
particles, only $\sim$ one $X$ particle is released from a
initially large loop in the SD case. This in turn makes the
time dependence of the rate of release of massive particles
$\propto t^{-4}$ in the SD case (see equation (\ref{N_SD_rate})), 
while it is $\propto t^{-3}$ in the QD case (see equation 
(\ref{N_QD_rate})). Thus, while the SD process dominates at 
sufficiently early times, the QD process can dominate at relatively 
late times and can potentially contribute to the non-thermal 
gamma ray background.

\section{Analytical estimation of baryon asymmetry}
\subsection{Decay of heavy right-handed Majorana neutrinos and
$L$-asymmetry}
We follow the same convention as given in section 2.3.1. 
In this scenario the lepton asymmetry is produced by the 
decay of right-handed heavy Majorana neutrino to $SM$ lepton 
($\ell$) and Higgs ($\phi$) through the Yukawa coupling. We 
assume a normal mass hierarchy in the right handed heavy 
Majorana neutrino sector, $M_{1} < M_{2} < M_{3}$. In this scenario 
it is reasonable to expect that the final lepton asymmetry 
is produced mainly by the decay of the lightest right handed 
neutrino $N_{1}$. As the Universe expands, the temperature of 
the thermal plasma falls. Below a temperature $T_F\sim M_1$, 
all $L$-violating scatterings mediated by $N_{1}$ freeze out, 
thus providing the out-of-equilibrium situation~\cite{sakharov.67} 
necessary for the survival of any net $L$-asymmetry generated by 
the decay of the $N_1$'s. The final $L$-asymmetry is, therefore, 
given essentially by the $CP$ asymmetry parameter 
\be
\epsilon_{1}=\frac{3}{16\pi}\frac{M_1 m_3}{v^2}\delta\,,
\label{epsilon1value}
\ee
where $v\simeq 174\gev$ is the electroweak symmetry breaking 
scale and $\delta\leq 1$.

An accurate calculation of the net $L$-asymmetry can only be 
done by numerically solving the full Boltzmann equations that 
include all lepton number violating interactions involving all 
the $N_1$'s present at any time, including the $N_1$'s of 
non-thermal origin such as the ones produced from the decaying 
cosmic string loops, as well as those of thermal origin. 
This will follow up our analytical estimations. For analytical 
calculation, first we shall simply assume that below the 
temperature $T_F=M_1$, all interactions except the decay of the 
$N_1$ are unimportant, so that each $N_1$ released from cosmic 
strings additively produces a net $L$-asymmetry when it decays. 

For the present purpose, we note that there is an upper
bound~\cite{davidson&ibarra.02,buch-bari-plum.02} on $\epsilon_1$, 
which is related to the properties of the light neutrino masses. In 
a standard hierarchical neutrino mass scenario with $m_3\gg m_2>
m_1$, this upper limit is given by~\cite{davidson&ibarra.02,
buch-bari-plum.02}
 
\be
|\epsilon_{1}|=\frac{3}{16\pi}\frac{M_1 m_3}{v^2}\,.
\label{epsilon_1_max}
\ee
The above upper limit is in fact {\it saturated}~\cite{buch-bari-plum.02} 
in most of the reasonable neutrino mass models, which we shall assume 
to be the case.
 
Assuming the standard light neutrino mass hierarchy, 
the heaviest light neutrino is given by $m_3\simeq 
\left(\Delta m^2_{\rm atm}\right)^{1/2}\simeq 0.05\ev$. In our 
calculations below, we shall use 
\be
\epsilon_1\simeq 9.86\times 10^{-4}\left(\frac{M_1}{10^{13}\gev}
\right)\left(\frac{\left(\Delta m^2_{\rm atm}\right)^{1/2}}
{0.05\ev}\right)\,.
\label{epsilon_1_value}
\ee
 
The $L$-asymmetry is partially converted to a $B$-asymmetry by the 
rapid nonperturbative sphaleron transitions which violate $B+L$ but 
preserve $B-L$. Assuming that sphaleron transitions are ineffective 
at temperatures below the electroweak transition temperature 
($T_{\rm EW}$), the $B$-asymmetry is related to $L$-asymmetry by 
the relation~\cite{har&tur.90}
\be
B=p\, (B-L)=\frac{p}{p-1}L\simeq -0.55 L\,,
\label{B_L_relation}
\ee
where we have taken $p=28/79$ appropriate for the particle
content in $SM$~\cite{har&tur.90}. 

The net baryon asymmetry of the Universe is defined as
\be
Y_B=\frac{n_B-n_{\bar{B}}}{s}\,,
\label{bau_def}
\ee
where $s=43.86 g_*T^3$ is the entropy density, with $g_*$ being 
the number of relativistic degrees of freedom contributing to the 
entropy at the temperature $T$. At temperatures in the early 
Universe relevant for the process of baryon asymmetry generation, 
$g_*\simeq 100$ in $SM$. 
 
Observationally, the BAU is often expressed in terms of the
baryon-to-photon ratio $\eta\equiv (n_B-n_{\bar{B}})/n_\gamma$, 
whose present-day-value $\eta_0$ is related to that of $Y_B$ 
through the relation
\be
\eta_0\simeq 7.0 Y_{B,0}\,.
\label{eta_0_def}
\ee
We now proceed to estimate the contribution to the $BAU$ from the two
cosmic string loop processes discussed in the previous section.
 
\subsection{Slow death case}
The contribution of the SD process to $\eta_0$ can be written as
\be
\eta_0^{\rm SD}\simeq 7.0 \times 0.55\, \epsilon_1
\int_{t_{F}}^{t_{0}}\frac{1}{s}
\left(\frac{dn_N}{dt}\right)_{\rm SD} dt\,,
\label{eta_0_SD_def}
\ee
where $t_F$ is the cosmic time corresponding to the temperature 
$T_F\simeq M_1$ and $t_0$ is the present age of the Universe. 
Using equations (\ref{N_SD_rate}) and the standard time-temperature 
relation in the early Universe,
\be
t\simeq 0.3 g_*^{-1/2}\frac{M_{\rm Pl}}{T^2}\,,
\label{t_T_rel}
\ee
where $M_{\rm Pl}\simeq 1.22\times 10^{19}\gev$ is the Planck mass, 
we see that the dominant contribution to the integral in equation
(\ref{eta_0_SD_def}) comes from the time $t_F\ll t_0$, i.e., from 
the epoch of temperature $T_F\simeq M_1$, giving
\bea
\eta_0^{\rm SD} & \simeq & 2.0\times 10^{-7} N_N
\left(\frac{M_{1}}{10^{13}\gev}\right)^{4}
\left(\frac{\eta_{\rm B-L}}{10^{13}\gev}\right)^{-2}\nonumber \\
&=& 2.0\times 10^{-7} N_N f_1^4
\left(\frac{\eta_{\rm B-L}}{10^{13}\gev}\right)^{2}\,,
\label{eta_0_SD_value}
\eea 
where we have defined the Yukawa coupling $f_1\equiv M_1/\eta_{\rm 
B-L}$,
used equation (\ref{epsilon_1_value}) for $\epsilon_1$
with $\left(\Delta m^2_{\rm atm}\right)^{1/2} = 0.05\ev$, and also
taken $x=0.5$, $\Gamma=100$, $K=1$ and $f_{\rm SD}=1$ in
equation (\ref{N_SD_rate}).
 
The Yukawa couplings are generally thought to be less than unity.
With $f_1\leq 1$, we see from (\ref{eta_0_SD_value}) and
(\ref{b-asy-wmap}) that the cosmic string loop slow death 
process can produce the observed $BAU$ only for $B-L$ phase 
transition scale
\be
\eta_{\rm B-L}^{\rm SD}\gsim 5.5\times 10^{11}N_N^{-1/2}\gev\,.
\label{eta_scale_lower_limit_SD}
\ee
Assuming $N_N\lsim 10$, say, we see that cosmic string loop SD 
process can contribute to BAU if 
$\eta_{\rm B-L}\gsim 1.7\times10^{11}\gev$; lower values
of $\eta_{\rm B-L}$ are relevant only if we allow $f_1>1$.
 
At the same time, for a given $\eta_{\rm B-L}$ satisfying
(\ref{eta_scale_lower_limit_SD}), in order that the contribution 
(\ref{eta_0_SD_value}) not exceed the highest allowed observed 
value of $\eta_0$ given by equation (\ref{b-asy-wmap}), the Yukawa 
coupling $f_1$ must satisfy the constraint
\be
f_1^{\rm SD}\lsim 0.24
\left(\frac{\eta_{\rm B-L}}{10^{13}\gev}\right)^{-1/2}N_N^{-1/4}\,,
\label{f_1_SD_constraint}
\ee
which, in terms of the lightest heavy right handed Majorana
neutrino mass $M_1$, reads
\be  
M_1^{\rm SD}\lsim 2.4 \times 10^{12} N_N^{-1/4}
\left(\frac{\eta_{\rm B-L}}{10^{13}\gev}\right)^{1/2}\gev\,.
\label{M1_constraint_SD}
\ee
Note the rather weak dependence of the above constraints on $N_N$.
Also, the 4th power dependence on $M_1$ of equation
(\ref{eta_0_SD_value}) and the rather narrow range of the observed 
value of $\eta_0$ given by equation (\ref{b-asy-wmap}) together 
imply that, in order for the SD process to explain the observed 
BAU, $M_1$ (and equivalently $f_1$) cannot be much smaller than 
their respective values saturating the above constraints. 

It is important to note that in deriving the above constraint on
$M_1$ we assume that there is a strong hierarchy in the low
energy neutrino sector. However, this limit becomes worse
in the case degenerate neutrinos~\cite{gu&ma.05} 
 
\subsection{Quick death case}
Replacing $\left(\frac{dn_N}{dt}\right)_{\rm SD}$ in equation
(\ref{eta_0_SD_def}) by $\left(\frac{dn_N}{dt}\right)_{\rm QD}$ 
given by equation (\ref{N_QD_rate}), and following the same 
steps as in the SD case above, we get the contribution of the QD 
process to $\eta_0$ as
\bea
\eta_0^{\rm QD} & \simeq & 5.17 \times 10^{-13}N_N f_{\rm QD}
\left(\frac{M_1}{10^{13}\gev}\right)^2
\left(\frac{\eta_{\rm B-L}}{10^{13}GeV}\right)\nonumber \\
&=&  5.17\times 10^{-13}N_N f_{\rm QD} f_1^2
\left(\frac{\eta_{\rm B-L}}{10^{13}GeV}\right)^3\,.
\label{eta_0_QD_value_1}
\eea
From (\ref{eta_0_QD_value_1}) and (\ref{b-asy-wmap}) we see that,
considering the most optimistic situation with $f_{\rm QD}=1$, the 
QD process is relevant for BAU only for 
\be
\eta_{\rm B-L}^{\rm QD}\gsim 1.1\times 10^{14}N_N^{-1/3}\gev\,;
\label{eta_scale_lower_limit_QD_1}
\ee
lower values of $\eta_{\rm B-L}$ are relevant only if we allow 
$f_1>1$. On the other hand, the constraint (\ref{f_QD_constraint}) 
allows $f_{\rm QD}=1$ {\it only if} $\eta_{\rm B-L}\leq 3.1
\times10^{13}\gev$. This can be reconciled with the above 
constraint (\ref{eta_scale_lower_limit_QD_1}) only for $N_N > 45$ 
or so. Such a large value of $N_N$ seems unlikely. 
 
In general, using the constraint (\ref{f_QD_constraint}) on
$f_{\rm QD}$ in (\ref{eta_0_QD_value_1}) we get
\bea
\eta_0^{\rm QD} & \lsim&  5.0 \times 10^{-12}N_N
\left(\frac{M_1}{10^{13}\gev}\right)^2
\left(\frac{\eta_{\rm B-L}}{10^{13}GeV}\right)^{-1}\nonumber \\
&=&  5.0\times 10^{-12}N_N f_1^2
\left(\frac{\eta_{\rm B-L}}{10^{13}GeV}\right)\,.
\label{eta_0_QD_value_2}
\eea
Comparing again with the observed value of $\eta_0$, we now
see that, for $f_1\leq 1$, the QD process can be relevant for 
$BAU$ only for
\be
\eta_{B-L}^{\rm QD}\gsim 1.2\times 10^{15}N_N^{-1}\gev\,.
\label{eta_scale_lower_limit_QD_2}
\ee
 
For values of $\eta_{B-L}$ satisfying the above constraint
(\ref{eta_scale_lower_limit_QD_2}), the QD process can produce the
observed value of $BAU$ for
\be
f_1^{\rm QD}\lsim 0.36
\left(\frac{\eta_{\rm B-L}}{10^{16}\gev}\right)^{-1/2}N_N^{-1/2}\,,
\label{f_1_QD_constraint}
\ee
which in terms of $M_1$ now reads
\be
M_1^{\rm QD}\lsim 3.6 \times 10^{15} N_N^{-1/2}
\left(\frac{\eta_{\rm B-L}}{10^{16}\gev}\right)^{1/2}\gev\,.
\label{M1_constraint_QD}
\ee
 
From the above discussions we see that, as far as their 
contributions to the $BAU$ is concerned, the QD process becomes 
important only at relatively higher values of the symmetry 
breaking scale $\eta_{\rm B-L}$ compared to the SD process. 

\section{Numerical calculation of baryon asymmetry}
In section 7.4, neglecting the washout effects we saw that 
the final lepton asymmetry, produced by the string emitted 
right handed neutrinos, depends on the density of $N_1$'s at 
the freeze out epoch, $T_F\simeq M_1$, as well as the amount 
of $CP$-violation due to the decays of $N_1$. Note that, the 
final lepton asymmetry does not depend on the source of 
$N_1$'s whether they are generated by any thermal or non-thermal 
processes. Therefore, we attempt to establish a complete 
analysis of baryon asymmetry by including the $N_1$'s of thermal 
origin as well of cosmic string origin . We then discuss the 
constraints on the scale of $B-L$ symmetry breaking phase 
transition from the observed baryon asymmetry.

\subsection{Boltzmann Equations}
Considering the lightest right handed heavy Majorana
neutrino ($N_1$) of cosmic string origin as well of thermal origin
at any epoch of temperature T (or equivalently Z) the total
rate of change of the abundance of $N_1$'s is given by
\be
\frac{dY_{N_1}}{dZ}=\left(\frac{dY_{N_1}}{dZ}\right)_{D,S}+
\left(\frac{dY_{N_1}}{dZ}\right)_{\rm injection}\,.
\label{eff-rate}
\ee
The first term on the right hand side of equation (\ref{eff-rate})
is given by the usual Boltzmann equation ~\cite{luty.92,plumacher.96}
\be
\left(\frac{dY_{N1}}{dZ}\right)_{D,S}=
-(D+S)\left(Y_{N1}-Y^{eq}_{N1}\right)\,,
\label{thermal-rate}
\ee
where $D$ and $S$ constitute the decay and $\Delta L=1$
lepton number violating scatterings which dilute the number
density of $N_1$, and $Y^{eq}_{N1}$ is the abundance of $N_1$
in the thermal equilibrium situation.

The second term on the right hand side of equation (\ref{eff-rate}) gives
the rate of injection of $N_1$'s, coming from the
disappearance of string loops, into a comoving volume at a rate
\be
\left(\frac{dY_{N_1}}{dZ}\right)_{\rm injection}=\left(
\frac{dY_{N_1}^{st}}{dZ}\right)_{SD}+\left(\frac{dY_{N_1}^{st}}{dZ}
\right)_{QD}\,,
\label{nonthermal-rate}
\ee
where the two terms on the right hand side are given by
the equations (\ref{slowdeath-rate}) and (\ref{quickdeath-rate})
respectively. One can see from the equations
(\ref{slowdeath-rate}) and (\ref{quickdeath-rate}) that while
the first term, i.e, the slow death of string loops
dominates at early times, the second term, i.e., the quick death
of cosmic string loops dominates at late times.

The two terms in equation (\ref{eff-rate}) compete with
each other. While the first term dilutes the density
of $N_1$'s in a comoving volume, the second term try to
enrich it due to the continuous injection of $N_1$'s from
the shrinkage of cosmic string loops. The dilution of
the density of $N_1$'s, due to the $CP$-violating decay, 
vis-a-vis produces a net $B-L$ asymmetry dynamically. This
can be predicted by solving the Boltzmann equation
\be
\frac{dY_{B-L}}{dZ} = -\epsilon_{1} D\left(Y_{N1}-Y^{eq}_
{N1}\right)-W Y_{B-L}\,.
\label{asymmetry-rate}
\ee 
The first term in equation (\ref{asymmetry-rate}),
involving the decay term $D$, produces an asymmetry while a part
of it gets erased by the wash out terms involved in $W$. Note that
$W$ includes the processes of inverse decay and the $\Delta L=1$,
$\Delta L=2$ lepton number violating scatterings. The different 
terms $D$, $S$ and $W$ used in equations (\ref{thermal-rate}) and 
(\ref{asymmetry-rate}) are explained in section 2.3.3 and 
a summary of all the lepton violating scattering densities 
are given in section 4.3. Essentially the scattering densities 
depend on the two parameters $M_1$ and $\tilde{m}_1$, the 
effective neutrino mass parameter. 

\subsection{Constraint on effective neutrino mass ($\tilde{m}_1$)}
In order to explain the present baryon asymmetry of the
Universe all the lepton number violating interactions mediated
by $N_1$ have to satisfy the Sakharov's criteria
~\cite{sakharov.67}. While the complex nature of Yukawa
coupling (\ref{yukawa_term}) violates $CP$ at the same
time the decay rate of $N_1$ has to be less than the
Hubble expansion parameter $H$ in order to satisfy the out 
of equilibrium condition. This imposes an immediate 
constraint (\ref{FGLPbound}) on the effective neutrino mass 
parameter $\tilde{m}_1\leq 2\times 10^{-3}$ eV as 
explained in section 2.3.3. Therefore, for all $\tilde{m}_1\leq 
10^{-3}eV$ at an epoch $T\leq M_1$ a net $B-L$ asymmetry can be 
generated dynamically. 

The above constraint on $\tilde{m}_1$ can be envisaged in a model 
as follows. We assume a charge-neutral lepton 
symmetry~\cite{model}. In this scenario we take the texture 
of the Dirac mass of the neutrino to be 
\be
m_D=\begin{pmatrix}
0 & \sqrt{m_e m_{\mu}} & 0\\
\sqrt{m_e m_{\mu}} & m_{\mu} & \sqrt{m_e m_{\tau}}\\
0 & \sqrt{m_e m_{\tau}} & m_{\tau}.
\end{pmatrix}
\label{dirac-texture}
\ee
Using (\ref{dirac-texture}) and (\ref{lower-bound-M1}) in
equation (\ref{eff-neu-mass}) we get the constraint
\be
\tilde{m}_1 \leq 0.525\times 10^{-5}eV,
\ee
which is in concordance with our assumption $\tilde{m}_1 < 
10^{-3}$ eV at an epoch $T\leq M_1$.
 
\subsection{Solutions of Boltzmann equations}
At an epoch $T\gg M_1$ the lepton number violating processes are
sufficiently fast as to set the $B-L$ asymmetry to zero. As the
temperature falls and
becomes comparable with $M_1$ a net $B-L$ asymmetry is generated through
the $CP$-violating decay of $N_1$. The resulting asymmetry can be
obtained by solving the Boltzmann equations. We solve the equations
(\ref{eff-rate}) and (\ref{asymmetry-rate}) numerically with the
following initial conditions
\be
Y^{in}_{N1}=Y_{N_1}^{eq}~~{\mathrm and}~~ Y^{in}_{B-L}=0.
\label{in_condn}
\ee
Using the first initial condition we solve equation
(\ref{eff-rate}) for $Y_{N_1}$, and the corresponding $B-L$ asymmetry
$Y_{B-L}$ is obtained from equation (\ref{asymmetry-rate}) by using
the second initial condition of equation (\ref{in_condn}).
\begin{figure}[htbp]
\epsfig{file=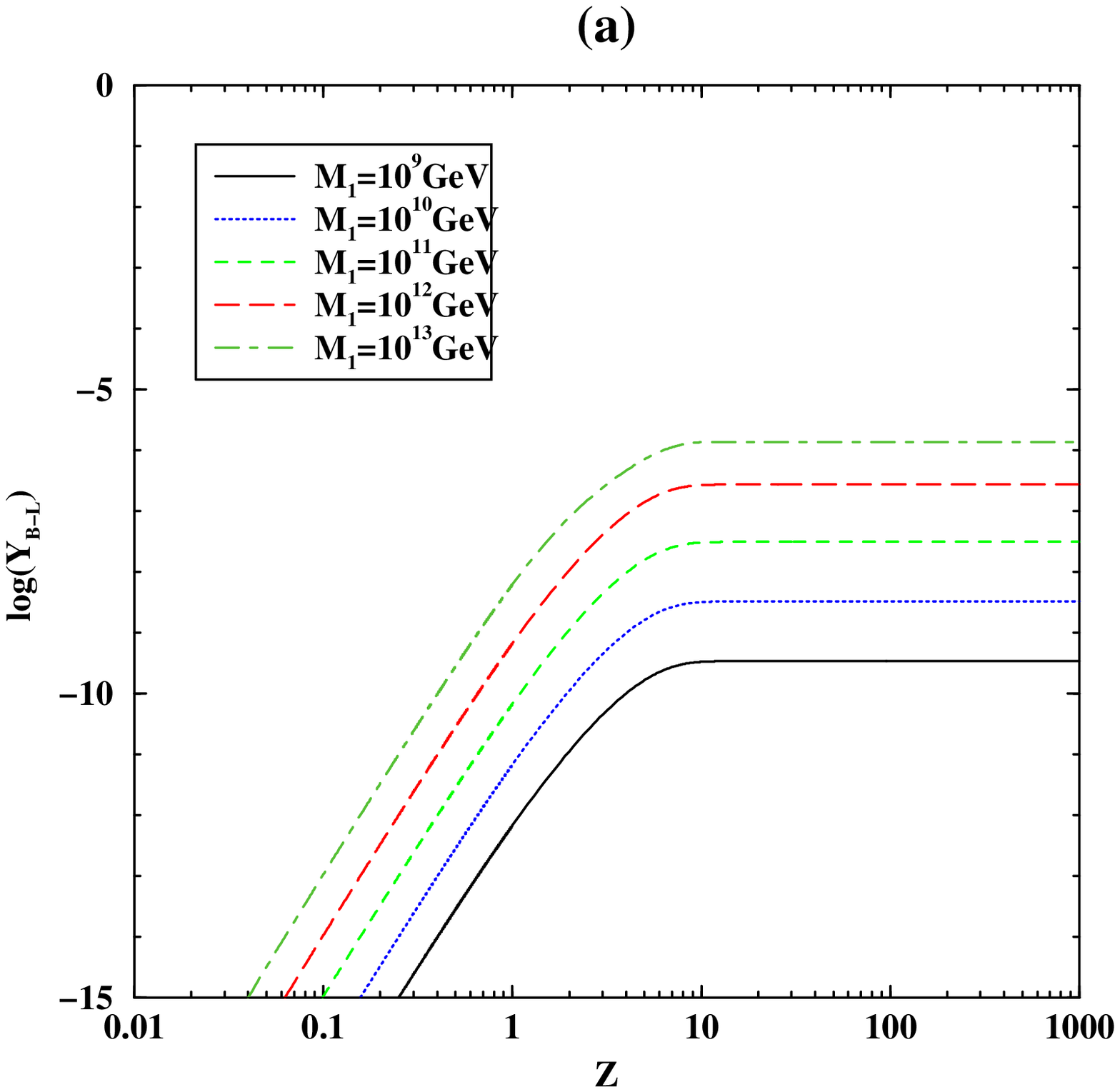, width=0.5\textwidth}
\epsfig{file=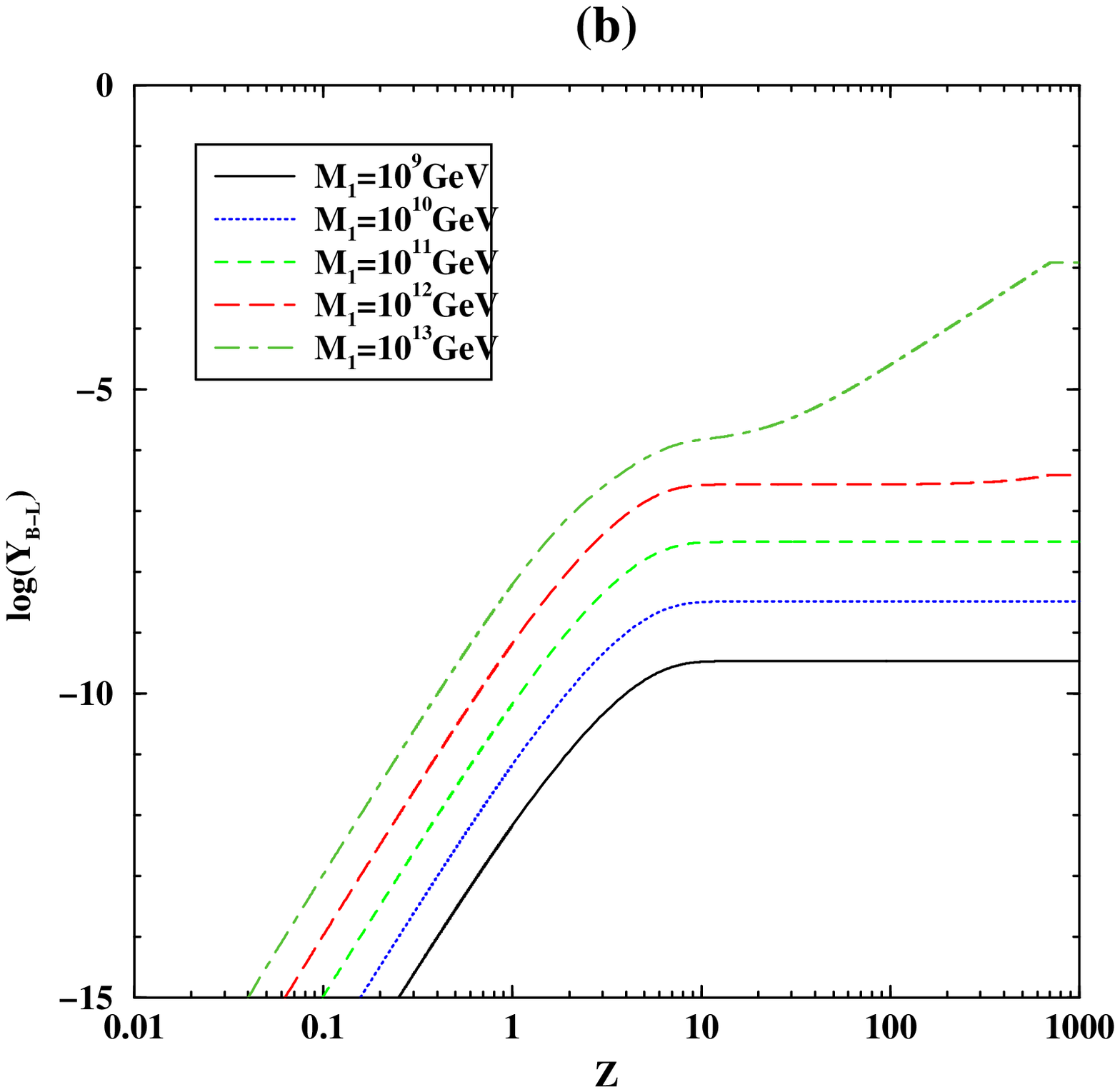, width=0.5\textwidth}
\caption{The evolution of B-L asymmetry with maximal allowed
$CP$ violation parameter $\epsilon_1$ for different
values of $M_1$, with
$\tilde{m}_1=10^{-4}eV$ and $\eta_{B-L}=10^{13}GeV$ (a) in absence
of cosmic strings, and (b) in presence of cosmic strings.}
\label{fig:eta=13-4}
\end{figure}

\begin{figure}[htbp]
\epsfig{file=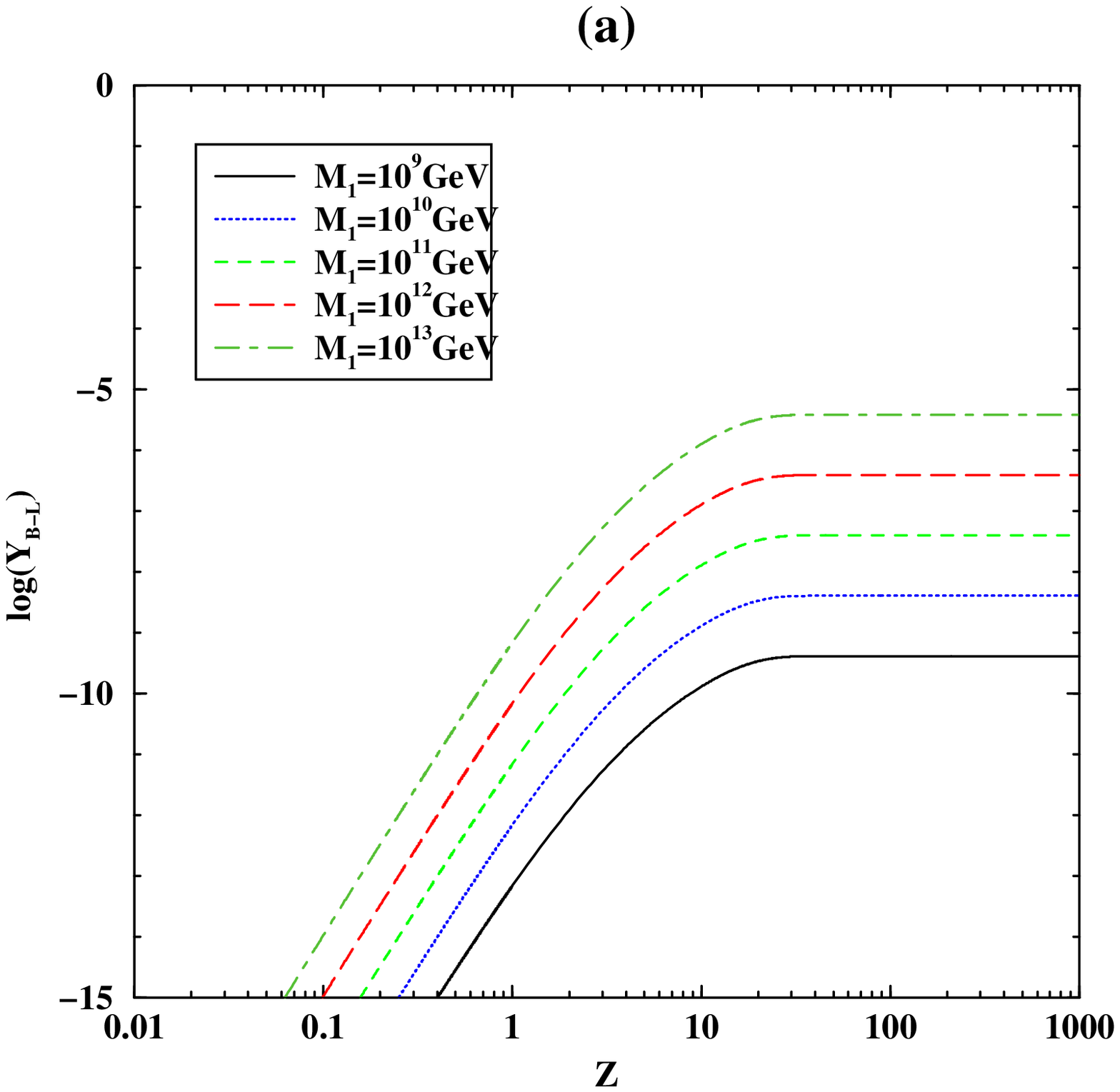, width=0.5\textwidth}
\epsfig{file=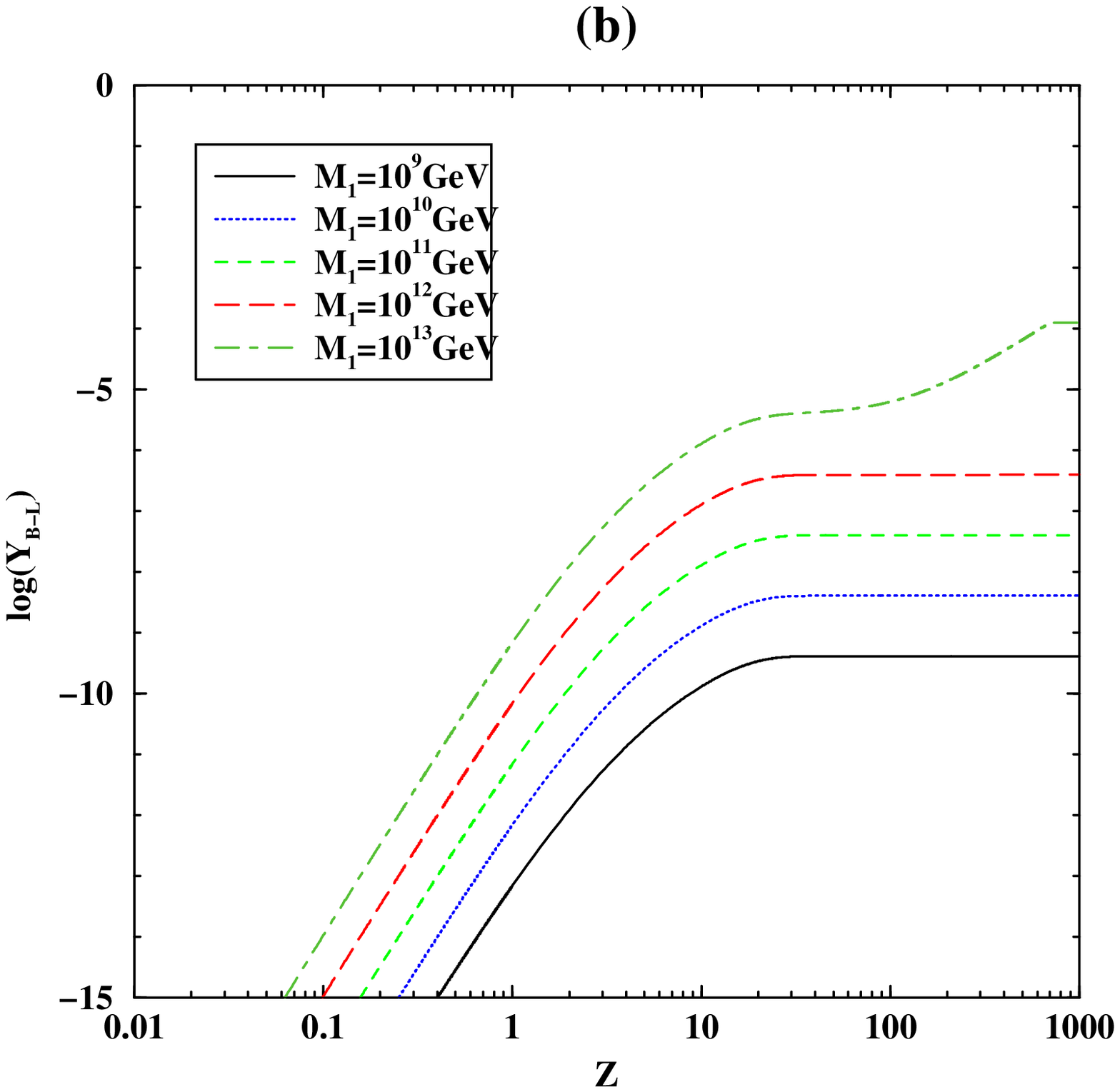, width=0.5\textwidth}
\caption{The evolution of B-L asymmetry with maximal allowed
$CP$ violation parameter $\epsilon_1$ for different
values $M_1$, with
$\tilde{m}_1=10^{-5}eV$ and $\eta_{B-L}=10^{13}GeV$ (a) in absence
of cosmic strings, and (b) in presence of cosmic strings.}
\label{fig:eta=13-5}
\end{figure}

In the usual thermal scenario the $B-L$ asymmetry depends on not
only $\tilde{m}_1$, but also $M_1$. In the present case the
$B-L$ asymmetry depends, additionally, on $\eta_{\rm B-L}$ since
the injection rate of $N_1$'s from the cosmic string loops strictly
depends on it. Depending on the $B-L$ symmetry breaking scale, the
effects of cosmic strings on the baryon asymmetry are shown in the
Figures \ref{fig:eta=13-4} and \ref{fig:eta=13-5}. In these Figures we
have taken the saturated value of the $CP$-asymmetry parameter
$\epsilon_1$.

In Figure \ref{fig:eta=13-4}(a) it can be seen that in a thermal bath
the $B-L$ asymmetry approaches the final value at around $M_1=10 T$
when all the wash out processes fall out of equilibrium. In contrast
to it, in the presence of cosmic strings the $B-L$ asymmetry continues
to build up until the injection rate of $N_1$'s in a comoving volume
is insignificant. For $f_1=1$ this happens around $M_1=6\times 10^{2} T$,
as shown in Figure \ref{fig:eta=13-4}(b), which is far larger than in the
purely thermal case. As a result of this, in presence of cosmic
strings, for a fixed value of the
symmetry breaking scale $\eta_{\rm B-L}=10^{13}\gev$, the final $B-L$
asymmetry is enhanced by three orders of magnitude for the effective
neutrino mass $\tilde{m}_1=10^{-4}eV$ (Fig.~\ref{fig:eta=13-4}(b)) and
by two orders of magnitude for $\tilde{m}_1=10^{-5}eV$
(Fig.~\ref{fig:eta=13-5}(b)). For $f_1<0.01$ the effect of
cosmic string essentially disappears. 

The above happenings, in particular
the dependence on the value of $\tilde{m}_1$, can be
understood as follows: In the absence of injection term when the wash out
processes fall out of equilibrium the asymmetry produced by the decay of
$N_1$'s does not get wiped out, and the produced $B-L$ asymmetry
remains as the final asymmetry. Since the decay rate of $N_1$'s depends
linearly on $\tilde{m}_1$ as inferred from equation (\ref{dilution}),
a relatively larger value of $\tilde{m}_1$ implies that the condition of
decay in out-of-equilibrium situation is satisfied at a relatively
later time when the abundance of thermal $N_1$'s is relatively smaller,
thus yielding a smaller final value of $Y_{B-L}$. And as expected, this
effect is larger for larger values of $M_1$.

Now let us see what happens when the injection term due to the string
loops is added. For values of $\eta_{B-L}$ for which the string
contribution remains subdominant to the thermal contribution, the
dependence on $\tilde{m}_1$ is essentially same as in the absence of the
strings as explained above. However, the situation is reversed for those
values of $\eta_{B-L}$ for which the string contributions are dominant:
Since the string-produced $N_1$'s dominate over those of thermal origin at
late times, the $B-L$ asymmetry produced by the decay of the
string-produced
$N_1$'s automatically satisfy the out-of-equilibrium condition. In such a
situation, a relatively larger value of $\tilde{m}_1$ (implying a
relatively larger rate of decay of the $N_1$'s) simply leads to a quicker
rate of development of the $Y_{B-L}$ at a relatively earlier time when the
injection rate of the $N_1$'s is relatively larger, thus yielding a
relatively higher final value of $Y_{B-L}$. Figures \ref{fig:eta=13-4}(b)
and \ref{fig:eta=13-5}(b) plotted for $\tilde{m}_1=10^{-4}eV$ and
$\tilde{m}_1=10^{-5}\ev$ simultaneously bear out these expectations.

Allowing a maximal CP-asymmetry, we see that for a reasonable
value of $\tilde{m}_1$ in certain models the string contribution
overproduces the baryon asymmetry in comparison to the thermal value
for $\eta_{\rm B-L}=10^{13}\gev$ and $f_1\gsim 0.1$. This suggests a
lower $\eta_{\rm B-L}$ or $\delta<1$ in equation (\ref{epsilon1value})
may be in better accord with the observed baryon asymmetry in the
presence of $B-L$ cosmic strings. Therefore, we have repeated
the above calculations for
$\eta_{\rm B-L}=10^{12}\gev$ and $\eta_{\rm B-L}=10^{11}\gev$.
The results are summarized in tables (\ref{table-1}) and (\ref{table-2}).
In both cases one can see that for Yukawa coupling $f_1<1$ and
$\tilde{m}_1=10^{-5}\ev$ the effect of cosmic strings disappear, i.e.
$(\Delta Y_{\rm B-L}/Y_{\rm B-L}^{th})\rightarrow 0$, where
$\Delta Y_{\rm B-L}\equiv Y_{\rm B-L}^{th+st}-Y_{\rm B-L}^{th}$. This
is in concordance with our assumption that $\tilde{m}_1\leq O(10^{-5})\ev$
in the given model in section 7.5.2. 

Studying the needed bound on $\delta$ we find that for
$\eta_{\rm B-L}=10^{13}\gev$ and $M_1=10^{13}\gev$ to
produce the final $B-L$ asymmetry, $Y_{\rm B-L}^{th+st}=O(10^{-10})$,
we need $\delta=O(10^{-7})$ and $O(10^{-6})$ for $\tilde{m}_1=10^{-4}$eV
and $\tilde{m}_1=10^{-5}$eV respectively. These values of $\delta$
diminish the purely thermal contributions, $Y_{\rm B-L}^{th}=O(10^{-13})$
and $O(10^{-12})$ respectively. Thus there exists ranges of parameter
values where, while the thermally abundant heavy neutrinos
are not sufficient to produce the required $B-L$ asymmetry, the cosmic
string contribution can produce the required asymmetry. 

\begin{table}[htbp]
\begin{center}
\caption{Effect of cosmic strings on final $B-L$ asymmetry is
shown for different values of the Yukawa coupling $f_1$ at
$\eta_{\rm B-L}=10^{12}\gev$}
\begin{tabular}{|c|c|c|c|c|c|c|}
\hline
$f_1$ & \multicolumn{3}{|c|} {$\tilde{m}_1=10^{-4}eV$} &
\multicolumn{3}{|c|} {$\tilde{m}_1=10^{-5}eV$} \\ \cline{2-7}
 & $Y^{th}_{B-L}$ & $Y^{th+st}_{B-L}$ & $\log(\delta_{reqd})$
& $Y^{th}_{B-L}$ & $Y^{th+st}_{B-L}$ & $\log(\delta_{reqd})$\\
[2mm]\hline
1 & $2.721\times 10^{-7}$ & $1.231\times 10^{-5}$ & $-5$ &
$3.903\times 10^{-7}$ & $1.594\times 10^{-6}$ & -4\\ [2mm]\hline
0.1 & $3.098\times 10^{-8}$ & $3.218\times 10^{-8}$ & $-2$
& $3.970\times 10^{-8}$ & $3.982\times 10^{-8}$ & $-2$\\
[2mm]\hline
0.01 & $3.254\times 10^{-9}$ & $3.254\times 10^{-9}$  & $-1$
& $4.038\times 10^{-9}$ & $4.038\times 10^{-9}$ & -1\\ [2mm]\hline
0.001 & $3.393\times 10^{-10}$ & $3.393\times 10^{-10}$ & 0 &
$4.107\times 10^{-10}$ & $4.107\times 10^{-10}$ & 0\\[2mm]\hline
\end{tabular}
\end{center} 
\label{table-1}
\end{table}

\begin{table}[htbp]
\begin{center}
\caption{Effect of cosmic strings on final $B-L$ asymmetry is
shown for different values of the Yukawa coupling $f_1$ at
$\eta_{\rm B-L}=10^{11}\gev$}
\begin{tabular}{|c|c|c|c|c|c|c|c|}
\hline
$f_1$ & \multicolumn{3}{|c|} {$\tilde{m}_1=10^{-4}eV$} &
\multicolumn{3} {|c|} {$\tilde{m}_1=10^{-5}eV $}\\ \cline{2-7}
 & $Y^{th}_{B-L}$ & $Y^{th+st}_{B-L}$ & $\log(\delta_{reqd})$
& $Y^{th}_{B-L}$ & $Y^{th+st}_{B-L}$ & $\log(\delta_{reqd})$\\
[2mm]\hline
1 & $3.098\times 10^{-8}$ & $1.514 \times 10^{-7}$ & -3 & $3.970\times
10^{-8}$ & $5.174\times {-8}$ & -2 \\ [2mm]\hline
0.1 & $3.254\times 10^{-9}$ & $3.266\times 10^{-9}$ & $-1$
& $4.038\times 10^{-8}$ & $4.039\times 10^{-9}$ & $-1$\\
[2mm]\hline
0.01 & $3.392\times 10^{-10}$ & $3.392\times 10^{-10}$ & 0 & $ 4.107\times
10^{-10}$ & $4.107\times 10^{-10}$ &  0\\ [2mm]\hline
\end{tabular}
\end{center}
\label{table-2}
\end{table} 

While the mechanism investigated here can be easily generalized
to a supersymmetric model, there are unlikely to be significant
changes in the quantitative aspects. In particular unless some
salient physics significantly modifies the scale of efficacy, viz.,
$10^{11}\gev$ of this mechanism, the supersymmetric generalization
is not helpful in resolving the issue raised by the gravitino
bound and the need for a mechanism at scales $<10^9\gev$. 

A more fruitful approach towards success of this scenario at
lower energy scales would be to also seek additional
sources for $CP$ violation for the cosmic string generated
heavy neutrinos. This is possible since they get produced
from the decay of a bosonic condensate consisting of the
string loops. However the strength of the present investigation
is the direct comparison between the two sources of heavy
neutrinos. If the mechanism applicable to the non-thermal
source is different, the model becomes less constrained and
its verifiability is sacrificed to a certain extent.

A useful conclusion from the rather constrained scenario
considered here is that in the regime $\eta_{\rm B-L}>10^{11}\gev$
and $h_1\gsim 0.01$ for $\epsilon_1\ll \epsilon_1^{max}$
(i.e. $\delta\ll 1$) while the thermal abundance
of right handed neutrinos is not sufficient to produce the required
$B-L$ asymmetry the cosmic strings can give rise to the observed level of
the asymmetry.

\chapter{Conclusions}
Baryogenesis via leptogenesis is an attractive scenario 
that links the physics of right handed neutrino sector with 
the low energy neutrino data. In the light of current neutrino 
oscillation data we studied the bounds on the mass scale 
of lightest right handed neutrino as well as their mass 
hierarchy from the leptogenesis constraint which we discuss 
in two different parts of the thesis. In part-I, we study the 
baryogenesis via leptogenesis in a {\it thermal} scenario, 
while part-II of the thesis is devoted to a study of the same 
in a {\it non-thermal} scenario. Moreover, we divide the 
phenomenological models into two categories, {\it type-I} 
and {\it type-II}, depending on the seesaw mechanism used to 
generate the light Majorana neutrino masses. 

In Part-I, we begin with a brief introduction to type-I models. 
As an example, we consider a minimal extension of the $SM$ with the 
gauge group $SU(2)_L\times U(1)_Y\times U(1)_{Y'}$, where $Y'$ is 
a linear combination of $Y$ and $B-L$. Since $B-L$ is a gauge 
symmetry of the model any pre-existing asymmetry is washed out. 
A net asymmetry is generated when $B-L$ symmetry breaks. In such 
models, the right handed heavy Majorana neutrinos, $N_i$ with 
$i=1,2,3$, are singlet under the $SM$ gauge group $SU(2)_L\times 
U(1)_Y$. The type-I seesaw mechanism then gives rise to the light 
Majorana neutrino mass matrix, $m_\nu=m_D^TM_R^{-1}m_D$, with $m_D$ 
is the Dirac mass matrix of the neutrinos. In these models, we 
briefly discussed the baryogenesis via the route of leptogenesis in 
a {\it thermal scenario}. The leptogenesis occurs via the out of 
equilibrium decay of thermally generated heavy Majorana neutrinos.
We assume that the mass spectrum of right handed Majorana neutrinos 
is in normal hierarchy. In this scenario, the final lepton asymmetry 
is given by the $CP$-violating decays of the lightest right handed 
neutrino $N_1$ to $SM$ Higgs and lepton. A part of this
asymmetry is then transformed to the $B$-asymmetry through
the equilibrated sphaleron processes.

The $L$-asymmetry predicted by a model is a measure of the
magnitude of $CP$-violation in that model. The theoretical
upper bound on the $CP$-asymmetry, produced by the decay of
$N_1$, in type-I models is given by $\epsilon_1$ 
$\leq (3M_1/16\pi v^2)$$\sqrt{\Delta m^2_{atm}}$, with 
$\Delta m^2_{atm}$ is the mass squared difference of light
neutrinos in atmospheric data. The analytical estimation of
$L$-asymmetry in these models recasts the upper bound on
$CP$-asymmetry in terms of a lower bound on the mass scale of
$N_1$ to be $M_1\geq 10^9 GeV$. In the light of current 
neutrino oscillation data we numerically check the compatibility of 
the analytical ound on $M_1$ by solving the required Boltzmann 
equations. 

The $B-L$ gauge symmetry in the given example of type-I model
is quite ad hoc. We therefore consider models where $B-L$
gauge symmetry emerges naturally. In particular, we consider
Left-Right symmetric model, a specific example of type-II
seesaw models. The scalar sector of the model is very rich
and consists of two triplets, namely $\Delta_L$ and $\Delta_R$,
and a bidoublet Higgs $\Phi$, which contains two copies of
$SM$ Higgs. The type-II seesaw mechanism in this model gives
rise the Majorana mass matrix of the light neutrinos of the
form $m_\nu=M_L-m_DM_R^{-1}m_D^T$. In contrast to type-I models,
in the present case the additional mass, $M_L$, is provided by the
vacuum expectation value of the triplet $\Delta_L$.
The two terms, $M_L\equiv m_\nu^{II}$ and
$m_DM_R^{-1}m_D^T\equiv m_\nu^I$, contributing to $m_\nu$ are
called type-II and type-I respectively. In these models,
irrespective of the magnitudes of type-I and type-II terms, we 
show that the lower bound on $M_1$ can be reduced by an order 
of magnitude in comparison to type-I case.

Within the Left-Right symmetric model, we consider a special case
in which the $CP$-violation arises through the spontaneous symmetry
breaking. The Lagrangian of the model is $CP$
invariant which demands that all the Yukawa couplings should be
real. In this scenario, the vacuum expectation values ($VEV$s) of
the neutral Higgses are complex and they lead to complex masses
for fermions and hence $CP$-violation. In the Left-Right symmetric
model, there are four complex neutral scalars which acquire VEVs.
However, the unbroken global $U(1)$ symmetries associated with
$SU(2)_L$ and $SU(2)_R$ gauge groups allow two of the phases to
be set to zero. Using the remnant $U(1)$ symmetry after the
breaking of $SU(2)_R$, one phase choice is made to make the
$VEV$ of $\Delta_R$, and hence the mass matrix of right handed
neutrinos, real. The phase associated with the other
$U(1)$ symmetry can be chosen to achieve two different types
of simplification of neutrino mass matrix. In the
{\it type-II choice}, the $m_\nu^I$ is made real leaving
the $CP$-violating phase purely with $m_\nu^{II}$.
In this phase convention, we derive a lower bound on the mass
scale of $N_1$ from the leptogenesis constraint by assuming a
normal mass hierarchy in the right handed neutrino sector.
It is shown that the mass scale of $N_1$ satisfy the
constraint $M_1\geq 10^8 GeV$, which is in
good agreement with the lower bound on $M_1$ in generic type-II
seesaw models. In the {\it type-I phase choice}, only the type-I 
term contains $CP$-violating phase leaving type-II term real. This 
allows us to derive an upper bound on the heavy neutrino mass 
hierarchy from the leptogenesis constraint. In order to achieve the
observed baryon asymmetry of the present Universe, it is
found that the mass hierarchy of right handed neutrinos
satisfy the constraint $M_2/M_1\leq 17$ and $M_3/M_1\leq 289$
simultaneously. Numerically we verified that these bounds are 
compatible with the low energy neutrino oscillation data for all 
values of $M_1\geq 10^8 GeV$ as implied by the lower bound on $M_1$ 
in type-II phase convention. The bound on $M_1$, in production of 
$L$-asymmetry through the $CP$-violating decays of thermally 
generated $N_1$,
is $\geq 10^8 GeV$ which is far above the current accelerator
energy range and beyond the reach of the next generation
accelerators. However, these scenarios are well motivated by
the current status of low energy neutrino oscillation data.

As an alternative, we consider mechanisms which work
at $TeV$ scale and be consistent with the low energy neutrino
data. We study a general scenario for $TeV$ scale leptogenesis
in a gauged $B-L$ symmetric model. By solving the Boltzmann 
equations we explore the viable regions in the plane 
of the effective neutrino mass arameter $\tilde{m}_1$ and 
the mass of lightest right handed neutrino $M_1$. We assume 
that the required lepton asymmetry of the present Universe is 
produced during the $B-L$ gauge symmetry breaking phase transition. 
The limited erasure by the processes mediated by $N_1$ requires 
that $\tilde{m}_1< 10^{-3}eV$. Solution of the relevant Boltzmann 
equations shows that for $\tilde{m}_1=10^{-4}eV$ the mass of $N_1$ 
has to be smaller than $10^{12}GeV$ and can be as low as $10 TeV$. 
In a more restrictive scenario where the neutrino Dirac mass matrix
is identified with that of the charged leptons it is necessary that 
$M_1>10^8 GeV$ in order to satisfy $\tilde{m}_1 < 10^{-3}eV$. In this 
scenario all values $M_1$, $10^8 GeV < M_1 < 10^{12}GeV$ can 
successfully create the required asymmetry. If the Dirac mass 
scale of neutrinos is less restricted, much lower values of 
$M_1$ are allowed. In particular, a right handed neutrino as 
low as $10 TeV$ is admissible if the Dirac mass scale is about 
a factor of $100$ smaller than the mass scale of charged 
leptons. We conjecture that the hypothesis of $TeV$ scale 
right handed neutrinos can be verified in the near future, thus 
providing an indirect evidence of baryon asymmetry creation at 
$TeV$ scale.

In part-II of the thesis, we discuss the formation and
evolution of topological defects, in particular cosmic strings, 
in the early Universe. Topological defects arise as the solitonic 
solutions in gauge theories. `Solitons' or `solitary waves' are 
the time independent solutions of non-linear wave equations in 
classical 
field theories. The prime among them is $\lambda \phi^4$ theory. In 
1+1-dimensions the solitonic solutions in $\lambda \phi^4$ theory are 
called `kinks'. Each solitonic solution is designated by a number
called the `topological charge' or `winding number'.
 
An inevitable feature of the early Universe phase transitions
is the formation of topological defects.
In particular, we deal with cosmic strings. These defects
are extended objects and are not distributed thermally.
Therefore, the decay of these objects can be a {\it non-thermal}
source of massive particles that constitutes them. Moreover, the
cosmic strings formed at a phase transition can also influence the
nature of a subsequent phase transition that may have important
implications for the generation of Baryon Asymmetry of the 
Universe ($BAU$).

In Quantum Field Theory ($QFT$), solutions of Dirac equation
in the presence of solitonic objects lead to a curious
phenomenon of `fractional fermion number'.
This is because of the existence of degenerate zero energy modes
of fermions while quantized in the background of a solitonic
vacuum. In contrast to it, in the translational invariant vacuum
there are no zero energy solutions of Dirac equation and therefore
fermions are quantized by integral unit. The fractional solitonic
states are therefore superselected from the normal vacuum and
are not allowed to decay in isolation.
 
An important feature of cosmic strings is that during their
formation they trap zero energy modes of fermions. These 
fermionic zero modes induce fractional
fermion number ($|n|/2$) on a string of winding number $n$.
If $n$ is odd then the induced fermion number on the string is
half-integral. Therefore, it is superselected from the translation 
invariant vacuum where
the eigenvalues of the number operator carry integral
fermion number. Thus, a string of half integer fermion number
can not decay in isolation because of the unavailability of
fractional fermion states in the translational invariant vacuum.
We construct examples where metastable infinitely long cosmic
strings enjoy stability due to this quantum mechanical phenomenon.

There exist both analytical as well as numerical studies of
the evolution of cosmic strings network in the early Universe.
These suggest that the strings network quickly enters a scaling
regime in which the energy density of the strings scales as a
fixed fraction of the energy density of radiation in the
radiation dominated epoch, or the energy density of matter in
the matter dominated epoch. In both cases the energy density
scales as $t^{-2}$. In this regime one of the fundamental physical
process that maintains the strings network to be in that
configuration is the formation of sub-horizon size {\it closed
loops} which are pinched off from the network whenever a
string segment curves over into a loop, intersecting itself. 

There have been many scenarios that consider decaying,
collapsing, or repeatedly self-intersecting closed loops of
such cosmic strings providing a non-thermal source of massive
particles that ``constitute'' the string. The decay of these
massive particles gives rise to the observed $B$-asymmetry or
a significant fraction of it. We have
estimated the contribution to $BAU$ from cosmic string loops which
disappear through the process of (a) slow shrinkage due to
energy loss through gravitational radiation --- which we call
slow death (SD), and (b) repeated self-intersections --- which
we call quick death (QD). We find that for reasonable values of
the relevant parameters, the SD process dominates over the QD
process as far as their contribution to $BAU$ is concerned.
We assume that the final demise of each string loop in the
SD process produces $O(1)$ right handed neutrino, $N_1$.
We demand that the $B$-asymmetry, produced by the decay of
non-thermally generated $N_1$'s through the leptogenesis route,
should not exceed the observed value predicted by WMAP. This
requires that the mass scale of $N_1$ satisfy the constraint
$M_1\leq 2.4 \times10^{12}\left(\eta_{\rm B-L}/10^{13}GeV\right)
^{1/2}GeV$, where $\eta_{\rm B-L}$ is the scale of $U(1)_{\rm B-L}$
gauge symmetry breaking phase transition. This in turn
constrains the $B-L$ symmetry breaking scale to be
$\eta_{\rm B-L}\geq 5.6\times 10^{11} GeV$ unless the Majorana
Yukawa coupling of $N_1$ is allowed to be greater than unity.
In this analytical approximation we have assumed that above 
the mass scale of $N_1$ there is no lepton asymmetry. A net 
lepton asymmetry has been produced just below the mass scale of 
$N_1$ by it's $CP$-violating decays to $SM$ Higgs and lepton. 
We then checked the analytical results against the numerical 
smulations. This is done by solving the relevant Boltzmann 
equations and including the effects of both thermal and string 
generated right handed neutrinos. We explored the parameter region 
spanned by the relevant light- and heavy neutrino mass parameters 
$\tilde{m}_1$ and $M_1$, and constrained the scale of $B-L$ symmetry 
breaking, $\eta_{B-L}$, as well as the Majorana Yukawa coupling 
$f_1$ of the lightest right handed neutrino. It is shown that for 
the values $\eta_{B-L}>10^{11}$GeV, where they can be effective, 
cosmic strings make a more dominant contribution than thermal 
leptogenesis. This, in turn, provides an upper bound on the 
$CP$ violating phase giving rise to the $L$-asymmetry.

\appendix
\chapter{The Boltzmann equations}
\section{Particle distribution in thermal plasma}
Let $f_{i}(\vec{p},t)$ be the density distribution of the particle 
species $i$ in phase space. Assuming that the Universe is isotropic 
$f_{i}$ can be chosen to be independent of the position vector $\vec{r}$. 
In the high temperature thermal bath it is assumed that all species
of particles in the Universe were initially in thermal equilibrium 
and spread homogeneously. The only deviation comes for the massive 
particles as the Universe expands. The expansion of the Universe causes 
the momenta of all particles to redshift, so that $p\sim \frac{1}{R}$. 
So long as the energy density of the Universe is dominated by the
ultra relativistic particle species the temperature of the
Universe will like wise redshift as $T\sim \frac{1}{R}$. Thus
the distribution function, $f_{i}$,  is simply given by
\begin{equation}
f_{i}(\vec{p})=e^{-\frac{|\vec{p_{i}}|}{T}}
\label{eA.1}
\end{equation}
where $i$ stands for the particle species. However, for massive 
particles the distribution function must be
\begin{equation}
f_{i}(\vec{p})=e^{-\frac{\sqrt{|\vec{p_{i}}|^{2}+M_{i}^{2}}}{T}}.
\label{eA.2}
\end{equation}
In the configuration space the number density of the of the
particle species $i$ is
\begin{equation}
n_{i}=g_{i}\int \frac{d^{3}\vec{p}}{(2\pi)^{3}}f_{i}(\vec{p})
\label{eA.3}
\end{equation}                                                        
where $g_{i}$ is the number of accessible spin states for the
particle species $i$. Thus for a massless thermal photon($\gamma$),
the density distribution, given by equation (\ref{eA.3}), is
\begin{equation}
n_{\gamma}=\frac{2}{\pi^{2}}T^{3}.
\label{eA.4}
\end{equation}
Far above their mass scales the massive particles are in thermal 
equilibrium. Therefore, the phase space distribution function is given 
by 
\begin{equation}
f_{i}^{eq}=e^{-\frac{E_{i}}{T}},
\label{eA.5}
\end{equation}
where $E_{i}=\sqrt{|p_{i}|^{2}+M_{i}^{2}}$. So the equilibrium
density distribution is
\begin{equation}
n_{i}^{eq}=g_{i}\int \frac{d^{3}p_{i}}{(2\pi)^{3}}f_{i}^{eq}
\label{eA.6}
\end{equation}
In terms of the dimensionless variables $x=\frac{\sqrt{|p_{i}|^{2}
+M_{i}^{2}}}{T}$ and $Z=\frac{M_{i}}{T}$, equation (\ref{eA.6}) can be 
rewritten as
\begin{eqnarray}
n_{i}^{eq} &=& g_{i}\frac{T^{3}}{2\pi^{2}}\int_{Z}^{\infty} x e^{-x}
\sqrt{x^{2}-Z^{2}}dx\nonumber\\
&=& g_{i}\frac{T^{3}}{2\pi^{2}} Z^{2}K_{2}(Z)
\label{eA.7}
\end{eqnarray}                                                        
where $K_{2}(Z)$ is modified Bessel function. As $Z\rightarrow 0$,
$z^{2}K_{2}(Z)\rightarrow 2$. In this limit, the density
distribution of massive particles resembles with mass less
particles. Thus the approximation that at a temperature
above their mass scale all particles are in thermally equilibrium
is a valid assumption. As the Universe expands the temperature falls 
($Z$ increases). Therefore, the density distribution of all particles
fall and is governed by the Boltzmann transport equations. 

\section{Boltzmann transport equations}
In this section, we give a pragmatic introduction to Boltzmann 
equations to study the evolution of the density of any particle 
species in the absence and presence of interactions.

The expansion of the Universe dilutes the number densities
of all types of particles even in the absence of interactions
at a rate
\begin{eqnarray}
\frac{dn_{i}}{dt} &=& -3\frac{\dot{R}}{R}n_{i}\nonumber\\
                  &=& -3Hn_{i}
\label{density-dil}
\end{eqnarray}
where $R(t)$ is scale factor in Freedman Robertson and Walker (FRW) 
Universe and $\dot{R}$ is derivative with respect to time. $H$ is Hubble 
expansion factor. Thus in the absence of any interaction the Boltzmann 
transport equation for the given particle species $i$ of density 
$n_{i}$ is
\begin{equation}
\frac{dn_{i}}{dt}+3Hn_{i}=0.
\label{boltzmann-eqn}
\end{equation}
Now we scale out the effect of the expansion of the Universe by 
considering the evolution of the number of particles in a comoving 
volume. This can be done by dividing the number density of the particle 
species $i$ with its entropy density, i.e.
\begin{equation}
Y_i=\frac{n_{i}}{s}.
\label{boltzmann-eqn-1}
\end{equation}
Using the conservation of entropy per comoving volume ($sR^{3}$ =
constant), equation (\ref{boltzmann-eqn-1}) can be written as
\begin{equation}
\frac{dn_{i}}{dt}+3Hn_{i}= s\dot{Y}=0.
\label{boltzmann-eqn-2}
\end{equation}                                                        

As the Universe expands the momentum $p_i$ of the particle species $i$
falls as $1/R$ and thus also the temperature $T$. Under rescaling the 
momenta of massless particles remain unchanged. So they keep themselves 
in equilibrium with the thermal plasma. Above the mass scale of any 
massive particle it will behave as a massless one. Below its mass scale 
the interaction rate decreases in comparison to the Hubble expansion rate 
and hence it falls out of equilibrium because it needs several collision 
times to keep it in equilibrium with the thermal photons. The departure 
of the density of any species $i$ from its thermal equilibrium value can 
be predicted by solving the Boltzmann transport equations.

For simplicity we consider the decay of any massive species $i$ to 
a set of particles $Y$. As a result the equation (\ref{boltzmann-eqn-2} 
modifies to
\be
\frac{dn_{i}}{dt}+3Hn_{i}=-\sum_{i\leftrightarrow Y}
\left[ \frac{n_i}{n_i^{eq}}\gamma(i\rightarrow Y)-
\frac{n_Y}{n_Y^{eq}}\gamma(Y\rightarrow i)\right],
\label{boltzmann-eqn-3}
\ee
where 
\be
\gamma(i\rightarrow Y)=\int d\Pi_i d\Pi_Y(2\pi)^4 
\delta^4(p_i-p_Y)f_i^{eq} |\mathcal{A}(i\rightarrow Y)|^2.
\label{sca-den}
\ee
In equation (\ref{sca-den}), $d\Pi=\frac{1}{2E}
\frac{d^3p}{(2\pi)^3}$. If we neglect $CP$-violation then 
$|\mathcal{A}(i\rightarrow Y)|^2=|\mathcal{A}(Y\rightarrow i)|^2$. 
Using (\ref{eA.6}) the above equation (\ref{boltzmann-eqn-3}) simplifies 
to 
\be
\frac{dn_{i}}{dt}+3Hn_{i}= -\Gamma_D(n_i-n_i^{eq}),
\label{boltzmann-eqn-4}
\ee
where we have used
\be
\Gamma_D=\frac{1}{2E_i}\int \frac{d^3 p_Y}{(2\pi)^3 2E_Y} (2\pi)^4 
\delta^4(p_i-p_Y) |\mathcal{A}|^2.
\ee
Note that in the above simplification we have assumed 
$n_Y=n_i^{eq}$ and it is true because the decay products 
$Y$ are massless till the later epochs of our interest. 
Substituting $Z=M_i/T$ and $Y_i=n_i/s$ in equation 
(\ref{boltzmann-eqn-3}) we get  
\bea
\frac{dY_i}{dZ} &=& -\frac{\Gamma_D}{ZH(Z)}(Y_i-Y_i^{eq})\nonumber\\
&=& -D(Y_i-Y_i^{eq}).
\label{boltzmann-eqn-5}
\eea 
Considering the $2\leftrightarrow 2$ scatterings involving 
the species $i$ equation (\ref{boltzmann-eqn-5}) can be 
extended to 
\be
\frac{dY_i}{dZ}=-(D+S)(Y_i-Y_i^{eq}),
\ee
where $S=\Gamma_s/ZH$. This is the final Boltzmann equation 
for the evolution of any species $i$ due to its decay and 
scatterings.
\chapter{Homotopy theory}
In the following we discuss some of the fundamental definitions 
which are useful for our purpose.\\
\underline{Path}: A path $f$ in a manifold $M$ is defined 
as a continuous function $f(t)$ of a real parameter 
$t$, so that each value of $t$ in the interval $0\le t 
\le 1$ corresponds to a point $f(t)$ in the manifold $M$.
If a path $f(t)$ connects the points $P$ and $Q$ we have 
$f(0)=P$, $f(1)=Q$. If $f(0)=f(1)=P$ we have a {\it closed} path. 
On the other hand, if $f(0)\neq f(1)$ then the path is 
${\it open}$.\\
\underline{Inverse} of a path $f$ is written as $f^{-1}$ 
and is defined as 
\begin{equation}
f^{-1}(t)=f(1-t), 
\label{eB.1}
\end{equation}
so that it corresponds to the same path traversed in 
the opposite direction.\\ 
\underline{Product} of two paths $f$ and $g$ is written as 
$h=fg$ and is given by 
\begin{eqnarray}
h(t) &=& f(2t)\hspace{1cm} for\hspace{5mm} 0\le t\le 
1/2 \nonumber\\
h(t) &=& f(2t-1) \hspace{1cm} for\hspace{5mm} 1/2 
\le t \le 1.
\label{eB.2}
\end{eqnarray}
Two paths $f(t)$ and $g(t)$, both starting at $P$ and ending 
at $Q$($ P\neq Q$), are said to be {\it homotopic} to each other 
if $f(t)$ is continuously deformed to $g(t)$ and is defined by 
$f \sim g$. More specifically we can define a function $L(s, t)$ 
such that $L(0, t)=f(t)$ and $L(1, t)= g(t)$.

Having defined path we shall construct a group by 
introducing a class of paths homotopic to $f$, denoted 
by $[f]$. They must, of course, have the same end-points. 
These homotopy classes may be multiplied, by defining 
the multiplication law
\begin{equation}
[f][g]=[fg].
\label{eB.3}
\end{equation}
It is easy to see that, this multiplication law defines a 
group, called the {\it fundamental} group or {\it first homotopy} 
group of the manifold $M$, and denoted by $\pi_{1}(M)$. This 
requires the  four properties that satisfied by a group. 
They are:\\
1.\hspace{2mm} Closure: If $[f]\in \pi_{1}(M)$ and $[g]
\in \pi_{1}(M)$, then adhering to (\ref{eB.3}) $[f][g]\in 
\pi_{1}(M)$. \\
2. \hspace{2mm} Associativity: Since $(fg)h \sim f(gh)$ 
we have $([f][g])[h]=[f]([g][h])$.\\
3. \hspace{2mm}Identity element: This is the class of 
paths[I] that can be shrunk to a point.\\
4.\hspace{2mm} Inverse: Since $[f^{-1}][f]=[I]$, so 
$[f]^{-1}= [f^{-1}]$

The manifold $M$ is said to be a {\it simply connected} if 
all the closed paths in that manifold can be shrunk to a point. 
On the other hand, if all paths can not be shrunk to 
a point then, the Manifold  is {\it path connected}.


\begin{thebibliography}{999}
\bibitem{sakharov.67} A.D.~Sakharov, JETP Lett. {\bf 5}, 24 (1967).

\bibitem{yoshimura}M.~Yoshimura, {\PRL {41}}, 281 (1978);
        \ erratum, 42, 740 (1979).

\bibitem{weinBgen} S.~Weinberg, {\PRL {42}}, 850 (1979)
                                                      
\bibitem{cknrev}A.G.~Cohen, D.B.~Kaplan and A.E.~Nelson
        \ Ann. \ Rev. \ Nucl. \ Part. \ Sci. 42:27-70, (1993).        

\bibitem{trodrev} M.~Trodden, \ Rev. \ Mod. \ Phys. 71, 1463 (1996).

\bibitem{yajbgen} U.A.~Yajnik, \ Pramana, 54, 471 (2000).

\bibitem{jansen.96} K. Jansen, Nucl.\ Phys.\ B (proc. suppl.) 
{\bf 47}, 196, 1996.  

\bibitem{fukugita.86}M.~Fukugita and T.~Yanagida, Phys.~Lett.
B {\bf 174}, 45 (1986).

\bibitem{luty.92} M.A.~Luty, Phys. ~Rev. D {\bf 45}, 455 (1992).    

\bibitem{mohapatra.92}R.N.~Mohapatra and X.~Zhang,
Phys.\ Rev.\ D {\bf 46}, 5331 (1992).  

\bibitem{plumacher.96} M.~Plumacher, Z.phy.C {\bf 74}, 549 (1997).   

\bibitem{krs.86}V.A.~Kuzmin, V.A.~Rubakov and M.E.~Shaposhnikov
        {\PLB {155}}, 36 (1985).

\bibitem{arn_mac.88} P.A.~Arnold and L.~Mc Lerran,
        {\PRD {36}}, 581 (1987); {\PRD {37}}, 1020 (1988)
 
\bibitem{aaps.91}
J.~Ambjorn, T.~Askgaard, H.~Porter and M.E.~Shaposhnikov, 
\PLB{244},  479 (1990); \NPB{353}, 346 (1991)

\bibitem{solar_data} S.N.~Ahmed et al (SNO collaboration),
arXiv:nucl-ex/0309004; Q.R.~Ahmed et al, Phys.Rev.Lett. {\bf 89},
011301-011302 (2002); J.N.~Bahcall and C.~Pena-Garay,
[arXiv:hep-ph/0404061]. 

\bibitem{atmos_data} S.~Fukuda et al. (Super-Kamiokande 
Collaboration) Phys.~Rev.~Lett. {\bf 86}, 5656 (2001).

\bibitem{kamland_data} K.~Eguchi et al (KamLAND collaboration),
Phys.~Rev.~Lett. {\bf 90}, 021802 (2003).                             

\bibitem{gel-ram-sla} M. ~Gell-Mann, P.~Ramond and R.~Slansky
in {\it Supergravity} (P. van Niewenhuizen and D.~Freedman, eds),
(Amsterdam), North Holland, 1979; T.~Yanagida in {\it Workshop
on Unified Theory and Baryon number in the Universe} (O. Sawada
and A. Sugamoto, eds), (Japan), KEK 1979; R. N. Mohapatra and
G.~Senjanovic, Phys.~Rev.~Lett. {\bf 44}, 912 (1980).

\bibitem{slansky_rep} R.~Slansky, Phys.~Rep.~{\bf 79},~1(1981).

\bibitem{magg-wet.80} M.~Magg and C.~Wetterich,
Phys.~Lett.~B {\bf 94},~61 (1980).
 
\bibitem{wett.81} C.~Wetterich,
Nucl.~Phys.~B {\bf 187}, 343 (1981).
 
\bibitem{moh-senj.81} R.N.~Mohapatra and G.~Senjanovic
Phys. Rev. D{\bf 23}, 165 (1981).
 
\bibitem{laz-shf-wet.81} G.~Lazarides, Q.~Shafi and C.~Wetterich,
Nucl. Phys. B {\bf 181}, 287 (1981).
 
\bibitem{moha-susy-book.92}
R.N.~Mohapatra,{\it Unification And Supersymmetry},
(Springer-Verlag, New-York, 1992).

\bibitem{smirnov_jhep}E.K.~Akhmedov, M.~Frigerio and A.Y.~Smirnov,
JHEP {\bf 0309}, 021 (2003)[arXiv:hep-ph/0305322], G.C. Branco,
R. Gonzalez Felipe, F.R. Joaquim, M.N. Rebelo, Nucl. Phys. B{\bf 640}
202-232,2002,[arXiv: hep-ph/0202030].
D.~Falcone,
  Phys.\ Rev.\ D {\bf 68}, 033002 (2003)
  [arXiv:hep-ph/0305229].

\bibitem{lep_phase}
D.~Falcone and F.~Tramontano,
  Phys.\ Rev.\ D {\bf 63}, 073007 (2001)
  [arXiv:hep-ph/0011053].
 A.~S.~Joshipura, E.~A.~Paschos and W.~Rodejohann,
  Nucl.\ Phys.\ B {\bf 611}, 227 (2001)
  [arXiv:hep-ph/0104228].
JHEP {\bf 0108}, 029 (2001)
  [arXiv:hep-ph/0105175].
 G.~C.~Branco, M.~N.~Rebelo and J.~I.~Silva-Marcos,
  arXiv:hep-ph/0510412.
 G.~C.~Branco, R.~Gonzalez Felipe, F.~R.~Joaquim, I.~Masina, M.~N.~Rebelo and C.~A.~Savoy,
  Phys.\ Rev.\ D {\bf 67}, 073025 (2003)
  [arXiv:hep-ph/0211001].
 S.~Pascoli, S.~T.~Petcov and W.~Rodejohann,
  Phys.\ Rev.\ D {\bf 68}, 093007 (2003)
 [arXiv:hep-ph/0302054].
\bibitem{davidson&ibarra.02} S.~Davidson and A.~Ibarra,
Phys. Lett. B {\bf 535}, 25 (2002).
[arXiv:hep-ph/0202239]                                                

\bibitem{buch-bari-plum.02} W.~Buchmuller, P.~Di Bari
and M.~Plumacher, Nucl. Phys. B {\bf 643}, 367 (2002).
[arXiv:hep-ph/0205349].                                 

\bibitem{hamaguchi-etal.02} K.~Hamaguchi, H.~Murayama and
T.~Yanagida, Phys.~Rev.~D {\bf 65}, 043512 (2002). 

\bibitem{hamby.03} T.~Hambye {\it et.al.}[arXiv:hep-ph/0312203].   

\bibitem{antusch.04} S.~Antusch and S.F.~King,
Phys.\ Lett.\ B {\bf 597}, 199 (2004)
[arXiv:hep-ph/0405093].

\bibitem{sahu&uma_prd.04} N.~Sahu and S.~Uma Sankar,
Phys. Rev. D{\bf 71}, 2005 (013006), [arXiv:hep-ph/0406065]. 

\bibitem{sahu&uma.05} N.~Sahu and S.~Uma Sankar,                          
Nucl.\ Phys.\ B {\bf 724},329 (2005), [arXiv: hep-ph/0501069].
 
\bibitem{scpv} G. Senjanovic, Nucl. Phys. B {\bf 153},
334 (1979); A.~Masiero, R.~N.~Mohapatra and R.~D.~Peccei,
  Nucl.\ Phys.\ B {\bf 192} (1981) 66,
J.~Basecq, J.~Liu, J.~Milutinovic and L.~Wolfenstein,
  Nucl.\ Phys.\ B {\bf 272}, 145 (1986).
 N.G.~Deshpande, J.~F.~Gunion, B.~Kayser
and F.~I.~Olness,
Phys.\ Rev.\ D {\bf 44}, 837 (1991); 
G.~Barenboim and J.~Bernabeu,
  Z.\ Phys.\ C {\bf 73} (1997) 321
  [arXiv:hep-ph/9603379],
Y.~Rodriguez and C.~Quimbay,
  Nucl.\ Phys.\ B {\bf 637}, 219 (2002)
  [arXiv:hep-ph/0203178].

\bibitem{susy_tev_group}
 L.~Boubekeur, T.~Hambye and G.~Senjanovic,
Phys.\ Rev.\ Lett.\  {\bf 93}, 111601 (2004)
[arXiv:hep-ph/0404038]; A.~Pilaftsis, T.E.J~Underwood,
Nucl.Phys. {\bf B692}, 303-345 (2004), [arXiv:hep-hep/0309342];
T.~Hambye, J.M.~Russel, S.M.~West, JHEP {\bf 0407}, 070 (2004),
[arXiv:hep-ph/0403183]. E.J.~Chun, [arXiv: hep-ph/0508050]
S.~Dar, Q.~Shafi and A.~Sil,
[arXiv:hep-ph/0508037].
A.~Abada, H.~Aissaoui and M.~Losada,
Nucl.\ Phys.\ B {\bf 728}, 55 (2005)[arXiv:hep-ph/0409343].

\bibitem{huber&schimdt.01} S.J.~Huber, M.G.~Schmidt, 
Nucl. Phys. B {\bf 606}, 83 (2001).
 
\bibitem{abel_sar_white.95} S.A.~Abel, S.~Sarkar, R.L.~White,
Nucl. Phys. B {\bf 454}, 663-681 (1995).
 
\bibitem{pan_tam_PLB1.99} C.~Panagiotakopoulos, K.~Tamvakis,
Phys. Lett. B {\bf 446},224 (1999).
 
\bibitem{pan_tam_PLB2.99} C.~Panagiotakopoulos, K.~Tamvakis,
Phys. Lett. B {\bf 469},145, (1999).
 
\bibitem{pan&pil_PRD.01} C.~Panagiotakopoulos, A.~Pilaftsis,
Phys. Rev. D {\bf 63} (2001) 055003.
 
\bibitem{ded_hug_mor_tam_PRD.01} A.~Dedes, C.~Hugonie, S.~Moretti,
K.~Tamvakis, Phys. Rev. D {\bf 63}, 055009 (2001).
 
\bibitem{men_mor_wag_prd.04} A.~Menon, D.E.~Morrissey, C.E.M.~
Wagner, Phys. Rev. D {\bf 70}, 035005 (2004),
[arXiv:hep-ph/0404184].

\bibitem{sahu&yajnik_prd.04} N.~Sahu and U.A.~Yajnik, 
Phy. Rev. D{\bf 71}, 023507 (2005),[arXiv:hep-ph/0410075]; 
[arXiv:hep-ph/0509285].                                                                                
\bibitem{har&tur.90}J.A.~Harvey and M.S.~Turner,
{\PRD {42}}, 3344 (1990).
 
\bibitem{fglp.91} W.~Fischler, G.F.~Guidice, R.G.~Leigh and 
S.~Pawan, {\PLB {258}} 45 (1991).
 
\bibitem{vilen&shell} A.~Vilenkin and E.P.S.~Shellard, {\it Cosmic
strings and other topological defects} (Cambridge University 
Press, 1994).                                                                  
\bibitem{jackiw&rebbi.76} R.~Jackiw and C.~Rebbi, Phys Rev. D 
{\bf 13}, 3398 (1976).

\bibitem{sbdanduay1} S.B.~Duari and U.A.~Yajnik,
{\PLB {326}} 21 (1994).
 
\bibitem{brandenetal} R.H.~Brandenberger, A.C.~Davis, and
M.~Trodden, {\PLB {335}} 123 (1994).

\bibitem{jackiw.81} R.~Jackiw and P.~Rossi, Nucl.~Phys. B 
{\bf 190}, 681 (1981).        

\bibitem{weinberg.81} E.J.~Weinberg, Phys.~Rev. D {\bf 24}, 
2669 (1981).                             

\bibitem{sahu&yajnik_plb.04} N.~Sahu and U.A.~Yajnik, 
Phys.~Lett. B {\bf 596}, 1 (2004).

\bibitem{pijush.82} P.~Bhattacharjee, T.W.B.~Kibble and N.~Turok,
Phys.~Lett. B {\bf 119}, 95 (1982). 

\bibitem{brandenberger&co.91} R.H.~Brandenberger, A.C.~Davis and
M.~Hindmarsh, Phys.~Lett. B {\bf 263}, 239 (1991).

\bibitem{riotto&lewis.94}H.~Lew and A.~Riotto, Phys.~Rev. D 
{\bf 49}, 3837 (1994).                     

\bibitem{jeannerot.96}  R.~Jeannerot, Phys.~Rev.~Lett. {\bf 77},
3292 (1996).            

\bibitem{pijush.98} P.~Bhattacharjee, Phys.~Rev.~Lett. {\bf 81}, 
260 (1998).              

\bibitem{bha_sahu_yaj_prd.04} P.~Bhattacharjee, N.~Sahu and 
U.A.~Yajnik; Phys.~Rev. D {\bf 70}, 083534 (2004), 
[arXiv:hep-ph/0406054]; 

\bibitem{strumia.04} G.F.~Giudice, A.~Notari,
M.~Raidal, A.~Riotto and A.~Strumia,
Nucl.\ Phys.\ B {\bf 685}, 89 (2004),
[arXiv:hep-ph/0310123]; 

\bibitem{sahu_bha_yaj_prep} N.~Sahu, P.~Bhattacharjee,
U.A.~Yajnik, [arXiv:hep-ph/0512350]. 

\bibitem{mns-matrix} Z.~Maki, M.~Nakagawa
and S.~Sakata, Prog.\ Theor.\ Phys.\  {\bf 28}, 870 (1962); 
B.~Pontecorvo,
  Sov.\ Phys.\ JETP {\bf 7}, 172 (1958)
  [Zh.\ Eksp.\ Teor.\ Fiz.\  {\bf 34}, 247 (1957)].
  B.~Pontecorvo,
  Sov.\ Phys.\ JETP {\bf 6}, 429 (1957)
  [Zh.\ Eksp.\ Teor.\ Fiz.\  {\bf 33}, 549 (1957)].

\bibitem{selfenergy_group}
M.~Flanz, E.~A.~Paschos, U.~Sarkar and J.~Weiss,
  Phys.\ Lett.\ B {\bf 389} (1996) 693
  [arXiv:hep-ph/9607310].
M.~Flanz, E.~A.~Paschos and U.~Sarkar,
  Phys.\ Lett.\ B {\bf 345} (1995) 248
  [Erratum-ibid.\ B {\bf 382} (1996) 447]
  [arXiv:hep-ph/9411366].
L.~Covi, E.~Roulet and F.~Vissani,
  Phys.\ Lett.\ B {\bf 384}, 169 (1996)
  [arXiv:hep-ph/9605319].

\bibitem{spergel.03}  D.N.~Spergel {\it et. al.}
Astrophys.J.Suppl. 148 (2003) 175 [astro-ph/0302209].
                                                                                
\bibitem{buc_gre_min.91}W.~Buchmuller, C. Greub and P. Minkwoski,
Phys. Lett. B{\bf 267}, 395 (1991).

\bibitem{donnell&sarkar.94} P. O'Donnell and U. Sarkar,
Phys. Rev. D {\bf 49} (1994) 2118.
 
\bibitem{lazarides&shafi.98} G.~Lazarides and Q.~Shafi,
Phys. Rev. D {\bf 58} (1998) 071702.

\bibitem{hamb-senj.03}T.~Hambye and G.~Senjanovic,
Phys. Lett. B {\bf 582}, 73 (2004).

\bibitem{ball.00} P.~Ball, J.M.~Frere and J.~ Matias,
Nucl. Phys. B {\bf 572}, 3 (2000).

\bibitem{mahanthapa.04}M.C.~Chen and K.T.~Mahanthappa,
[arXiv:hep-ph/0411158].

\bibitem{fritzsch.79}H.~Fritzsch, Nucl. Phys. {\bf B155},
189 (1979).

\bibitem{bilenkyetal.80} S.M.~Bilenky, J.~Hosek and S.T.~Petcov,
Phys. Lett. B{\bf 94}, 495 (1980).
 
\bibitem{rodejohan_npb.01} W.~Rodejohann, Nucl. Phys. B 
{\bf 597}, 110 (2001).
 
\bibitem{gonzalez-garcia_prd.03} M.C.~Gonzalez-Garcia and
C.~Pena-Garay, Phys. Rev.D {\bf 68}, 093003 (2003).

\bibitem{nimai.04} N.~Nimai Singh, M.~Patagiri, Mrinal Das,
[arXiv:hep-ph/0406075].

\bibitem{plumacher.97} M.~Plumacher,
Nucl.\ Phys.\ B {\bf 530}, 207 (1998).

\bibitem{cynr.02} J.M.~Cline, U.A.~Yajnik, S.N.~Nayak and
M.~Rabikumar, {\PRD  {66}}, 65001 (2002); 
J.~M.~Frere, L.~Houart, J.~M.~Moreno, J.~Orloff and M.~Tytgat,
  Phys.\ Lett.\ B {\bf 314}, 289 (1993)
  [arXiv:hep-ph/9301228].

\bibitem{leftright_group}J.C.~Pati and A.~Salam, Phys.~Rev.~D
{\bf 10}, 275 (1974); R.N.~Mohapatra and J.C.~Pati,
Phys.~Rev.~D{\bf 11}, 566(1975); Phys.~Rev.~D {\bf 11}, 2558 (1975);
R.N.~Mohapatra and G.~Senjanovic, Phys.~Rev.~D{\bf 12}, 1502
(1975).

\bibitem{buch_bari_plum.03} W.~Buchmuller, P.~Di Bari and
M.~Plumacher,
Nucl.\ Phys.\ B {\bf 665}, 445 (2003)
[arXiv:hep-ph/0302092].
R.~Barbieri, P.~Creminelli, A.~Strumia and N.~Tetradis,
  Nucl.\ Phys.\ B {\bf 575}, 61 (2000)
  [arXiv:hep-ph/9911315];

\bibitem{buch-bari-plum-ped.04} W.~Buchmuller, P.~Di Bari
and M.~Plumacher,
Annals Phys.\  {\bf 315}, 305 (2005),[arXiv:hep-ph/0401240]

\bibitem{ywmmc} U.A.~Yajnik, H.~Widyan, S.~Mahajan,
A.~Mukherjee and D.~Choudhuri {\PRD{59}}  (1999) 103508.

\bibitem{lazar}  T.W.B.~Kibble, G.~Lazarides and Q.~Shafi,
        {\PRD {26}}, 435 (1982),
        G.~Lazarides and Q.~Shafi, {\PLB {159}}, 261 (1985).
 
\bibitem{lew-rio} H.~Lew and A.~Riotto, {\PLB {309}}, 258 (1993).

\bibitem{jpt} M.~Joyce, T.~Prokopec and N.~Turok,
        {\PRL {75}}, 1695 (1995); [erratum-ibid 75,3375 (1995)].
 
\bibitem{clijokai}  J.M.~Cline, M.~Joyce and K.~Kainulainen
        {\PLB {417}} 79 (1998) \ [erratum: arXiv:hep-ph/0110031].
 
\bibitem{clikai} J.M.~Cline and K.~Kainulainen
        {\PRL {85}}, 5519 (2000).

\bibitem{rajaraman.82} R.~Rajaraman{\it Solitons and
Instantons} (Amestradam, North Holland, 1982).

\bibitem{sushreefer&hegger.80} W.P.~Su, J. R.~Schrieffer,
Nucl. Phys. B{\bf 190}, 253 (1981).
 
\bibitem{rice.79} M.J.~Rice, Phys. Lett. A{\bf 71}, 152 (1979).
 
\bibitem{rajaraman.01} R.~Rajaraman, [arXiv:cond-mat/0103366].
 
\bibitem{jackiw.99} R.~Jackiw, hep-th/9903255.
 
\bibitem{witten.85}E.~Witten, Nucl. Phys. B {\bf 249},557(1985).

\bibitem{wicketal.52} G.C.~Wick, A.S.~Wightman and 
E.P.~Wigner, Phys.~Rev.~{\bf 88}, 101 (1952).

\bibitem{pres&vil.92} J.~Preskill and A.~Vilenkin,
Phys.\ Rev.\ D {\bf 47}, 2324 (1993).

\bibitem{devega.78} H.~de Vega, Phys. Rev. D {\bf 18}, 
2932 (1978).
 
\bibitem{stern_prl.83} A.~Stern, Phys. Rev. Lett.{\bf 52}, 
2118(1983).
 
\bibitem{stern&yaj.86} A.~Stern and U.A.~Yajnik, Nucl. Phys. 
B {\bf 267}, 158 (1986), S. R. Das, Nucl. Phys. {\bf B227}, 462 (1983).

\bibitem{ganoulis.88} N.~Ganoulis, G.~Lazarides,
Phys.~Rev.~D {\bf 38} 547 (1988).

\bibitem{DHN_74} R.F.~Dashen, B.~Hasslacher and A.~Neveu
Phys.\ Rev.\ D {\bf 10}, 4138 (1974).
 
\bibitem{nohl.75} C.R.~Nohl, Phys.\ Rev.\ D  {\bf  12}, 1840 
(1975) .    

\bibitem{jackiw.77} R.~Jackiw, Rev. Mod. Phys. 49, 681(1977).

\bibitem{sud&yaj.86} E.C.G.~Sudarshan and U.A.~Yajnik, 
Phys. Rev. D {\bf 33}, 1830 (1986).

\bibitem{Fuk&Suz_book} See for instance M.~Fukugita 
and T.~Yanagida, in "\textsl{Physics and Astrophysics of 
Neutrinos}", M.~Fukugita and A.~Suzuki, (eds.), Springer-Verlag, 
(1994), pp. 1-248.

\bibitem{wilczek.84} R.~MacKenzie and F.~Wilczek,
Phys.\ Rev.\ D {\bf 30}, 2194 (1984).

\bibitem{Sweinberg_book} S. Weinberg, {\it The Quantum Theory 
of fields}, Vol. I, sec. 3.3,  Cambridge University Press, 1996.

\bibitem{Davis:1999ec}
S.C.~Davis, W.B.~Perkins and A.C.~Davis,
Phys.\ Rev.\ D {\bf 62}, 043503 (2000)

\bibitem{jeannerot&davis.95}  R.~Jeannerot and A.C.~Davis,
Phys.~Rev. D {\bf 52}, 7220 (1995).
 
\bibitem{jeannerot_prd.96}  R.~Jeannerot Phys.~Rev. D {\bf 53},
5426 (1996). 

\bibitem{sdavis.00}S.C.~Davis, Int.~J.~Theor.~Phys. {\bf 38}, 
2889 (1999); G.~Starkman, D.~Stojkovic and
T.~Vachaspati, Phys.\ Rev.\ D{\bf 63}, 085011 (2001),
Phys.~Rev.~D{\bf 65}, 065003 (2002).

\bibitem{pijushreport}P.~Bhattacharjee and G.~Sigl, Phys. Rept.
{\bf 327}, 109 (2000).

\bibitem{landriau-shellard} M.~Landriau and E.P.S.~Shellard, 
Phys.~Rev.D {\bf 69}, 023003 (2004).     

\bibitem{td_cmb} N.~Bevis, M.~Hindmarsh and M.~Kunz,
Phys.\ Rev.\ D {\bf 70}, 043508 (2004); R.~Durrer, M.~Kunz and
A.~Melchiorri, Phys.\ Rept.\ {\bf 364}, 1 (2002); C.~Contaldi,
M.~Hindmarsh and J.~Magueijo, Phys.\ Rev.\ Lett.\ {\bf 82}, 679 (1999).

\bibitem{vah} G.~Vincent, N.~Antunes and M.~Hindmarsh, 
Phys.~Rev.~Lett. {\bf 80}, 2277 (1998).

\bibitem{moore-shell} J.N.~Moore, E.P.S.~Shellard and 
C.J.A.P.~Martins, Phys.~Rev. D {\bf 65}, 023503 (2002).

\bibitem{kibble&turok-82}T.W.B.~Kibble and N.~Turok, Phys.~Lett.B 
{\bf 116}, 141 (1982).     

\bibitem{turok-84}N.~Turok, Nucl. Phys. B {\bf 242}, 520 (1984).     

\bibitem{chendicarlo&hotes-88}A.L.~Chen, D.A.~DiCarlo and
S.A.~Hotes, Phys.~Rev. D {\bf 37}, 863 (1988). 

\bibitem{delaneyetal-90} D.~DeLaney, K.~Engle and X.~Scheick,
Phys.~Rev. D {\bf 41}, 1775 (1990).
                                     
\bibitem{siemens&kibble.95} X.A.~Siemens and T.W.B.~Kibble,
Nucl.~Phys. B {\bf 438}, 307 (1995).

\bibitem{pijush&rana}  P.~Bhattacharjee and N.C.~Rana, Phys.~Lett. 
B {\bf 246}, 365 (1990).      

\bibitem{uhecr_obs} M. Takeda et al (AGASA Collaboration), 
Astropart.Phys. 19 (2003) 447; T. Abu-Zayyad et al 
(HiRes Collaboration),arXiv:astro-ph/0208243, astro-ph/0208301.   

\bibitem{sreekumar} P.~Sreekumar et al, Astrophys.~J. {\bf 494}, 
523 (1998).               

\bibitem{gu&ma.05} Pei-Hong Gu, Hong Mao, Phys.\ Lett.\ B
{\bf 619}, 226 (2005).

\bibitem{model}
B.~R.~Desai and A.~R.~Vaucher,
  Phys.\ Rev.\ D {\bf 63}, 113001 (2001);
  [arXiv:hep-ph/0007233].
J.~L.~Chkareuli and C.~D.~Froggatt,
  Phys.\ Lett.\ B {\bf 450}, 158 (1999)
  [arXiv:hep-ph/9812499].
see also D.~Falcone and F.~Tramontano in ref.~\cite{lep_phase}
\end{thebibliography}
\end{document}